\documentclass{aa}  
\usepackage{graphicx}
\usepackage{txfonts}
\usepackage{hyperref}
\begin{document} 

\title{Modeling the Ly$\alpha$ transit absorption of the hot Jupiter HD~189733b}

\author{P. Odert\inst{1,2}
\and
N.~V. Erkaev\inst{3,4}
\and
K.~G. Kislyakova\inst{5,1}
\and
H. Lammer\inst{1}
\and
A.~V. Mezentsev\inst{4}
\and
V.~A. Ivanov\inst{4}
\and \\
L. Fossati\inst{1}
\and
M. Leitzinger\inst{2}
\and
D. Kubyshkina\inst{1}
\and
M. Holmstr\"om\inst{6}
}

\institute{Space Research Institute, Austrian Academy of Sciences, Schmiedlstra{\ss}e 6, A-8042 Graz, Austria
\and
Institute of Physics/IGAM, University of Graz, Universit\"atsplatz 5, A-8010 Graz, Austria\\
\email{petra.odert@uni-graz.at}
\and
Institute of Computational Modelling of the Siberian Branch of the Russian Academy of Sciences, 660036 Krasnoyarsk, Russian Federation
\and
Siberian Federal University, 660041 Krasnoyarsk, Russian Federation
\and
Institute for Astronomy, University of Vienna, T\"urkenschanzstra{\ss}e 17, A-1180 Vienna
\and
Swedish Institute of Space Physics, PO Box 812, SE-98128 Kiruna, Sweden
}

\date{Received ???; accepted ???}

 
\abstract
{Hydrogen-dominated atmospheres of hot exoplanets expand and escape hydrodynamically due to the intense heating by the X-ray and extreme ultraviolet (XUV) irradiation of their host stars. Excess absorption of neutral hydrogen has been observed in the Ly$\alpha$ line during transits of several close-in gaseous exoplanets, indicating such extended atmospheres.}
{For the hot Jupiter HD~189733b, this absorption shows temporal variability. We aim to study if variations in stellar XUV emission and/or variable stellar wind conditions may explain this effect.}
{We applied a 1D hydrodynamic planetary upper atmosphere model and a 3D magnetohydrodynamic stellar wind flow model to study the effect of variations of the stellar XUV irradiation and wind conditions at the planet's orbit on the neutral hydrogen distribution. This includes the production of energetic neutral atoms (ENAs) and the related Ly$\alpha$ transit signature.}
{We obtain comparable, albeit slightly higher Ly$\alpha$ absorption than that observed in 2011 with a stellar XUV flux of $1.8\times10^4$\,erg\,cm$^{-2}$\,s$^{-1}$, rather typical activity conditions for this star. Flares with parameters similar to that observed eight hours before the transit are unlikely to have caused a significant modulation of the transit signature. We find that the resulting Ly$\alpha$ absorption is dominated by atmospheric broadening, whereas the contribution of ENAs is negligible, as they are formed inside the bow shock from decelerated wind ions that are heated to high temperatures. Thus, within our modeling framework and assumptions, we find an insignificant dependence of the absorption on the stellar wind parameters.}
{Since the transit absorption can be modeled with typical stellar XUV and wind conditions, it is possible that the nondetection of the absorption in 2010 was affected by less typical stellar activity conditions, such as a very different magnitude and/or shape of the star's spectral XUV emission, or temporal and/or spatial variations in Ly$\alpha$ affecting the determination of the transit absorption.}


\maketitle
%

\section{Introduction}

The first observational evidence of an expanding and escaping exoplanetary atmosphere was the Ly$\alpha$ transit absorption detected for the hot Jupiter HD~209458b \citep{Vidal-Madjar03}. Subsequently, more observations of excess absorption in Ly$\alpha$ and UV lines of other elements succeeded, the most prominent ones being the hot Jupiters HD~209458b \citep{Vidal-Madjar04,  Vidal-Madjar08, Vidal-Madjar13, Ben-Jaffel07, Ben-Jaffel08, Ballester07, Ehrenreich08, Linsky10, Ben-Jaffel10, Ballester15}, HD~189733b \citep{LecavelierdesEtangs10, LecavelierdesEtangs12, Bourrier13, Ben-Jaffel13}, WASP-12b \citep{Fossati10, Haswell12}, 55~Cnc~b \citep{Ehrenreich12}, and the warm Neptune GJ~436b \citep{Kulow14, Ehrenreich15, Lavie17}. Numerous hydrodynamic models for hot Jupiter upper atmospheres were developed, aiming to explain the observations \citep{Yelle04, Tian05, GarciaMunoz07, Penz08b, Stone09, Murray-Clay09, Guo11, Guo13, Koskinen13, Shaikhislamov14, Salz16a, Guo16, Erkaev16, Khodachenko17, Debrecht18}. Other studies employed Direct Simulation Monte Carlo (DSMC) models, which include the generation of energetic neutral atoms (ENAs) due to interaction with the stellar wind and take into account the effects of radiation pressure on the shaping of the large hydrogen clouds \citep{Holmstroem08, Ekenbaeck10, Bourrier13a, Kislyakova14b, Bourrier15, Bourrier16}.

It is therefore important to study the interaction of the expanding planetary atmosphere with the magnetized stellar wind. The escaping atmospheric particles mixed with stellar wind plasma can have a strong influence on the wind plasma parameters in the vicinity of the planet. The main effect of an intrinsic planetary magnetic field on atmospheric escape is to suppress the outflow and to make it highly anisotropic \citep{Adams11, Trammell11, Trammell14, Owen14, Khodachenko15a}. Many studies investigated the interaction between a close-in planet and its host star's wind, but some of them neglected magnetic fields and just considered a purely hydrodynamic interaction \citep{Stone09, Bisikalo13a, Tremblin13, Christie16}. Other studies applied magnetohydrodynamic (MHD) models instead \citep{Cohen11, Matsakos15, Tilley16}, but employed mostly simplified descriptions of the planetary wind. \citet{Shaikhislamov16} used a 2D multi-fluid code to study the interaction of a non-magnetized HD~209458b-like hot Jupiter with the stellar wind, taking into account heating by the stellar X-ray and extreme ultraviolet (XUV) flux and hydrogen photochemistry to self-consistently model the planetary outflow. However, they did not include the interplanetary magnetic field (IMF) and its effect on the formation of the planetary obstacle. This was addressed by \citet{Erkaev17}, who used a 1D hydrodynamic planetary upper atmosphere model in combination with a 3D MHD stellar wind flow model to investigate the build-up of a planetary obstacle by the interaction of the partially ionized planetary wind with the plasma flow of a magnetized stellar wind.

The hot Jupiter HD~189733b was discovered by \citet{Bouchy05} based on both radial velocity and photometric transit observations. The most recent determinations of its mass and radius are $1.13\,M_\mathrm{Jup}$ and $1.13\,R_\mathrm{Jup}$, based on parallaxes from the first data release of \textit{Gaia} \citep{Stassun17}. The host star HD~189733 has a spectral type of K2V \citep{Gray03} and is the primary of a double system with a mid-M dwarf companion located at a projected separation of $\sim$216\,AU \citep{Bakos06}. The activity level and rotation period \citep[11.95\,d;][]{Henry08} of HD~189733, comparable to the similar K star $\epsilon$~Eri, indicate an age of 1-2\,Gyr \citep{Poppenhaeger14}. However, the accompanying M dwarf is rather inactive and has an estimated age of ${>}5$\,Gyr. It was therefore suggested that HD~189733's high rotation rate and associated strong magnetic activity may be caused by interactions with its hot Jupiter \citep{Poppenhaeger14}. Independent of the reasons for the star's high activity level, HD~189733b is exposed to the intense stellar XUV emission, and possibly also to a dense and fast stellar wind. This may lead to enhanced atmospheric losses, which is also indicated by UV transit observations. Moreover, \citet{Pillitteri15} suggested that the observed FUV variability could stem from accretion of matter from the planet onto the star. \citet{Route19} analyzed multiwavelength data of HD~189733 to find possible indications of star-planet interaction, but could not identify a clear relation between the stellar activity characteristics and the planetary orbital phases.

\citet{LecavelierdesEtangs10} observed three planetary transits and detected an absorption of $5.05\pm0.75$\% in the unresolved Ly$\alpha$ line, significantly higher than the absorption by the planetary disk at optical wavelengths. Later, \citet{LecavelierdesEtangs12} obtained Ly$\alpha$ observations with higher spectral resolution at two different epochs (April 2010, September 2011). While during the first epoch no excess absorption could be detected (just absorption of the total flux comparable to the 2.4\% absorption by the planetary disk at optical wavelengths, but no spectrally resolved absorption), they found absorption of $14.4\pm3.6$\% (i.e., $12.3\pm3.6$\% excess absorption) in the blue wing of the line in a velocity range of $-230$ to $-140$\,km\,s$^{-1}$, indicating absorbing material moving away from the star. They also found absorption in the red wing from 60 to 110\,km\,s$^{-1}$ of $7.7\pm2.7$\%, meaning $5.5\pm2.7$\% excess, indicating absorbing material moving towards the star, but this detection was not statistically significant. Using Monte Carlo simulations, \citet{Bourrier13a} were able to reproduce the excess absorption observed in 2011 with a neutral hydrogen outflow rate of $4\times10^8-10^{11}$\,g\,s$^{-1}$, an ionizing flux of 6-23 times the solar value, a stellar wind with a temperature of $3\times10^4$~K, a velocity of $200\pm20$\,km\,s$^{-1}$, and a density between $10^3-10^7$\,cm$^{-3}$. \citet{LecavelierdesEtangs12} suggested that the absence of excess absorption in 2010 could be due to a much lower escape rate or a less dense stellar wind. They explained the discrepancy between the observations in 2010 and 2011 by the influence of a flare that occurred $\sim$8\,h prior to the transit in 2011. \citet{Bourrier13} presented a more detailed analysis of the 2011 observations, reaching similar conclusions.

Further attempts were made to explain the temporal variability of the Ly$\alpha$ transit absorption of HD~189733b. \citet{Guo16} investigated the effect of the stellar XUV spectral energy distribution (SED) on atmospheric profiles and mass-loss rates. They found that the mass-loss rate is mainly determined by the total absorbed energy, whereas the ionization is strongly affected by the SED. For SEDs dominated by the low-energy part of the spectrum, the H/H$^+$ transition moves closer to the planet, and the amount of H atoms at a certain altitude can differ by 1-2 orders of magnitude, in comparison to SEDs dominated by the high-energy part. They used the method of \citet{Ben-Jaffel08} to investigate the differences in absorption signature depending on the stellar XUV SED and found that they can explain the differences in observations between 2010 and 2011 by assuming a harder stellar spectrum in 2011. The model of \citet{Ben-Jaffel08} assumes an extended thermosphere with an absorption profile broadened by natural broadening and does not include nonthermal H populations like ENAs. Therefore, the resulting transmission spectrum is symmetric around the Ly$\alpha$ line center and does not reproduce the localized absorption at higher velocities in the blue wing. Recently, \citet{Chadney17} studied the influence of flares on the upper atmospheres and escape rates of hot Jupiters. They found a maximum mass-loss enhancement of about a factor of two, and much less if the limited duration of the radiation enhancement during a flare is taken into account. However, they suggested that an extreme proton event associated with the flare could have led to sufficiently enhanced escape.

\citet{Ben-Jaffel13} reported a $6.4\pm1.8$\% absorption in \ion{O}{i} and a marginal detection of early-ingress \ion{C}{ii} absorption. However, they could not exclude that the latter had a stellar or instrumental origin. Transit absorption in the H$\alpha$ line was also detected in several observations, as well as a pre-transit signature \citep{Jensen12, Cauley15, Cauley16, Cauley17a}. However, both in- and pre-transit absorption signatures were highly variable in time and strongly affected by stellar activity, making their interpretation difficult \citep{Barnes16a, Cauley17b, Kohl18}. Furthermore, a 6-8\% transit absorption in X-rays was reported by \citet{Poppenhaeger13}, but more observations with higher sensitivity are needed to confirm this result \citep{Marin17}.

In this paper, we investigate the possible causes for the variations of the Ly$\alpha$ transit absorption of HD~189733b. To achieve this, we modeled the structure of the upper atmosphere and the associated planetary mass loss, taking into account stellar XUV heating  at both quiescent and flaring conditions. Then, we modeled the interaction between the stellar wind and the planetary atmosphere, including the related production of ENAs, and its effect on the UV transit signature. In Section~\ref{sec:hd}, we describe the hydrodynamic model and show the resulting upper atmosphere profiles and planetary mass-loss rates in Section~\ref{sec:res}. In Section~\ref{sec:mhd}, we describe the MHD flow model and show the stellar wind flow maps. In Section~\ref{sec:abs}, we compute the Ly$\alpha$ absorption for the various scenarios and compare them with the observations. In Section~\ref{sec:disc}, we discuss our findings and compare them with other studies. Conclusions are presented in Section~\ref{sec:conc}.

\section{Hydrodynamic upper atmosphere modeling}\label{sec:hd}

\subsection{Model description}\label{sec:hdmodel}
The hydrodynamic model solves the time-dependent system of the equations of mass, momentum and energy conservation in 1D spherical geometry along the star-planet line,
\begin{align}
\frac{\partial \rho r^2}{\partial t} + \frac{\partial \rho u r^2}{\partial r} &= 0, \label{eq:hd1} \\
\frac{\partial \rho u r^2}{\partial t} + \frac{\partial \left(\rho u^2 + p\right) r^2}{\partial r} &= g\rho r^2 + 2pr, \label{eq:hd2} \\
\frac{\partial E_\mathrm{th} r^2}{\partial t} + \frac{\partial E_\mathrm{th} u r^2}{\partial r} &= Qr^2 + \frac{\partial}{\partial r} \left(\kappa r^2 \frac{\partial T}{\partial r}\right) - p\frac{\partial ur^2}{\partial r} \label{eq:hd3}.
\end{align}
The gas parameters $\rho$, $u$, $T$, $E_\mathrm{th}=p/(\gamma-1)$ are the mass density, velocity, temperature, and thermal energy of the upper atmosphere. The gas pressure is $p=\rho R T/\mu$, with the mean molecular weight $\mu$ and the gas constant $R$. The distance from the planet's center is $r,$ and $t$ is the time. For the specific heat ratio $\gamma,$ we adopted 5/3 for monatomic gas. The gravitational force $g = -\partial\Phi/\partial r$ was derived from the Roche potential $\Phi$ along the star-planet line:
\begin{equation}
\Phi = -\frac{GM_\mathrm{p}}{r} - \frac{GM_*}{a-r} - \frac{G\left(M_\mathrm{p}+M_*\right)}{2a^3} \left(a\frac{M_*}{M_\mathrm{p}+M_*}-r\right)^2
\end{equation}
\citep[e.g.,][]{GarciaMunoz07}. Here, $G$ is the gravitational constant, $M_\mathrm{p}$ the planet's mass, $M_*$ the stellar mass, and $a$ the star-planet separation. The right-hand side of Eq.~\ref{eq:hd3} includes the net volume heating rate $Q$ and thermal conduction, which are both described in Section~\ref{sec:hc}.

The simulations assumed an atmosphere composed entirely of hydrogen, meaning we neglected helium and other heavier elements, which are minor compared to hydrogen. For HD~189733b, we also neglected molecular species present in the lower atmosphere, but the strong ionization of the upper atmosphere is expected to destroy them efficiently \citep[cf.,][and Appendix~\ref{sec:h2}]{Guo16}. Therefore, we only considered hydrogen atoms (H) and protons (H$^+$). The production of H$^+$ is calculated as
\begin{equation}
\frac{\partial n_\mathrm{H+}}{\partial t} + \frac{1}{r^2}\frac{\partial n_\mathrm{H+} u r^2}{\partial r} = \alpha_\mathrm{ion}n_\mathrm{H} - \alpha_\mathrm{rec}n_\mathrm{H+}n_\mathrm{e} + \alpha_\mathrm{col}n_\mathrm{H}n_\mathrm{e},
\end{equation}
where $\alpha_\mathrm{ion}$ is the photoionization rate, $\alpha_\mathrm{rec}$ the radiative recombination rate, and $\alpha_\mathrm{col}$ the collisional ionization rate (all quantities in cgs units). The parameters $n_\mathrm{H}, n_\mathrm{H+}, n_\mathrm{e}$ are the number densities of neutral hydrogen atoms, protons, and electrons, respectively. We assumed quasi-neutrality, $n_\mathrm{H+} = n_\mathrm{e}$. The photoionization rate is given by
\begin{equation}\label{eq:phion}
\alpha_\mathrm{ion}(r) = \sum_{\lambda} \sigma_\mathrm{ion,\lambda}\frac{F_\mathrm{XUV,\lambda}\ \mathrm{e}^{-\tau_\lambda(r)}}{E_{\lambda}},
\end{equation}
where $\sigma_\mathrm{ion,\lambda}$ is the photoionization cross-section, $F_\mathrm{XUV,\lambda}$ the spectral XUV flux at the planet's orbit outside of the atmosphere, $\tau_\lambda(r)$ the optical depth at distance $r$ from the planet's center, and $E_{\lambda}$ the photon energy \citep[e.g.,][]{Murray-Clay09}. We computed the average photoionization cross-sections per spectral bin from the fits of \citet{Verner96a} and adopted photon energies corresponding to the central wavelengths of the bins. The recombination and collisional ionization rates were taken from \citet{Glover07}. For the former, we adopted the case~B\footnote{The case B recombination coefficient takes into account that photons emitted by recombinations to the ground level immediately ionize a nearby atom and do not effectively contribute to recombination.} coefficient
\begin{equation}
\alpha_\mathrm{rec} = 2.753\times10^{-14} \left(\frac{315614}{T}\right)^{1.5} \left[1 + \left(\frac{115188}{T}\right)^{0.407}\right]^{-2.242},
\end{equation}
while the latter is given by
\begin{align}
\alpha_\mathrm{col} &= \exp\left(-3.271396786\times10^{1} + 1.35365560\times10^{1}\theta \right. \\
&- 5.73932875\times10^{0}\theta^2 + 1.56315498\times10^{0}\theta^3 \\
&- 2.87705600\times10^{-1}\theta^4 + 3.48255977\times10^{-2}\theta^5 \\
&- 2.63197617\times10^{-3}\theta^6 + 1.11954395\times10^{-4}\theta^7 \\
&- \left. 2.03914985\times10^{-6}\theta^8\right),
\end{align}
where $\theta = \ln T_\mathrm{e}$ with the electron temperature $T_\mathrm{e}$ in eV.

The system of equations is normalized as described in \citet{Erkaev16} and solved using the MacCormack scheme \citep{MacCormack69}. We slightly modified the code in comparison with the version described in \citet{Erkaev16}. Mainly, we extended the scheme with the total variation diminishing (TVD) property. The code extensions are described in detail in Appendix~\ref{sec:tvd}. Another modification is introduced here. Our initial modeling attempts of HD~189733b resulted in unphysical behavior close to the lower boundary of the computational domain. Specifically, the velocity was increasing towards $r_0$, which violated the mass conservation. This happened mainly because the outflow of HD~189733b's upper atmosphere remains subsonic up to the L1 point, and is therefore highly subsonic close to the lower boundary. Schemes like the MacCormack method are not well suited to describe such flows. As the kinetic energy is very small for highly subsonic flows, we modified the code to solve for the thermal energy instead of the total energy. We thereby discretized the third term on the right-hand side of Eq.~\ref{eq:hd3} equally to the flux terms in both MacCormack steps.

The code evolves the system of equations until a steady-state solution is achieved. We assumed that this condition is fulfilled when the mass flux throughout the simulation domain is approximately constant, meaning mass conservation is fulfilled to 1\%. Moreover, we compared the total heating and cooling rates to confirm that energy conservation is fulfilled as well (cf. Appendix~\ref{sec:cons}).

\subsection{Heating and cooling processes} \label{sec:hc}
The main source of heating is the stellar XUV radiation. The XUV volume heating rate can be written as
\begin{equation}\label{eq:qxuv}
Q_\mathrm{XUV}(r) = \sum_\lambda \eta_\mathrm{ph} \sigma_\mathrm{ion,\lambda} n_\mathrm{H}(r) \frac{(E_\lambda{-}13.6\mathrm{eV})}{E_\lambda} F_\mathrm{XUV,\lambda}\frac{\exp{(-\tau_\lambda(r))}}{1+\alpha\tau_\lambda(r)},
\end{equation}
where the term $1+\alpha\tau_\lambda(r)$ takes into account 2D effects of the energy absorption in an approximate way \citep{Sekiya80a}. The optical depth is calculated as
\begin{equation}
\tau_\lambda(r) = \int_r^\infty \sigma_\mathrm{ion,\lambda} n_\mathrm{H}(r)\,dr.
\end{equation}
We adopted a constant photoelectron heating efficiency $\eta_\mathrm{ph}$ of 50\%. This is intermediate between the values calculated for Jupiter \citep[63\%;][]{Waite83} and HD~209458b \citep[20-40\%;][]{Shematovich14}, and was adopted by studies similar to ours \citep{Shaikhislamov14, Shaikhislamov16}. The heating rate from Eq.~\ref{eq:qxuv} yields very similar results to the 2D method described in previous papers \citep{Erkaev13, Erkaev16} for $\alpha=4$. We reverted to a 1D calculation of $Q_\mathrm{XUV}$, because we introduced the usage of XUV spectra here, motivated by the previously demonstrated effects of the assumed SED \citep{Guo16}. This also allowed inclusion of X-ray heating, which is important for hot Jupiters. The calculation of a non-gray 2D heating function would be computationally more expensive, but the resulting $Q_\mathrm{XUV}$ is very similar to the 1D method shown in Eq.~\ref{eq:qxuv} (see Appendix~\ref{sec:1d2d} for a comparison). The adopted XUV spectra are described in Section~\ref{sec:params}.

We included several cooling processes, namely Ly$\alpha$ cooling (H excitation), radiative recombination, free-free emission (Bremsstrahlung), and collisional ionization. All cooling rates were taken from \citet{Glover07} and are given in cgs units. The Ly$\alpha$ cooling rate
\begin{equation}
\Lambda_\mathrm{Ly\alpha} = 0.1\times7.5\times10^{-19} \left(1+\sqrt{\frac{T}{10^5}}\right)^{-1} \exp\left(-\frac{118348}{T}\right) n_\mathrm{e}n_\mathrm{H}
\end{equation}
was multiplied by a factor of 0.1, as suggested by \citet{Koskinen13} based on detailed calculations of the photon escape probability in the upper atmosphere of HD~209458b \citep{Menager13}, to account for the optical thickness. The cooling rate by radiative recombination is
\begin{equation}
\Lambda_\mathrm{rec} = 1.38\times10^{-16} T \alpha_\mathrm{rec} n_\mathrm{e}n_\mathrm{H+},
\end{equation}
by collisional ionization
\begin{equation}
\Lambda_\mathrm{col} = 2.179\times10^{-11} \alpha_\mathrm{col} n_\mathrm{e}n_\mathrm{H},
\end{equation}
and by free-free emission
\begin{equation}
\Lambda_\mathrm{ff} = 1.426\times10^{-27} T^{1/2} g_\mathrm{ff} n_\mathrm{e}n_\mathrm{H+},
\end{equation}
with $g_\mathrm{ff} = 0.79464 + 0.1243\log_{10}(T)$. The net heating rate is then computed via
\begin{equation}\label{eq:qnet}
Q_\mathrm{net} = Q_\mathrm{XUV} - (\Lambda_\mathrm{Ly\alpha}+\Lambda_\mathrm{rec}+\Lambda_\mathrm{col}+\Lambda_\mathrm{ff}).
\end{equation}

Thermal conduction is also included in the model. The conductivity coefficient is calculated as $\kappa = n^{-1}\sum_{j}n_j\kappa_j$ \citep[e.g.][]{GarciaMunoz07}, where $n_j$ are the number densities of the constituents and $\kappa_j = A_jT^{s_j}$, where $A_j$ and $s_j$ are the fitting parameters for the individual conductivities. We adopted $A_\mathrm{H}=379$, $A_\mathrm{H+}=7.37\times10^{-8}$, $A_\mathrm{e}=1.2\times10^{-6}$, $s_\mathrm{H}=0.69$, and $s_\mathrm{H+}=s_\mathrm{e}=2.5$ \citep{GarciaMunoz07}. We note that the contribution of H$^+$ is negligible compared to that of the electrons.

We computed the total heating and cooling rates to confirm the energy conservation in our code (Appendix~\ref{sec:cons}). The total cooling rate consists of the four explicitly included cooling processes described above, in addition to adiabatic cooling, which is implicitly included in Eq.~\ref{eq:hd3}. Adiabatic cooling includes the contributions of advection
\begin{equation}
\Lambda_\mathrm{ad} = \frac{1}{r^2} \dfrac{\partial E_\mathrm{th}ur^2}{\partial r}, 
\end{equation}
and expansion
\begin{equation}
\Lambda_\mathrm{ex} = \frac{p}{r^2} \dfrac{\partial(ur^2)}{\partial r} 
\end{equation}
\citep[e.g.][]{Salz16a}. The total heating rate is mostly dominated by XUV heating (Eq.~\ref{eq:qxuv}), but locally, advection can also contribute significantly, especially at larger heights. We note that conduction can also both cool and heat the gas, but we found it to be negligible, because its contribution to the energy balance is significantly smaller than the other processes on this specific planet.

\subsection{Initial and boundary conditions} \label{sec:bc}
The simulation domain extends from the planet's optical transit radius $r_0=R_\mathrm{p}$ to $r_1=4.5R_\mathrm{p}$. The upper boundary lies near the L1 point $R_\mathrm{L1} = (\mu-\mu^2/3)a\sim4.3R_\mathrm{p}$, where $\mu = (M_\mathrm{p}/(3M_*))^{1/3}$ \citep[e.g.,][]{Erkaev07}. We note that due to the high gravity of HD~189733b, the distance between $R_\mathrm{p}$ and typical mesopause pressure levels is very small. We used a nonuniform grid $r_i = r_1^{i/(N-1)}$ with typically $N=5000$ grid points. This high number is necessary because of the steep density gradient and the small velocities near the inner boundary that are present on this planet.

At the lower, subsonic inflow boundary, we fixed the values for temperature $T_0$ and number density $n_0$. For the temperature, we adopted the equilibrium temperature of 1200\,K (cf., Section~\ref{sec:params}) and for the number density a value of $10^{15}\,\mathrm{cm}^{-3}$. This corresponds to a pressure of 0.16\,mbar. Changing $T_0$ to 800\,K or 1600\,K decreases or increases, respectively, the mass-loss rate by only $\sim$10\%, and affects the atmospheric profiles negligibly (cf., Section~\ref{sec:profiles}). We checked that the assumed density $n_0$ is chosen large enough so that the optical depth in all XUV spectral bins exceeds 10. Only then is the stellar XUV flux completely absorbed in the simulation domain, and the resulting mass-loss rates do not strongly depend on the choice of $n_0$. Smaller values of $n_0$ would lead to significant ionization below the optical transit radius, which would be inconsistent with its measurement (see Section~\ref{sec:compare}). For runs that resulted in very subsonic flows, we set $u{=}0$ at the lower boundary. This does not affect the results, but improves the speed of convergence and the quality of the atmospheric profiles \citep[e.g.,][]{Shaikhislamov14}. At the upper boundary, we adopted free outflow (i.e., zero gradient) conditions on all parameters.

As initial conditions, we used a constant temperature and a monotonically increasing velocity profile $u(r) = 0.1(r-1)$. For the number density, we assumed hydrostatic conditions $n(r) = \exp(\Phi(r)-\Phi(r_0))$, but modified it in the upper region to $n(r)\propto 1/r^{2}$. This was found to be necessary because of the steep density gradient from the hydrostatic solution for this high-gravity planet. Initially, we assumed that the atmosphere consists only of atomic H, meaning $n_\mathrm{H+}(r) = 0$. Practically, most runs were started from previous solutions with similar parameter sets to reduce the computing time. We considered a run as properly converged when the mass flux $\rho u r^2$ is spatially constant within 1\% throughout the computational domain (excluding a small region close to the lower boundary where the mass flux may drop to zero for numerical reasons; cf., Fig.~\ref{fig:cons}). We checked that the adopted initial conditions and the chosen resolution of the grid do not affect the results. Moreover, we verified that the chosen location of the upper boundary does not affect the results.

\subsection{Adopted physical parameters}\label{sec:params}
The adopted planetary and stellar parameters (Table~\ref{tab:param}) were taken from \citet{Stassun17} and are based on the accurate parallaxes from the first data release of \textit{Gaia} \citep{GaiaCollaboration16a}. The distance $d$, stellar mass $M_*$ and radius $R_*$, stellar bolometric flux at Earth $F_\mathrm{bol}$, planetary mass $M_\mathrm{p}$ and radius $R_\mathrm{p}$, as well as the orbital inclination $i$ were taken from their study. The orbital separation $a$ was calculated from their quoted value of the parameter $a/R_*=8.84\pm0.27$ and the stellar radius. The average orbital velocity is therefore 151\,km\,s$^{-1}$, consistent with the maximum orbital radial velocity of $154^{+4}_{-3}$\,km\,s$^{-1}$ measured by \citet{deKok13}. Using the measured bolometric stellar flux, we find an equilibrium temperature $T_\mathrm{eq}=\left(F_\mathrm{bol}(1-A)/(f\sigma)\right)^{1/4}$ of 1200\,K, assuming full redistribution ($f=4$) and zero albedo ($A=0$). Here, $\sigma$ is the Stefan-Boltzmann constant. This is similar to, but slightly higher than, the apparent effective dayside temperature of $1163\pm37$~K \citep{Schwartz17}. Such small differences in the adopted $T_\mathrm{eq}$ do not affect our results.

\defcitealias{Linsky14}{L14}
\defcitealias{Sanz-Forcada11}{SF11}

\begin{figure}
\centering
\includegraphics[width=\columnwidth]{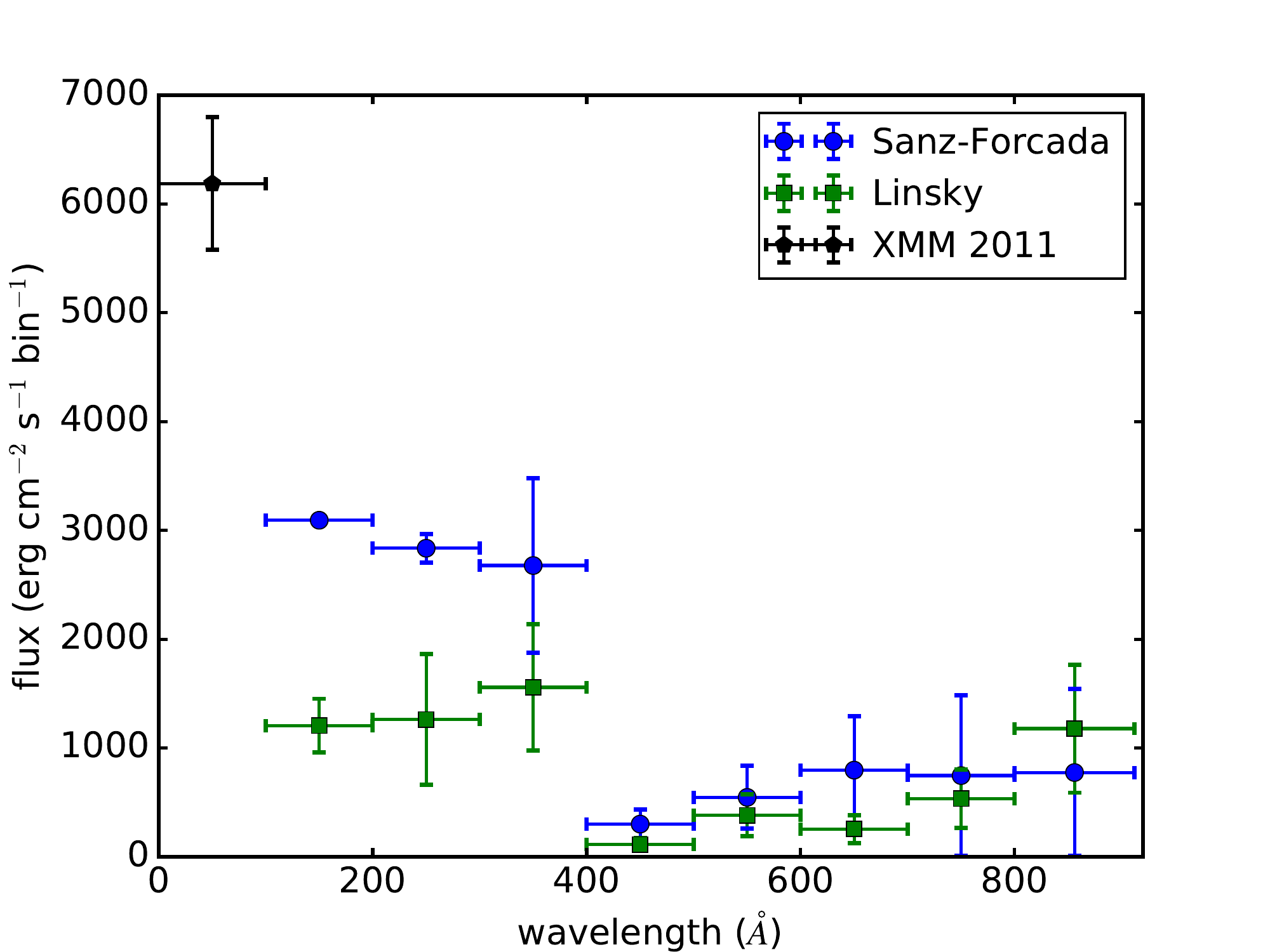}
\caption{XUV (5-912\AA) spectral flux at the orbit of HD~189733b in bins of 100\,\AA, obtained with the methods of \citetalias{Linsky14} and \citetalias{Sanz-Forcada11}. The X-ray flux is an average value outside of flares obtained with XMM-Newton in 2011 \citep{Pillitteri14}.}
\label{fig:xuv}
\end{figure}

The XUV flux of HD~189733 is strongly variable due to the high activity of this star. Observed X-ray luminosities are in a range of $1.1{-}2.8\times10^{28}$\,erg\,s$^{-1}$, which were partly obtained from different instruments with slightly different bandpasses, but were also measured at different epochs \citep{Huensch99, Pillitteri10, Pillitteri11, Pillitteri14, LecavelierdesEtangs12, Poppenhaeger13}. Since the extreme ultraviolet (EUV) part of the stellar spectrum is largely unobservable due to absorption by the interstellar medium (ISM), it has to be inferred indirectly. We compared two approaches to estimate the unobservable EUV spectrum. Firstly, we used the scaling relations from \citet[][hereafter L14]{Linsky14} based on the intrinsic Ly$\alpha$ flux, as derived from observations of the short-wavelength part of the EUV range ($<$400\,\AA) in nearby stars, and solar models for $>$400\,\AA. Since Ly$\alpha$ fluxes are also affected by the ISM absorption, the intrinsic stellar line profile needs to be reconstructed. From HST observations of HD~189733 in 2010, the intrinsic Ly$\alpha$ flux at Earth amounts to $7.5\times10^{-13}$\,erg\,cm$^{-2}$\,s$^{-1}$ with quoted uncertainties of 15-30\% \citep{France13}. The reconstructed line profile from \citet{Bourrier13} for the 2011 observations corresponds to a flux of $6.8\times10^{-13}$\,erg\,cm$^{-2}$\,s$^{-1}$, slightly lower, but consistent with \citet{France13}, considering the typical uncertainties of the reconstruction. We adopted the latter value, because we are focusing on modeling the 2011 observations. This flux value, together with the scalings from \citetalias{Linsky14}, gives a total EUV (100-912\,\AA) flux at the planet's orbit of $6.5\times10^3$\,erg\,cm$^{-2}$\,s$^{-1}$ with a spectral energy distribution shown in Fig.~\ref{fig:xuv}.

Secondly, we examined the synthetic spectrum from the X-exoplanets\footnote{\url{http://sdc.cab.inta-csic.es/xexoplanets/jsp/homepage.jsp}} database \citep[][hereafter SF11]{Sanz-Forcada10a, Sanz-Forcada11}. These authors used observed X-ray and UV spectra, reconstructed the emission measure distribution, and used it as an input for coronal models to infer the unknown EUV part of the spectrum. We show their synthetic spectrum with the same binning as for \citetalias{Linsky14} in Fig.~\ref{fig:xuv}. The errorbars in $x$-direction give the width of the bins (100\,\AA), those along the $y$-axis the estimated flux errors based on the  uncertainties quoted in the respective studies. It is apparent that the \citetalias{Sanz-Forcada11} fluxes are systematically higher than the \citetalias{Linsky14} ones (except for the 800-912\,\AA\ bin, which is likely due to the peak of the Lyman continuum, which is included in the solar models used by \citetalias{Linsky14}, but not in \citetalias{Sanz-Forcada11}). The $\lambda{>}300$\,\AA\ range is consistent within the estimated uncertainties. The $\lambda{<}300$\,\AA\ range differs significantly. This could be due to shortcomings in the models of either study, and/or the fact that the X-ray spectrum used by \citetalias{Sanz-Forcada11} was taken at a different epoch (2007) than the Ly$\alpha$ flux we adopted for the \citetalias{Linsky14} method (2011). Some intrinsic stellar variability could therefore also be a cause of the differences. \citet{Pillitteri14} gave a comparison of the X-ray fluxes obtained with XMM-Newton at different epochs (2007=\citetalias{Sanz-Forcada11}, 2009, 2011, 2012) and the associated temperatures and emission measures. In 2011, the temperatures were comparable, but the emission measures and X-ray fluxes were higher than in 2007. However, the X-ray observations in 2011 were not taken at the same time as the Ly$\alpha$ observations. Simultaneous observations with \textit{Swift}/XRT during the 2011 transit yielded an X-ray (0.3-3\,keV) flux of $3.6\times10^{-13}$\,erg\,cm$^{-2}$\,s$^{-1}$ \citep{LecavelierdesEtangs12}, whereas the XMM-Newton (0.12-2.48\,keV) flux in 2007 was $3.4\times10^{-13}$\,erg\,cm$^{-2}$\,s$^{-1}$ \citepalias{Sanz-Forcada11}, and $3.2{-}3.9\times10^{-13}$\,erg\,cm$^{-2}$\,s$^{-1}$ in 2011 \citep[excluding the flare;][]{Pillitteri14}. Hereafter, we adopted the average XMM-Newton value (excluding the flare) from 2011, $3.55\times10^{-13}$\,erg\,cm$^{-2}$\,s$^{-1}$ \citep{Pillitteri14}, which amounts to $6.2\times10^{3}$\,erg\,cm$^{-2}$\,s$^{-1}$ at the orbit of HD~189733b.

The total EUV fluxes (100-912~\AA) at the planet's orbit obtained with the two different methods are $6.5\times10^3$\,erg\,cm$^{-2}$\,s$^{-1}$ for the \citetalias{Linsky14} method and $1.2\times10^4$\,erg\,cm$^{-2}$\,s$^{-1}$ for \citetalias{Sanz-Forcada11}. As a comparison, we also calculated the total EUV flux using the method of \citet{Chadney15}, which uses a scaling with X-ray flux. With our adopted X-ray flux, we obtain an orbital EUV flux of $1.1\times10^4$\,erg\,cm$^{-2}$\,s$^{-1}$, comparable to \citetalias{Sanz-Forcada11}. We note, however, that the \citet{Chadney15} scaling is based on the \citetalias{Sanz-Forcada11} coronal models. On the basis of the discrepancies between the different reconstruction methods, we estimate that the uncertainty in EUV flux is at least a factor of two. The total XUV (5-912\,\AA) fluxes at the planet's orbit from each method amount to $1.3\times10^4$ and $1.8\times10^4$\,erg\,cm$^{-2}$\,s$^{-1}$ for \citetalias{Linsky14} and \citetalias{Sanz-Forcada11}, respectively (cf., Table~\ref{tab:res}). We explore the influence of XUV flux variations further in Section~\ref{sec:flare}, where we model the effects of a flare.

\begin{table}
\caption{Stellar and planetary parameters from \citet{Stassun17}. The orbital distance $a$ was calculated from their quoted value of $a/R_*$ and $R_*$.}
\label{tab:param}
\centering
\begin{tabular}{ll}
\hline\hline
Parameter & Value \\
\hline
$d$ (pc) & $19.84\pm0.09$ \\
$M_*$ (M$_{\sun}$) & $0.79\pm0.08$ \\
$R_*$ (R$_{\sun}$) & $0.75\pm0.01$ \\
$F_\mathrm{bol}$ (erg\,cm$^{-2}$\,s$^{-1}$) & $2.68\times10^{-8}\pm3.94\times10^{-10}$ \\
$M_\mathrm{p}$ (M$_\mathrm{Jup}$) & $1.13\pm0.08$ \\
$R_\mathrm{p}$ (R$_\mathrm{Jup}$) & $1.13\pm0.01$ \\
$a$ (AU) & $0.031\pm0.001$ \\
$i$ (\degr) & 85.71 \\
\hline
\end{tabular}
\end{table}

\section{Hydrodynamic modeling results}\label{sec:res}

\subsection{Atmospheric profiles and mass-loss rate} \label{sec:profiles}
\begin{figure*}
$\begin{array}{cc}
\includegraphics[width=\columnwidth]{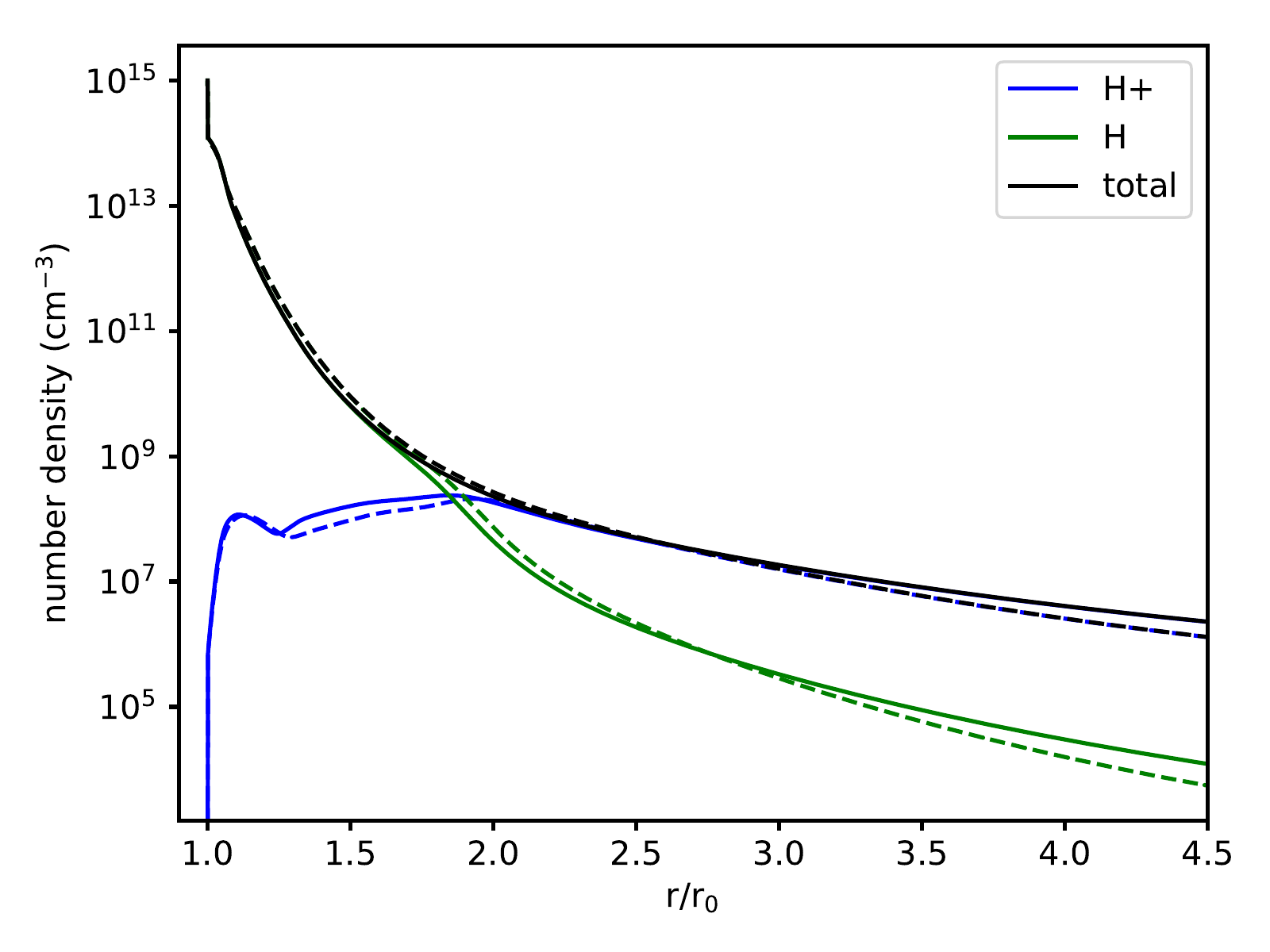} &
\includegraphics[width=\columnwidth]{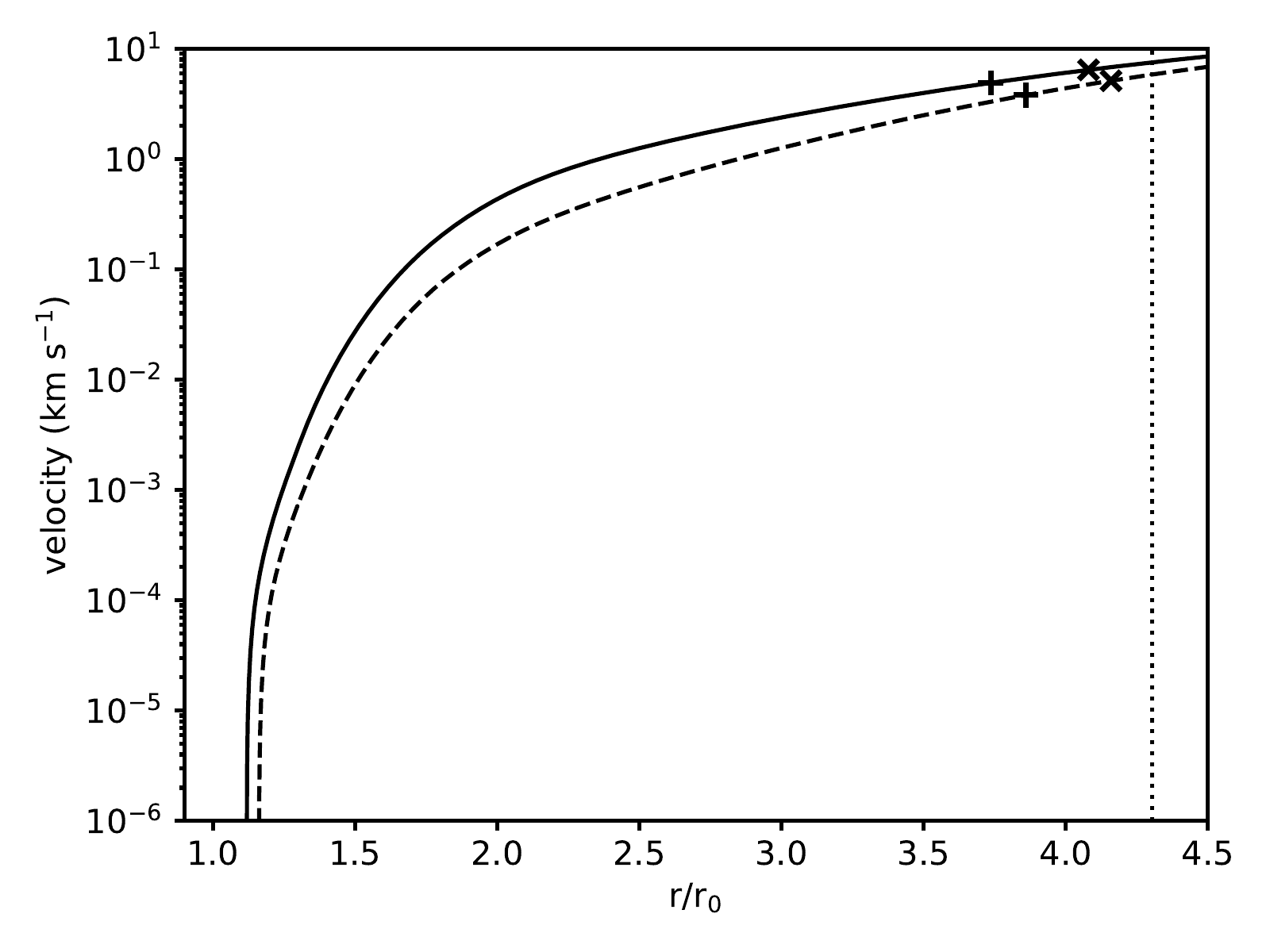} \\
\includegraphics[width=\columnwidth]{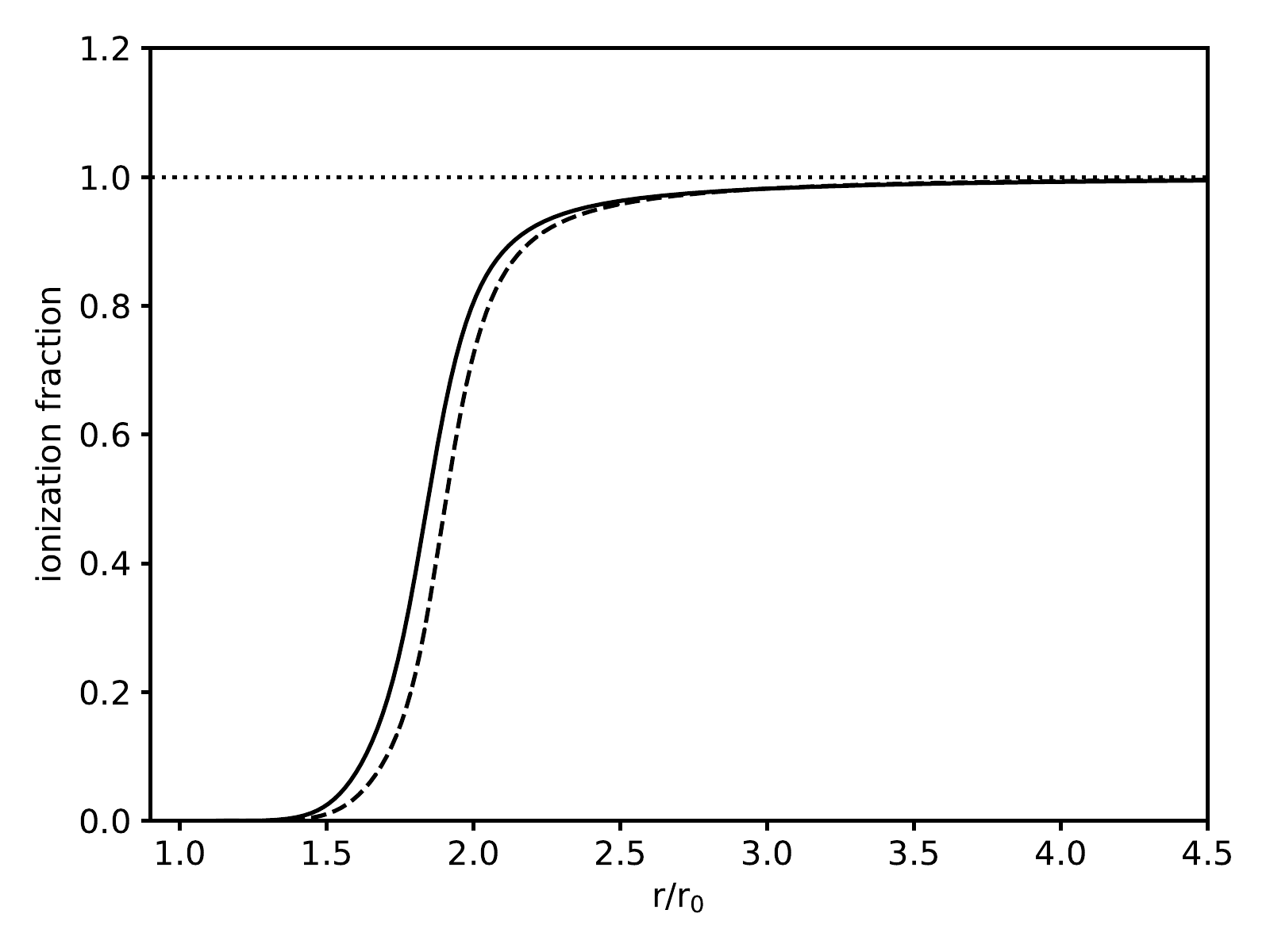} &
\includegraphics[width=\columnwidth]{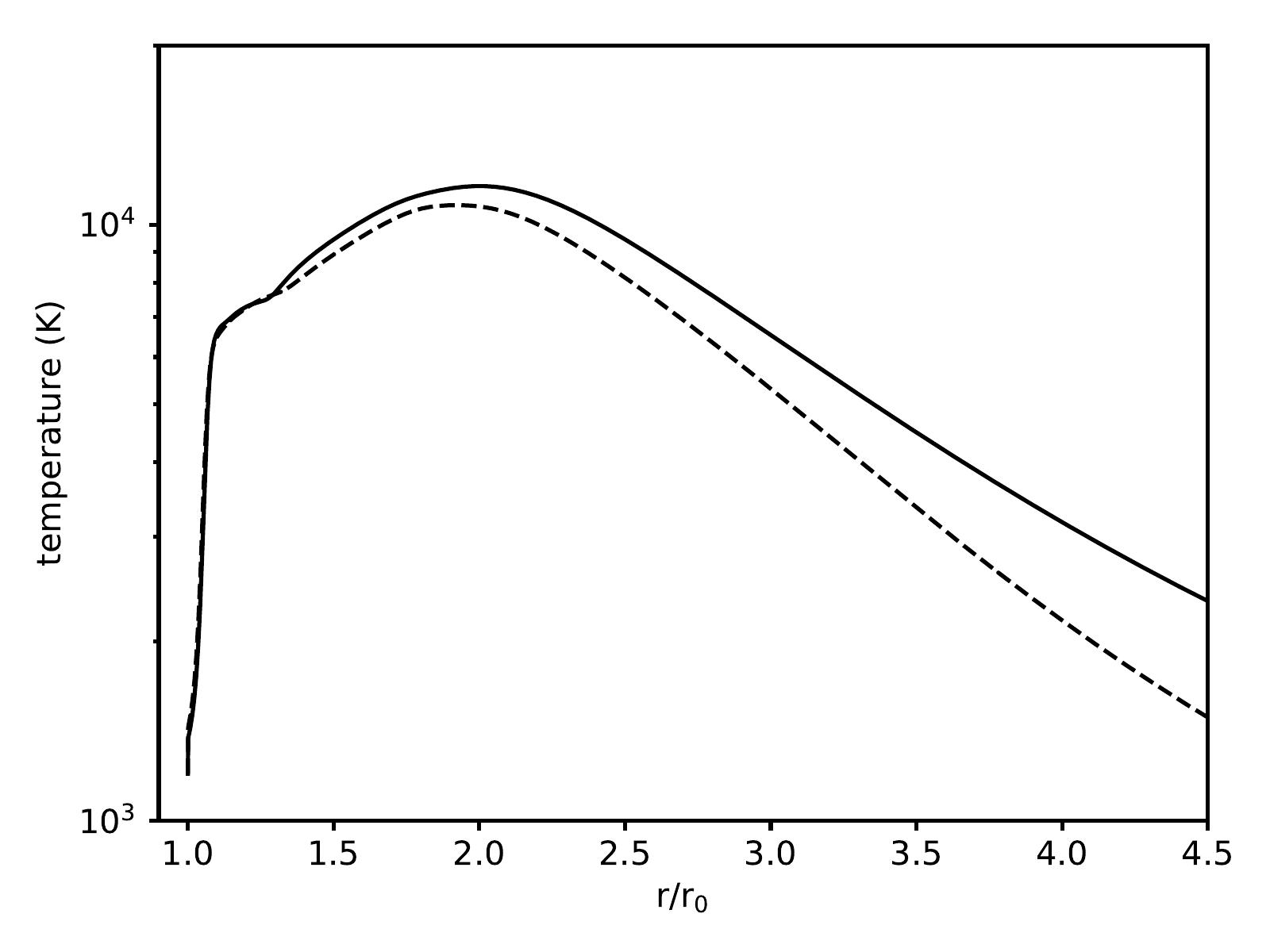}
\end{array}$
\caption{Planetary atmospheric profiles adopting the XUV spectra of \citetalias{Sanz-Forcada11} (solid) and \citetalias{Linsky14} (dashed). Shown are the number densities (total, atomic H, and protons), outflow velocity, ionization fraction, and temperature. In the velocity plot, crosses indicate the sonic points and plus signs denote the locations where the outflow velocity exceeds the escape velocity. The vertical dotted line indicates the location of the L1 point.}
\label{fig:profiles}
\end{figure*}

Here, we show the upper atmosphere profiles obtained using both the \citetalias{Sanz-Forcada11} and \citetalias{Linsky14} XUV spectra (Fig.~\ref{fig:profiles}). The isotropic (i.e., maximum) mass-loss rates $\dot{M}=4\pi r^2 \rho u$ (assuming that the star-planet line value is representative for entire atmosphere) are $5.4\times10^{10}$ and $2.5\times10^{10}$\,g\,s$^{-1}$ for \citetalias{Sanz-Forcada11} and \citetalias{Linsky14}, respectively (Table~\ref{tab:res}).

\begin{table}
\caption{Modeled mass-loss rates $\dot{M}$ for different stellar XUV fluxes/spectra and lower boundary parameters (number density $n_0$, temperature $T_0$). Rows 3 and 4 give the results for the X-ray and XUV flares, respectively (see Section~\ref{sec:flare}).}
\label{tab:res}
\centering
\begin{tabular}{lllll}
\hline\hline
$F_\mathrm{XUV}$ & XUV & $n_0$ & $T_0$ & $\dot{M}$ \\
(erg\,cm$^{-2}$\,s$^{-1}$) & & (cm$^{-3}$) & (K) & (g\,s$^{-1}$) \\
\hline
$1.80\times10^4$ & \citetalias{Sanz-Forcada11} & $10^{15}$ & 1200 & $5.4\times10^{10}$ \\
$1.27\times10^4$ & \citetalias{Linsky14} & $10^{15}$ & 1200 & $2.5\times10^{10}$ \\
$3.65\times10^4$ & \citetalias{Sanz-Forcada11}+X & $10^{15}$ & 1200 & $8.7\times10^{10}$ \\
$7.18\times10^4$ & \citetalias{Sanz-Forcada11}+XUV & $10^{15}$ & 1200 & $1.2\times10^{11}$ \\
$1.80\times10^4$ & \citetalias{Sanz-Forcada11} & $10^{15}$ & 800 & $4.8\times10^{10}$ \\
$1.80\times10^4$ & \citetalias{Sanz-Forcada11} & $10^{15}$ & 1600 & $6.1\times10^{10}$ \\
$1.80\times10^4$ & \citetalias{Sanz-Forcada11} & $5\times10^{15}$ & 1200 & $1.7\times10^{11}$ \\
\hline
\end{tabular}
\end{table}

In Fig.~\ref{fig:profiles}, we show the atmospheric profiles for number density, velocity, temperature, and ionization fraction ($n_\mathrm{H+}/n$) for both the \citetalias{Sanz-Forcada11} and \citetalias{Linsky14} reconstructions of the stellar XUV spectra. The results are rather similar for both spectra, although all parameters are slightly lower for the \citetalias{Linsky14} spectra because of the lower total XUV flux. The number density is dominated by neutral H below about $1.85R_\mathrm{p}$ and by H$^+$ above. The transition from a H-dominated atmosphere to one dominated by H$^+$ occurs where the ionization fraction exceeds 0.5. The outflow reaches the sonic speed just slightly below the L1 point at $4.3R_\mathrm{p}$ in both cases, whereas it already exceeds the escape velocity at points that are ${\sim}0.3R_\mathrm{p}$ closer to the planet. The temperature reaches a maximum of ${\sim}1.1\times10^4$~K at ${\sim}2R_\mathrm{p}$ for \citetalias{Sanz-Forcada11}, whereas the temperature maximum is slightly lower for \citetalias{Linsky14} and located closer to the planet.

The resulting atmospheric profiles are only considered to be valid well within the Roche lobe, and less reliable close to or even above the Roche lobe where 3D effects become significant \citep[e.g.,][]{Bisikalo13a}. Moreover, it is important to check if the outflow remains collisional within our computational domain so that the hydrodynamic treatment can be justified. The transition level to the collisionless regime is commonly taken as Knudsen number $Kn=\Lambda/X=1$, where $\Lambda=1/(n\sigma_\mathrm{col})$ is the mean free path and $X$ is an appropriate system scale, for example, the scale height $H=kT/(m_\mathrm{p}g)$ for regions close to the planet, or the planetary radius for more distant regions \citep{Shaikhislamov14}. For the collision cross-section $\sigma_\mathrm{col}$, we adopted the H-H$^+$ charge exchange cross-section \citep[${\sim}2\times10^{-15}$~cm$^{2}$; e.g.,][]{Lindsay05}, which is the appropriate choice for partially ionized hydrogen atmospheres \citep{Guo11, Salz16a, Shaikhislamov16}. Taking for $X$ either $H$ or $R_\mathrm{p}$, we find that $Kn{\ll}1$ inside our computational domain, justifying the hydrodynamic approach.

\begin{figure}
\includegraphics[width=\columnwidth]{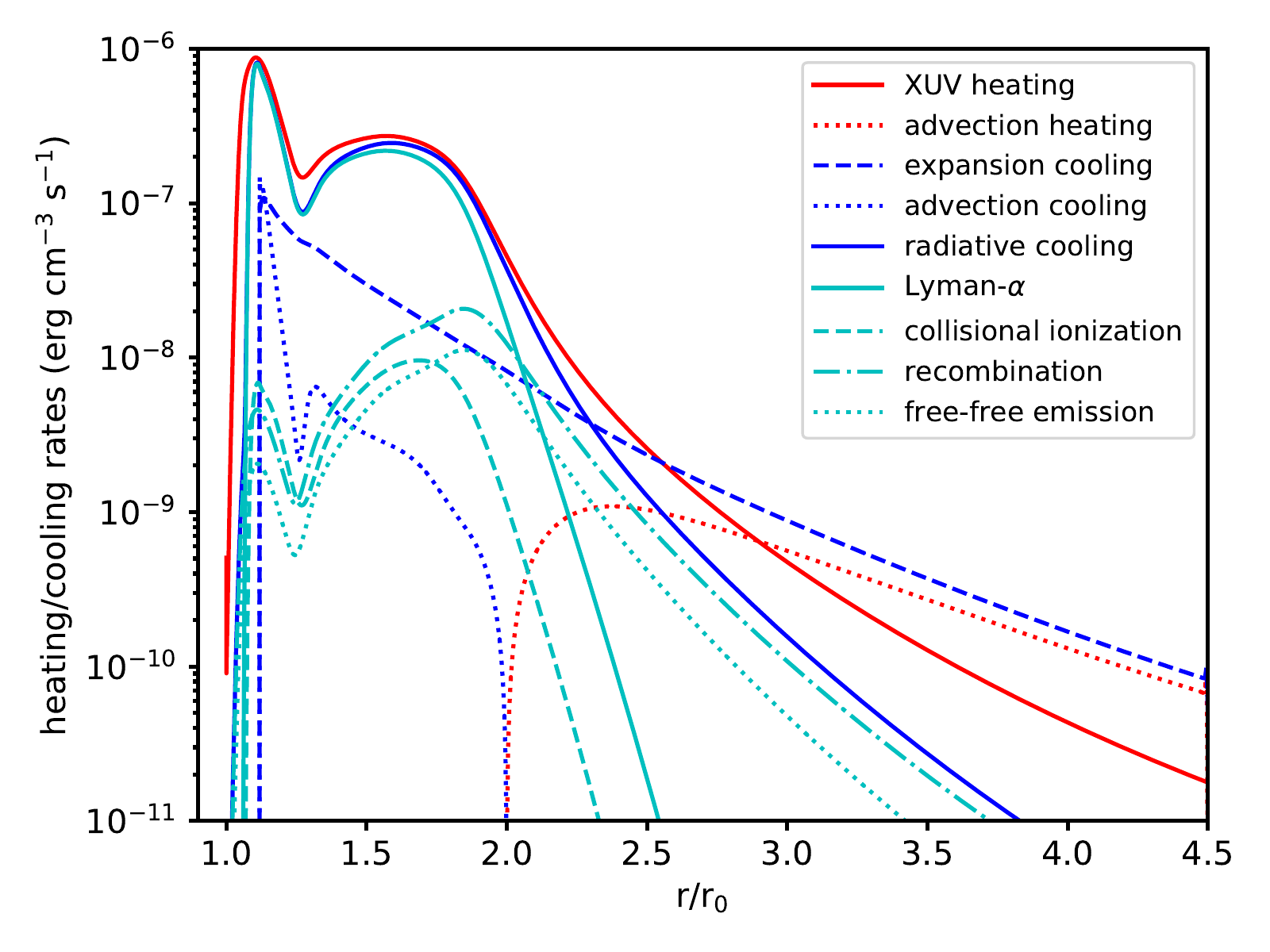}
\caption{All heating (red) and cooling (blue) processes included in our model, shown for the \citetalias{Sanz-Forcada11} run. The cyan lines detail the individual contributions to the radiative cooling rate (solid blue line).}
\label{fig:allq}
\end{figure}

Figure~\ref{fig:allq} details the individual heating and cooling processes in our model for the \citetalias{Sanz-Forcada11} run. One can see that the main heating source is the stellar XUV radiation, although advection contributes to heating in the uppermost regions. Advection cools the gas at lower heights, although its contribution is much smaller than the cooling from expansion. Expansion is even the dominant cooling mechanism above about $2.5R_\mathrm{p}$. Radiative cooling is also very important for HD~189733b and is dominated by Ly$\alpha$ emission close to the planet, whereas, in the upper layers, recombination radiation and free-free emission are more important. Both collisional ionization and conduction (not shown) are negligible cooling mechanisms.

\begin{figure}
\includegraphics[width=\columnwidth]{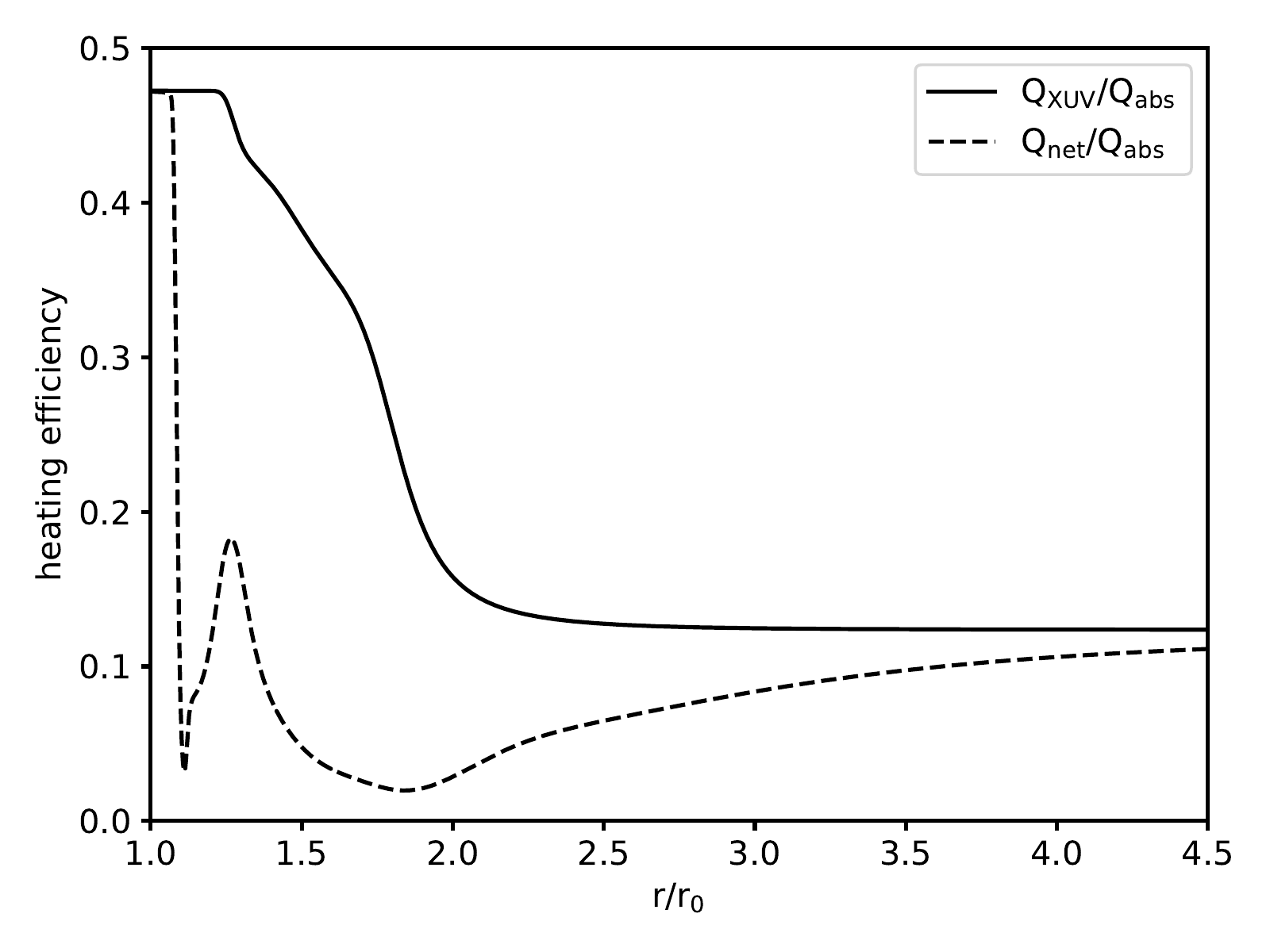}
\caption{Heating efficiencies $\eta_\mathrm{XUV}(r)$ (solid) and $\eta_\mathrm{net}(r)$ (dashed).}
\label{fig:eta}
\end{figure}

We calculated the heating efficiency using two different definitions commonly found in the literature. Firstly, we calculated the heating efficiency $\eta_\mathrm{XUV}(r)$ defined as the ratio of the XUV volume heating rate $Q_\mathrm{XUV}$ (Eq.~\ref{eq:qxuv}) to the locally absorbed XUV radiation
\begin{equation}
Q_\mathrm{abs}(r) = \sum_\lambda \sigma_\mathrm{ion,\lambda} n_\mathrm{H}(r) F_\mathrm{XUV,\lambda}\frac{\exp{(-\tau_\lambda(r))}}{1+\alpha\tau_\lambda(r)}.
\end{equation}
Secondly, we calculated $\eta_\mathrm{net}$ using the net local heating rate $Q_\mathrm{net}(r)$ (Eq.~\ref{eq:qnet}) divided by $Q_\mathrm{abs}(r)$ \citep[e.g.,][]{Salz16a}. A comparison is shown in Fig.~\ref{fig:eta}. The corresponding mean heating efficiencies ($\bar{\eta}_\mathrm{XUV}$, $\bar{\eta}_\mathrm{net}$) in the atmosphere, obtained by integrating $Q_\mathrm{XUV}(r)$ and $Q_\mathrm{net}(r)$, respectively, over $r$ and dividing by the stellar XUV flux at the planet's orbit (Table~\ref{tab:res}) amount to 12\% and 3\%. The former is in good agreement with detailed studies of the hot Jupiter HD~209458b \citep{Shematovich14, Ionov15}. We note that these heating efficiencies are different quantities than the photoelectron heating efficiency $\eta_\mathrm{ph}$ described in Section~\ref{sec:hc}.

The effective XUV absorption radius
\begin{equation}\label{eq:rxuv}
R_\mathrm{XUV} = \left(\frac{4\pi\int_{r_0}^{r_1}Q_\mathrm{XUV} r^2 dr}{\pi\bar{\eta} F_\mathrm{XUV}}\right)^{1/2}
\end{equation}
\citep{Erkaev07, Erkaev15} is about $2.6R_\mathrm{p}$ for \citetalias{Sanz-Forcada11} and $2.1R_\mathrm{p}$ for \citetalias{Linsky14}. Using $Q_\mathrm{net}$ instead of $Q_\mathrm{XUV}$ in Eq.~\ref{eq:rxuv} yields 1.15 \citepalias{Sanz-Forcada11} and $0.78R_\mathrm{p}$ \citepalias{Linsky14}, respectively. We note that this radius can be smaller than 1 if $Q_\mathrm{net}$ is used in Eq.~\ref{eq:rxuv} and radiative cooling is very efficient, and/or a too large value is chosen for the mean heating efficiency $\bar{\eta}$ in the denominator. Evaluating the energy-limited mass-loss rate
\begin{equation}\label{eq:eltd}
\dot{M}_\mathrm{el} = \frac{\pi R_\mathrm{p}R_\mathrm{XUV}^2\bar{\eta} F_\mathrm{XUV}}{G M_\mathrm{p} K}
\end{equation}
\citep{Erkaev07} with these parameters yields $1.5{-}3.1\times10^{11}$\,g\,s$^{-1}$ if adopting a typical mean heating efficiency for hot Jupiters of $\bar{\eta}=15$\% \citep[e.g.,][]{Shematovich14} and calculating the tidal enhancement factor $K$ as described in \citet{Erkaev07}. This overestimates the mass-loss rate of HD~189733b by almost an order of magnitude. Adopting the $R_\mathrm{XUV}$ values obtained with $Q_\mathrm{net}$ yields $2.1{-}6.2\times10^{10}$\,g\,s$^{-1}$, in much better agreement with the hydrodynamic rates. We note that the simple assumption of $R_\mathrm{XUV}{\sim} R_\mathrm{p}$ would coincidentally yield, for this specific planet, mass-loss rates of $3.4{-}4.7\times10^{10}$\,g\,s$^{-1}$, which are similar to the hydrodynamic results.

\subsection{The effect of a flare}\label{sec:flare}
The star HD~189733 is magnetically active, and flares have been detected frequently \citep{Pillitteri10, Pillitteri11, Pillitteri14}. An excess absorption of $\sim$14\% in Ly$\alpha$ was reported to have occurred during a planetary transit about eight hours after a strong X-ray flare was detected \citep{LecavelierdesEtangs12}. Here, we study if the enhanced XUV flux emitted by such a flare could have increased the planetary mass-loss rate enough to have caused this absorption.

The X-ray flux in the 0.3-3\,keV band of \textit{Swift}/XRT was increased by an average factor of four for a duration of $\sim$27\,min, which corresponds to one bin of the temporal resolution of the observations. Together with the average pre-flare flux of $3.6\times10^{-13}$\,erg\,cm$^{-2}$\,s$^{-1}$, this yields an estimated energy of ${\sim}8\times10^{31}$\,erg in this bandpass, and is thus a lower limit to the total radiated energy of this flare.

We tested the effect of such a flare on the planetary mass-loss rate and the atmospheric profiles. Two cases were considered: firstly, we increased the total XUV flux by the factor of four found in X-rays; secondly, we assumed that only the X-ray flux increased by a factor of four and the EUV flux remained unchanged, since the hot flare plasma could have radiated predominantly at shorter wavelengths, changing the shape of the XUV SED \citep[like in solar flares which radiate mostly in X-rays;][]{Chadney17}. We did not take into account the duration of the flux enhancement, but instead calculated the steady-state solution with these enhanced fluxes. If this solution does not show a significant increase of the neutral mass-loss rate, a shorter pulse of enhanced radiation would not either. The same conclusions were reached by \citet{Chadney17}, who investigated both the time evolution and steady-state enhancement effects of flares. We note that the true peak flux of the flare is unknown due to the low time resolution of the observations. The duration of the observed flare was approximately 27\,min, since there was a significant enhancement in only one temporal bin, and the flux in the following bin was comparable to (even slightly lower than) pre-flare levels \citep[see Fig.~4 in ][]{LecavelierdesEtangs12}. Thus, even though the true peak flux could have been higher, its duration cannot have exceeded more than a few minutes, limiting its effect on the planet.

\begin{figure*}
$\begin{array}{cc}
\includegraphics[width=\columnwidth]{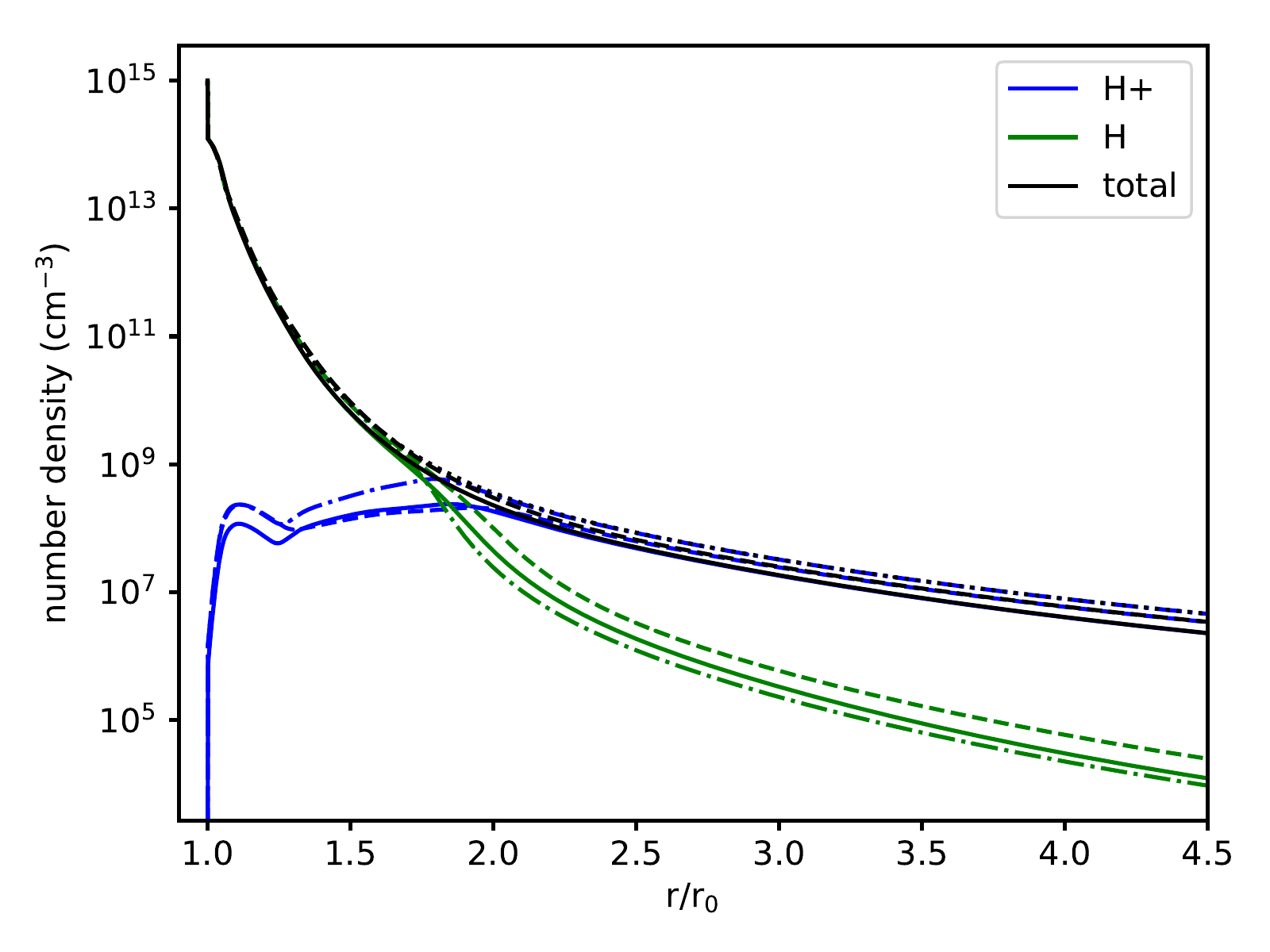} &
\includegraphics[width=\columnwidth]{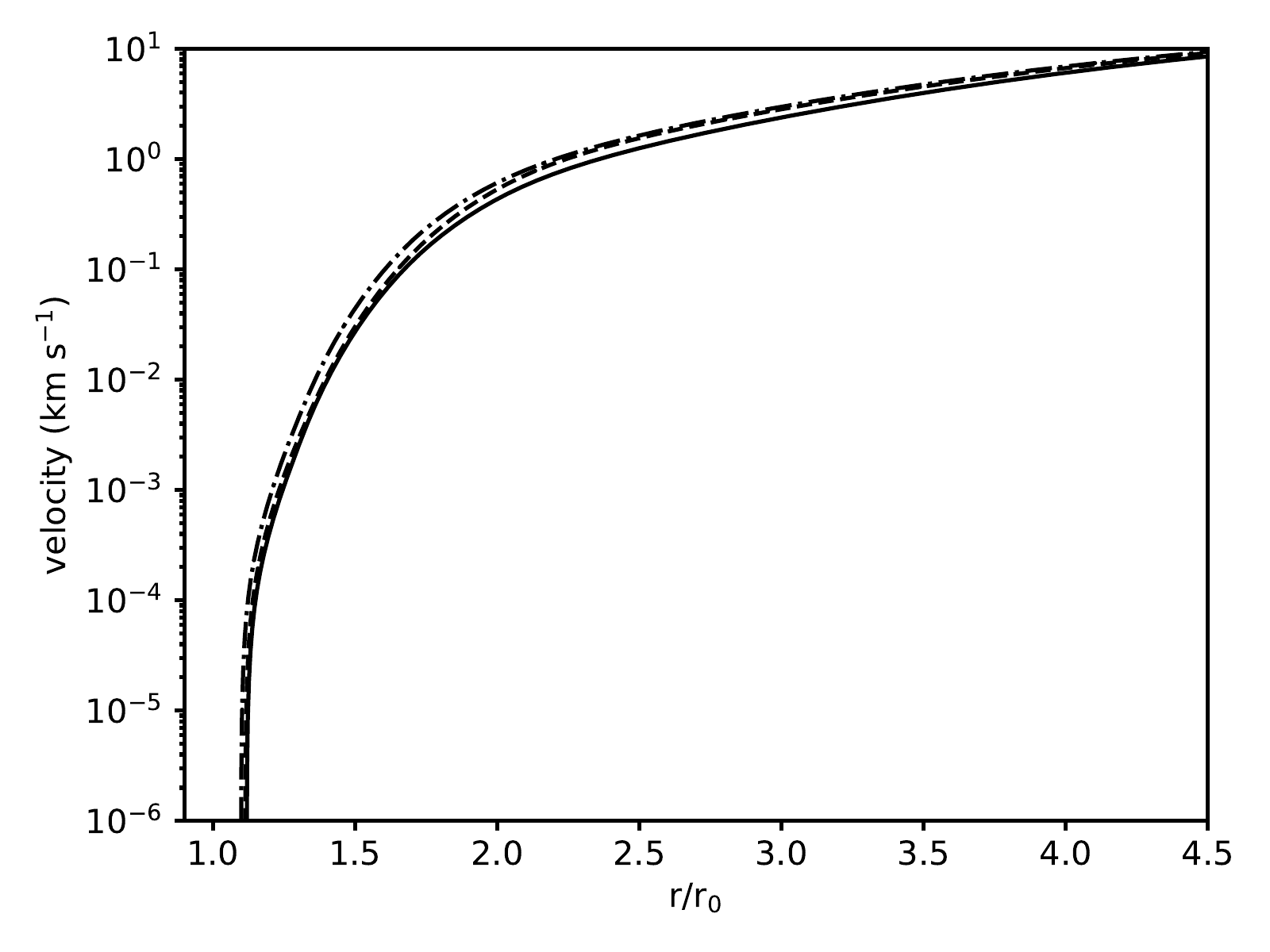} \\
\includegraphics[width=\columnwidth]{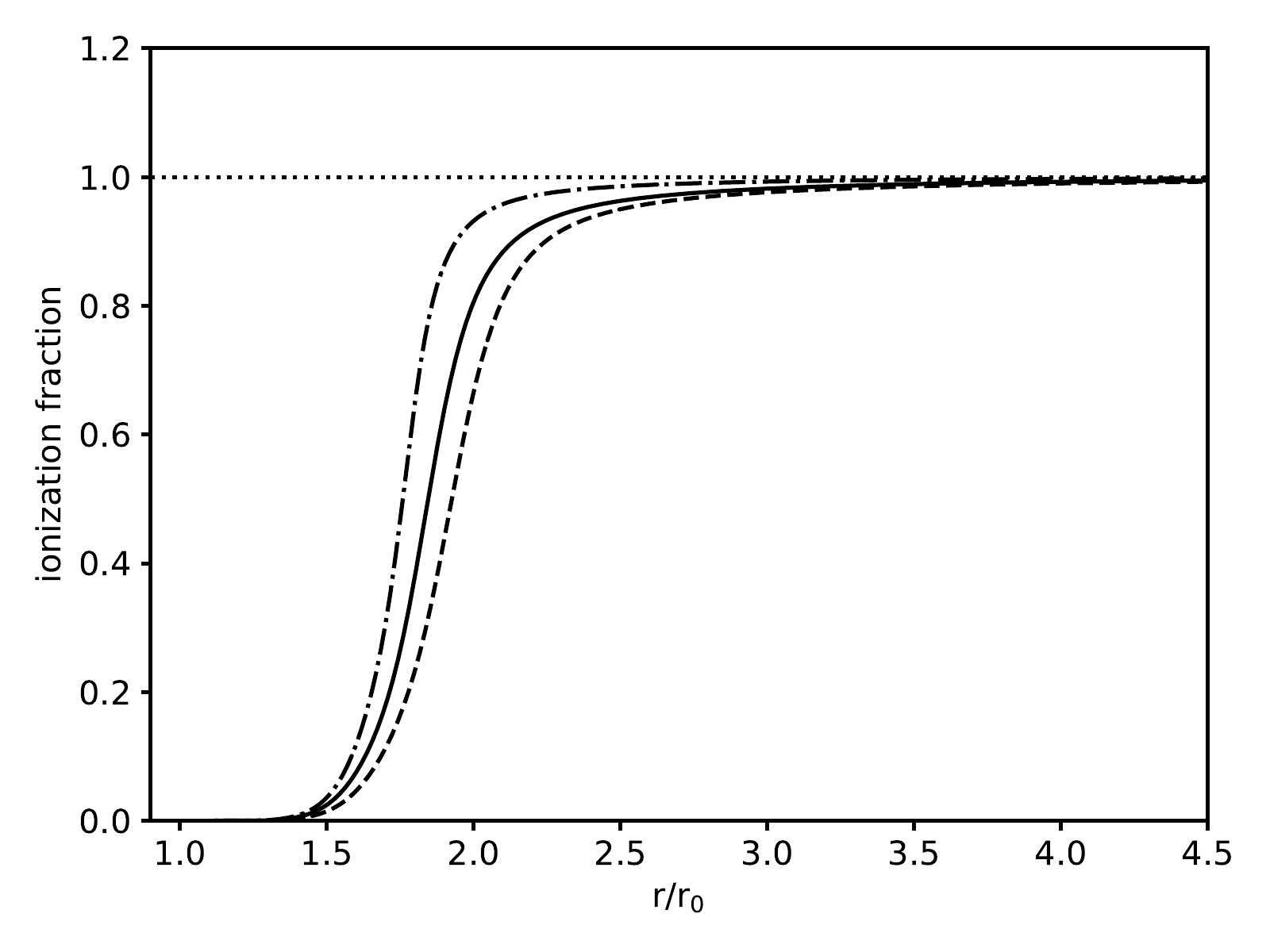} &
\includegraphics[width=\columnwidth]{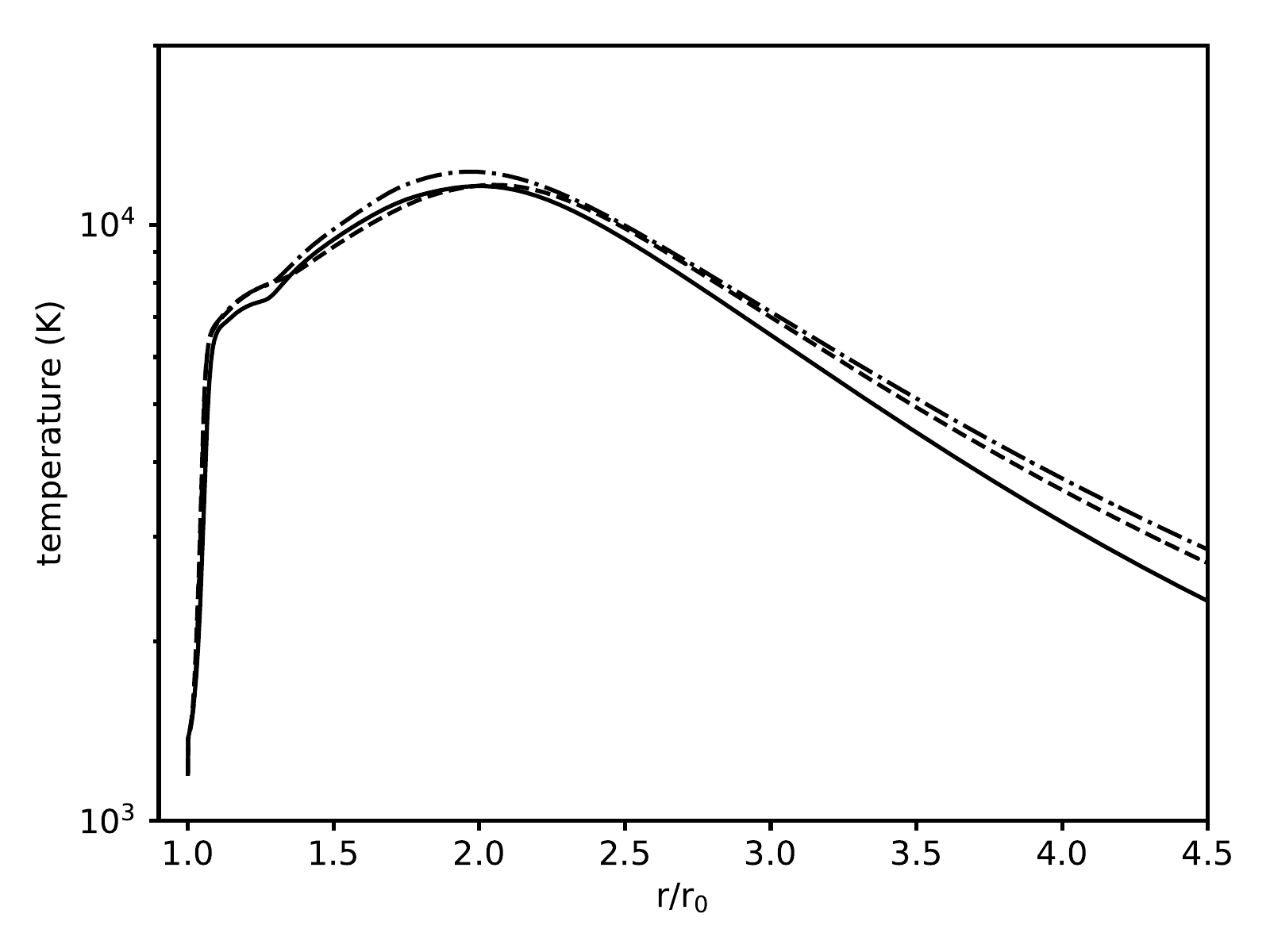}
\end{array}$
\caption{Atmospheric profiles for average XUV (solid), X-ray flare (dashed), and XUV flare (dash-dotted) cases based on the \citetalias{Sanz-Forcada11} spectrum.}
\label{fig:flare}
\end{figure*}

Figure~\ref{fig:flare} shows the atmospheric profiles under flaring conditions compared to those for the average XUV (\citetalias{Sanz-Forcada11}) spectrum. The profiles for all cases are rather similar. The most apparent differences are the stronger ionization for the XUV flare case and the slightly lower ionization for the X-ray flare case. Moreover, the temperatures in the uppermost regions are slightly higher for the flare cases than for the average XUV flux, but the peak is only slightly increased for the XUV flare. Other than that, the profiles are not strongly altered by exposure to the flare radiation, similar to what was found in the study of \citet{Chadney17}.

The total mass-loss rates are $8.7\times10^{10}$ (X-ray flare) and $1.2\times10^{11}$\,g\,s$^{-1}$ (XUV flare), higher by factors of 1.6 and 2.2, respectively, compared to the non-flaring state. Our flare-related mass-loss rate enhancements are very similar to the range of 1.8-2.2 found by \citet{Chadney17}. To compare with the modeling of \citet{Bourrier13a}, the neutral mass-loss rates at $2.95R_\mathrm{p}$ (the particle launch radius in their model) are $2.2\times10^{9}$ (X-ray flare) and  $8.9\times10^{8}$\,g\,s$^{-1}$ (XUV flare), meaning a factor of two higher (X-ray flare) and about 10\% lower (XUV flare) than in the non-flaring state ($10^{9}$\,g\,s$^{-1}$). This indicates that the enhanced radiation from the flare does not increase the neutral H densities and outflow fluxes sufficiently to account for the observed drastically different atmospheric absorption. More interestingly, although the total mass-loss rate increases for the XUV flare compared to both the non-flaring state and the X-ray flare because of the larger energy input, the neutral loss rate is actually smaller due to increased ionization. We note that the neutral outflow rates we obtain are in agreement with those required for the detected absorption ($5\times10^{8}{-}1.5\times10^{9}$\,g\,s$^{-1}$) in the model of \citet{Bourrier13a}, but for both flaring and non-flaring states. This indicates that the occurrence of a flare is likely insufficient to produce such strong variability in the planetary atmospheric absorption, especially if recalling that the actual flare duration was neglected. The nondetection in 2010 would require a reduction of the neutral escape rate by a factor of 5-20 compared to 2011 \citep{Bourrier13a}. This could only happen if the star were much less active in 2010 compared to our adopted XUV fluxes (reducing the total escape rate), or if it were much more active to increase the ionization sufficiently to reduce the neutral escape rate, despite increasing the total one. From existing observations, the variability in X-rays is about a factor of three (cf., Section~\ref{sec:params}), which is likely not sufficient to account for the variability of atmospheric absorption. However, this contradicts the necessity of having similar ionizing fluxes at both epochs in the \citet{Bourrier13a} model. Therefore, other processes may be responsible for, or contribute to, the variable planetary absorption, like variations of the stellar wind. Since strong X-ray flares on the Sun are frequently accompanied by coronal mass ejections (CMEs), we also consider the effect of a possible CME impact in Section~\ref{sec:cme}. We note that other upper atmosphere models of this planet find higher ionization (and therefore lower H densities) at $2.95R_\mathrm{p}$, yielding much lower neutral loss rates \citep{Guo11, Salz16a, Chadney17}. This is mainly related to the smaller number densities assumed at the lower boundary in these studies.

\section{MHD flow modeling}\label{sec:mhd}

\subsection{3D MHD flow model}
We used the 3D MHD flow model described in \citet{Erkaev17} to calculate the plasma flow around HD~189733b (assuming no intrinsic planetary magnetic field) and to obtain the plasma parameters and magnetic field in the region around the planet. This model takes into account radiation and charge-exchange processes acting on the hydrodynamically expanding upper planetary atmosphere penetrating into the stellar wind plasma.

Magnetic field and plasma parameters are determined by the following system of equations for mass, momentum and energy conservation, which are completed by the magnetic induction equation
\begin{align}
\frac{\partial(\rho \vec{V})}{\partial t} &+ \nabla\cdot\left[\rho\vec{V}\vec{V} + \vec{I}
\left(P + \frac{B^2}{8\pi}\right) - \frac{\vec{B}\vec{B}}{4\pi} \right] = \nonumber \\
&= Q_\mathrm{i}\vec{V}_\mathrm{H}  - Q_\mathrm{ex}(\vec{V} - \vec{V}_\mathrm{H}), \\
\nabla\cdot\vec{B} &= 0, \\
\frac{\partial\rho}{\partial t} &+ \nabla\cdot(\rho\vec{V}) = Q_\mathrm{i}, \\
\frac{\partial\vec{B}}{\partial t}& - \nabla\times(\vec{V}\times\vec{B}) = 0, \\
\frac{\partial W}{\partial t}  &+ \nabla\cdot\left(\frac{1}{2}\rho V^2\vec{V} +
\frac{\gamma}{\gamma-1}P\vec{V} + \frac{1}{4\pi}\vec{B}\times(\vec{V}\times\vec{B})\right) = \nonumber \\
&= \left(Q_\mathrm{i} + Q_\mathrm{ex}\right)\left(\frac{1}{2}V_\mathrm{H}^2 + \frac{3 k T_\mathrm{H}}{2 m_\mathrm{p}} \right) - Q_\mathrm{ex}\left(\frac{1}{2}V^2 + \frac{3 k T}{2 m_\mathrm{p}} \right), \\
W &= \frac{1}{2}\rho V^2 + \frac{1}{\gamma -1} P  + \frac{B^2}{8\pi},
\end{align}
where $\rho$, $\vec{V}$, $P$, and $\vec{B}$ are the mass density, velocity, plasma pressure, and magnetic field of the stellar wind, respectively. The parameter $\gamma$ is the polytropic index (assumed to be equal to 5/3) and $\vec{I}$ is the identity matrix. The parameters $\vec{V}_\mathrm{H}$ and $T_\mathrm{H}$ are the velocity and temperature of the atmospheric neutral hydrogen atoms escaping from the planet.

The mass conservation equation for ions includes an interaction source term, which is related to photoionization
\begin{equation}
Q_\mathrm{i} = \alpha_\mathrm{i}  N_\mathrm{H} m_\mathrm{p}
\end{equation}
and charge exchange ionization
\begin{equation}
Q_\mathrm{ex} = \rho\langle V_\mathrm{rel}\rangle N_\mathrm{H} \sigma_\mathrm{ex}
\end{equation}
of the hydrogen atoms. Here, $N_\mathrm{H}$ is the number density of the neutral planetary hydrogen atoms, $m_\mathrm{p}$ the particle mass, $\sigma_\mathrm{ex}$ ($\sim$10$^{-15}$\,cm$^2$) the charge exchange cross-section, $\langle V_\mathrm{rel}\rangle$ the average relative speed of the stellar wind and atmospheric particles, and $\alpha_\mathrm{i} = 5.9\times10^{-8}F_\mathrm{XUV}$ is the ionization rate proportional to the XUV flux $F_\mathrm{XUV}=1.8\times10^{4}$\,erg\,cm$^{-2}$\,s$^{-1}$ (cf., Table~\ref{tab:res}). The numerical scheme for the solution of this system of equations is described in \citet{Erkaev17}.

The neutral hydrogen atoms can be ionized via photoionization or  charge exchange processes. However, the latter is a dominating ionization mechanism for the atmospheric atoms in the stellar wind region, which leads to the generation of low-energy planetary ions and high-energy neutral hydrogen atoms (ENAs) originating from the stellar wind. These newly born ions are immediately accelerated by the local electric field and start to move together with the stellar wind plasma around the planetary obstacle. An acceleration of the picked up ions is accompanied by deceleration of the stellar wind plasma, with conservation of the total momentum.

The streamlined obstacle is considered to be a semi-sphere. The position of the stellar wind stagnation point ($R_\mathrm{s}$) is determined by the pressure balance condition, which means that the total
pressure of the external stellar wind has to be equal to the momentum flux of the internal ionized atmospheric particles at the boundary. The ratio of the curvature radius of the obstacle to the distance between the stagnation point and the planetary center was taken as 1.3, similar to the value used by \citet{Erkaev17}.

The calculation domain for the MHD stellar wind flow is bounded  by the external semi-sphere related to the undisturbed stellar wind region and the internal semi-sphere corresponding to the planetary obstacle. At the outer boundary, we applied the undisturbed stellar wind parameter values of density, velocity, temperature, and magnetic field. At the obstacle boundary, we assumed that the normal components of the stellar wind velocity and magnetic field vanish. Finally, we obtained a stationary solution for the stellar wind flow as a result of time relaxation of the nonsteady MHD solution. As initial conditions, we set the undisturbed stellar wind parameters in the whole computational domain. The final stationary solution is unique and does not depend on the particular initial conditions.

\subsection{Stellar wind interaction}
Table~\ref{tab:sw} summarizes the adopted stellar wind parameters used for the flow model. The mean and maximum values were taken from a 3D stellar wind model \citep{Llama13} based on the measured magnetic field map of the star \citep{Fares10}. Since all wind parameters vary along the orbit depending on longitude, we considered two cases. For the first case (mean wind), we computed the mean values of number density $N_\mathrm{sw}$, velocity $V_\mathrm{sw}$, temperature $T_\mathrm{sw}$, and magnetic field strength along the orbit. We computed both the parallel $B_\mathrm{sw,p}$ (parallel to $V_\mathrm{sw}$) and normal $B_\mathrm{sw,n}$ (normal to $V_\mathrm{sw}$) components of its magnitude, as well as the total field strength $B_\mathrm{sw,tot}$. The velocity is dominated by its radial component, and the magnetic field at the orbit is also dominated largely by its radial (i.e., parallel) field component. The angle $\theta_B$ is the angle of the total magnetic field vector and the star-planet line. For the second case (maximum wind), we used the parameter values where the ram pressure of the wind along the orbit is maximal. This corresponds to higher values compared to the mean wind (except for $B$), but not to the maximum values of the individual parameters. We chose this approach because the parameter maxima usually lie at different longitudes (e.g., the maximum velocity corresponds to the minimum density; cf., Fig.~\ref{fig:sw}). We ignored the velocity of the planetary orbital motion in the following calculations, as the ram pressure of the orbital motion is a factor of five smaller than that of the radial wind for the mean wind conditions, and even correspondingly smaller for the other scenarios.

\begin{table}
\caption{Adopted stellar wind parameters at the planet's orbit.}
\label{tab:sw}
\centering
\begin{tabular}{llll}
\hline\hline
Parameter & mean & max & CME \\
\hline
$N_\mathrm{sw}$ (cm$^{-3}$) & $4.4\times10^5$ & $4.9\times10^5$ & $4.4\times10^6$ \\
$V_\mathrm{sw}$ (km\,s$^{-1}$) & 326 & 472 & 1000 \\
$T_\mathrm{sw}$ (K) & $1.3\times10^6$ & $2\times10^6$ & $2\times10^6$ \\
$B_\mathrm{sw,p}$ (mG) & 44 & 23 & 0 \\
$B_\mathrm{sw,n}$ (mG) & 12 & 9.5 & 100 \\
$B_\mathrm{sw,tot}$ (mG) & 46 & 25 & 100 \\
$\theta_B$ (\degr) & 15 & 22 & 90 \\
$P_\mathrm{dsw}$ (dyn\,cm$^{-2}$) & $7.82\times10^{-4}$ & $1.83\times10^{-3}$ & $7.36\times10^{-2}$ \\
\hline
\end{tabular}
\end{table}

We adopted the hydrodynamic solution of the planetary upper atmosphere obtained with the \citetalias{Sanz-Forcada11} XUV spectrum as an input for the MHD flow modeling. Considering the stellar wind parameters given in Table~\ref{tab:sw}, we applied the 3D MHD flow model to calculate the spatial distribution of the magnetic field and plasma parameters in the vicinity of the planet. By solving the nonsteady MHD equations, a stationary solution is established as a result of time relaxation. As an initial condition, we assumed the uniform undisturbed stellar wind flow, which suddenly stops at the planetary obstacle. This leads to the appearance of a shock-like wave front propagating outwards from the obstacle. Since we have a super-Alfvenic stellar wind flow, this shock approaches its stationary position located at some distance from the obstacle.

In the case of the mean wind, the radial distance to the magnetopause is about $2.6R_\mathrm{p}$. Here, the ion and neutral densities of the atmospheric particles are $3.9\times10^7$\,cm$^{-3}$ and $1.25\times10^6$\,cm$^{-3}$, respectively, and the temperature is $8.8\times10^3$\,K. In the maximum wind case, the magnetopause position is slightly closer, at $2.4R_\mathrm{p}$, since the stellar wind dynamic pressure is larger. The corresponding atmospheric ion and neutral densities are $6.1\times10^7$\,cm$^{-3}$ and $2.9\times10^6$\,cm$^{-3}$, respectively, and the temperature is $1\times10^4$\,K (cf. Fig.~\ref{fig:profiles}).

\begin{figure}
\includegraphics[width=\columnwidth]{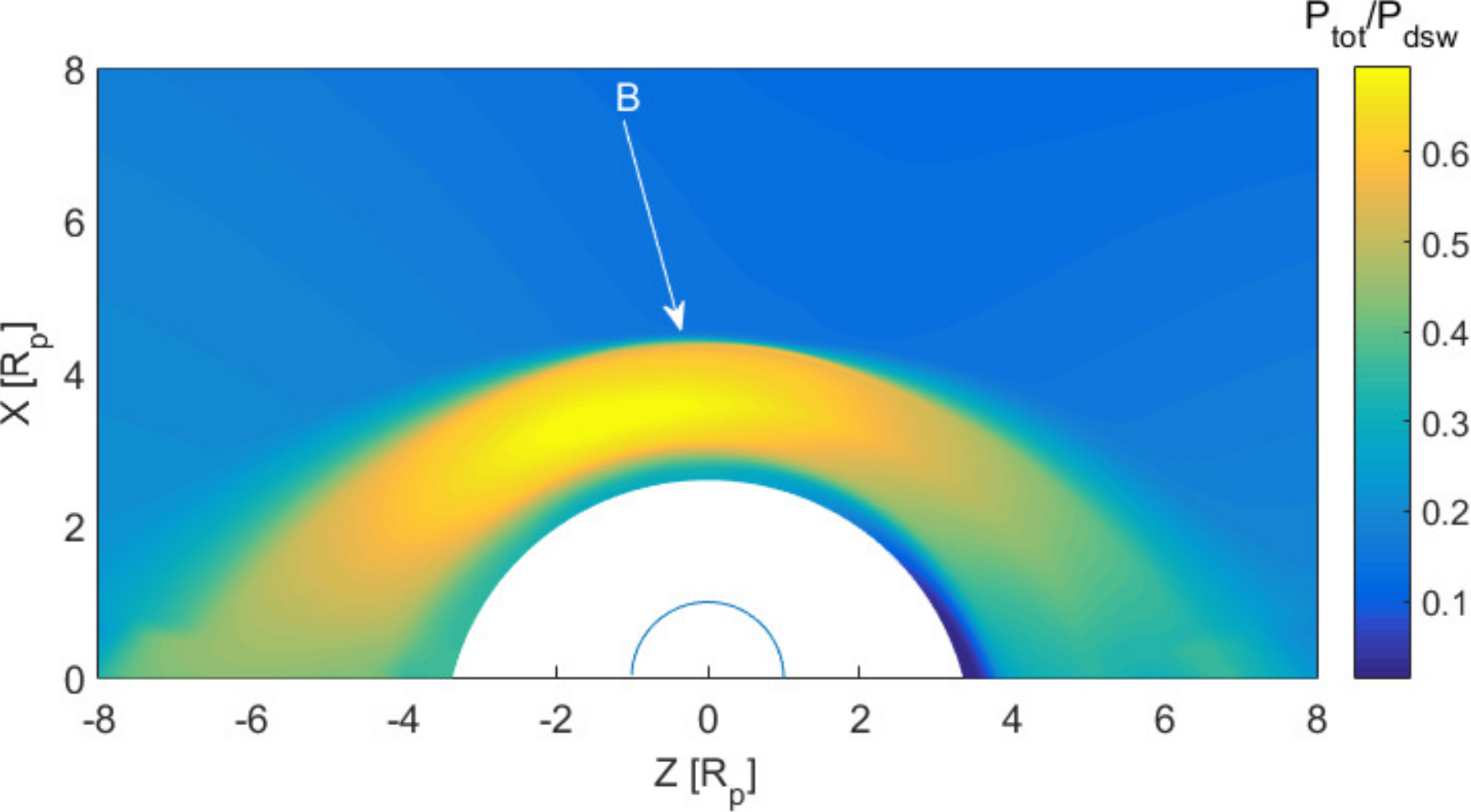}
\caption{Cut at $Y{=}0$ of the simulation showing the distribution of the total pressure (sum of the magnetic and gas pressures) around HD~189733b normalized to the stellar wind dynamic pressure for the mean stellar wind parameters. The white area close to the origin indicates the atmospheric region around the planet, while the semicircle indicates the planetary optical radius. The star is located along the $X$-axis. The arrow shows the direction of the interplanetary magnetic field.}
\label{fig:mhd1}
\end{figure}

\begin{figure}
\includegraphics[width=\columnwidth]{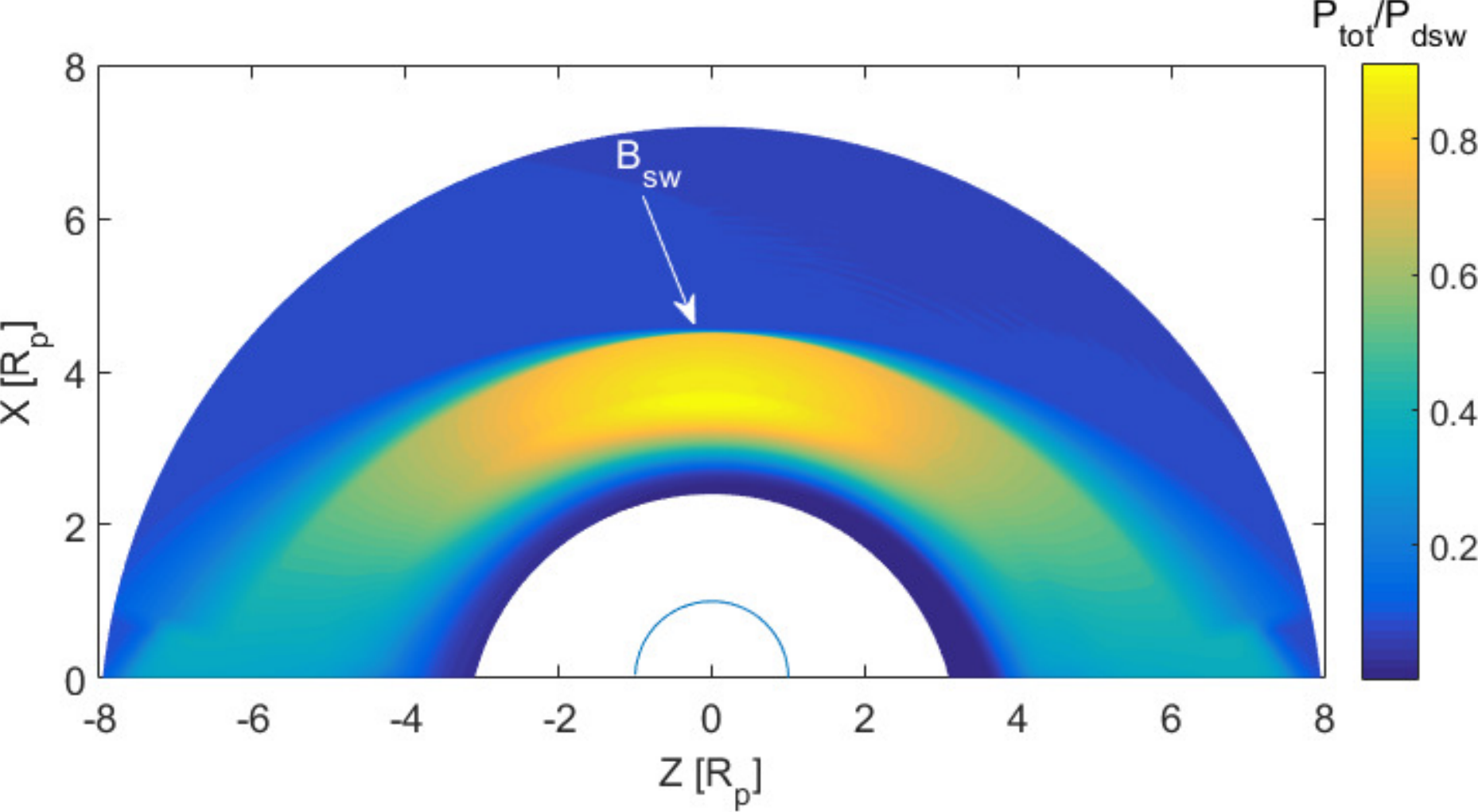}
\caption{Same as Fig.~\ref{fig:mhd1}, but for the maximum stellar wind.}
\label{fig:mhd2}
\end{figure}

Figure~\ref{fig:mhd1} shows the spatial distribution of the total pressure $P_\mathrm{tot}$ (sum of the magnetic and gas pressures) obtained from the MHD model employing the mean stellar wind parameters (see Table~\ref{tab:sw}). Here, the total pressure is normalized to the stellar wind dynamic pressure $P_\mathrm{dsw}$ = $N_\mathrm{sw} m_\mathrm{p} V_\mathrm{sw}^2 = 7.8\times10^{-5}$\,Pa. The origin of the coordinate system is placed at the planet's center, and the star is located along the positive $X$-axis. The direction of the $Z$-axis is chosen to have coplanarity between the $XZ$ plane and the interplanetary magnetic field vector $\vec{B}_\mathrm{sw}$ (the arrow in Fig.~\ref{fig:mhd1}). The white area around the center of the coordinate system indicates the region filled by atmospheric particles exclusively, while the embedded dark blue semicircle indicates the optical planetary radius $R_\mathrm{p}$. Figure~\ref{fig:mhd2} is similar to Fig.~\ref{fig:mhd1}, but corresponds to the maximum stellar wind parameters. The total pressure in this case is also normalized to the corresponding unperturbed stellar wind dynamic pressure of $P_\mathrm{dsw}$ = $1.8\times10^{-4}$\,Pa.

\begin{figure}
\includegraphics[width=\columnwidth]{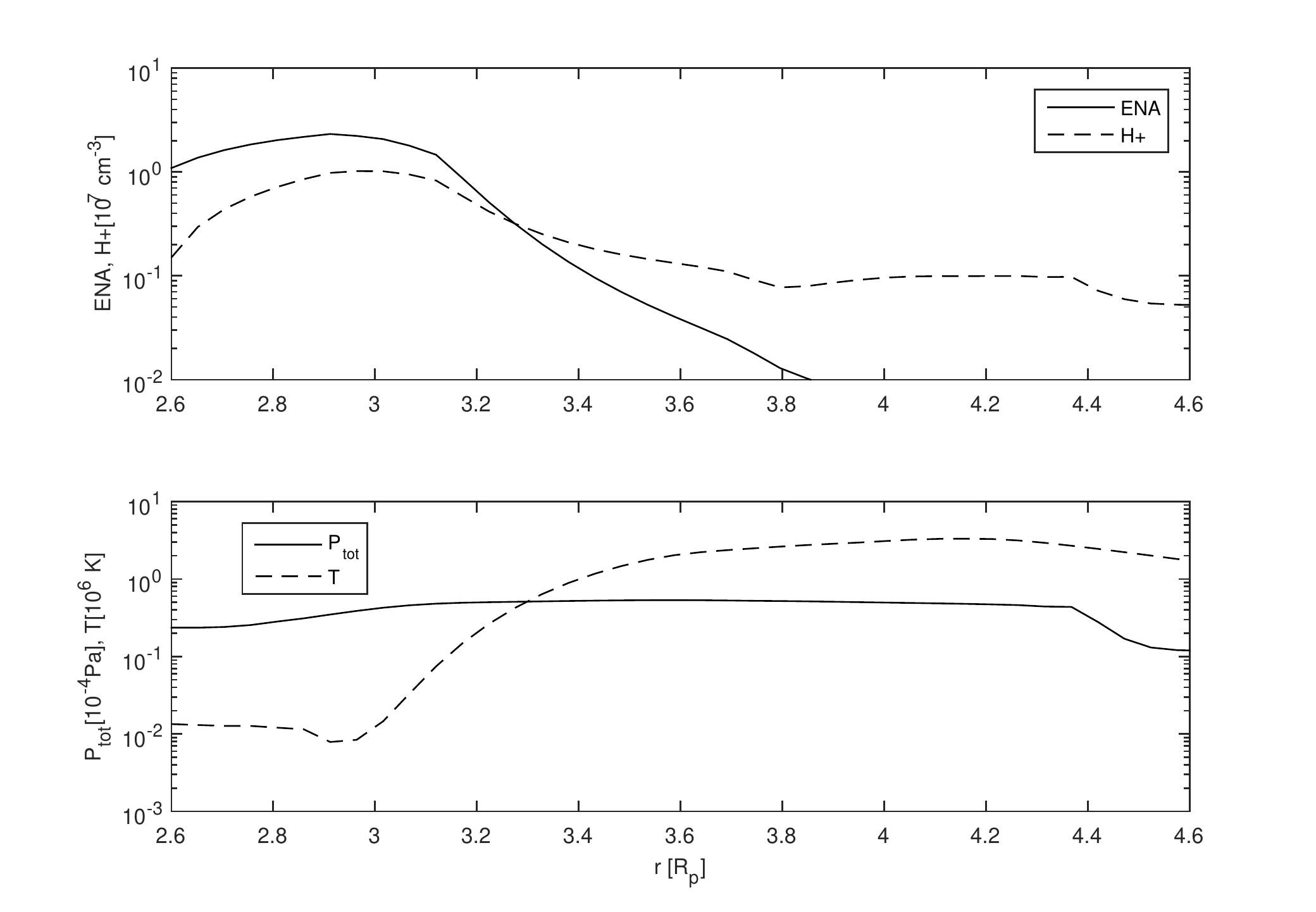}
\caption{Profiles of the parameters along the stagnation line from the magnetopause to the bow shock for the mean stellar wind conditions. The top panel shows the ENA and ion densities, the bottom panel the total pressure and temperature.}
\label{fig:ena1}
\end{figure}

\begin{figure}
\includegraphics[width=\columnwidth]{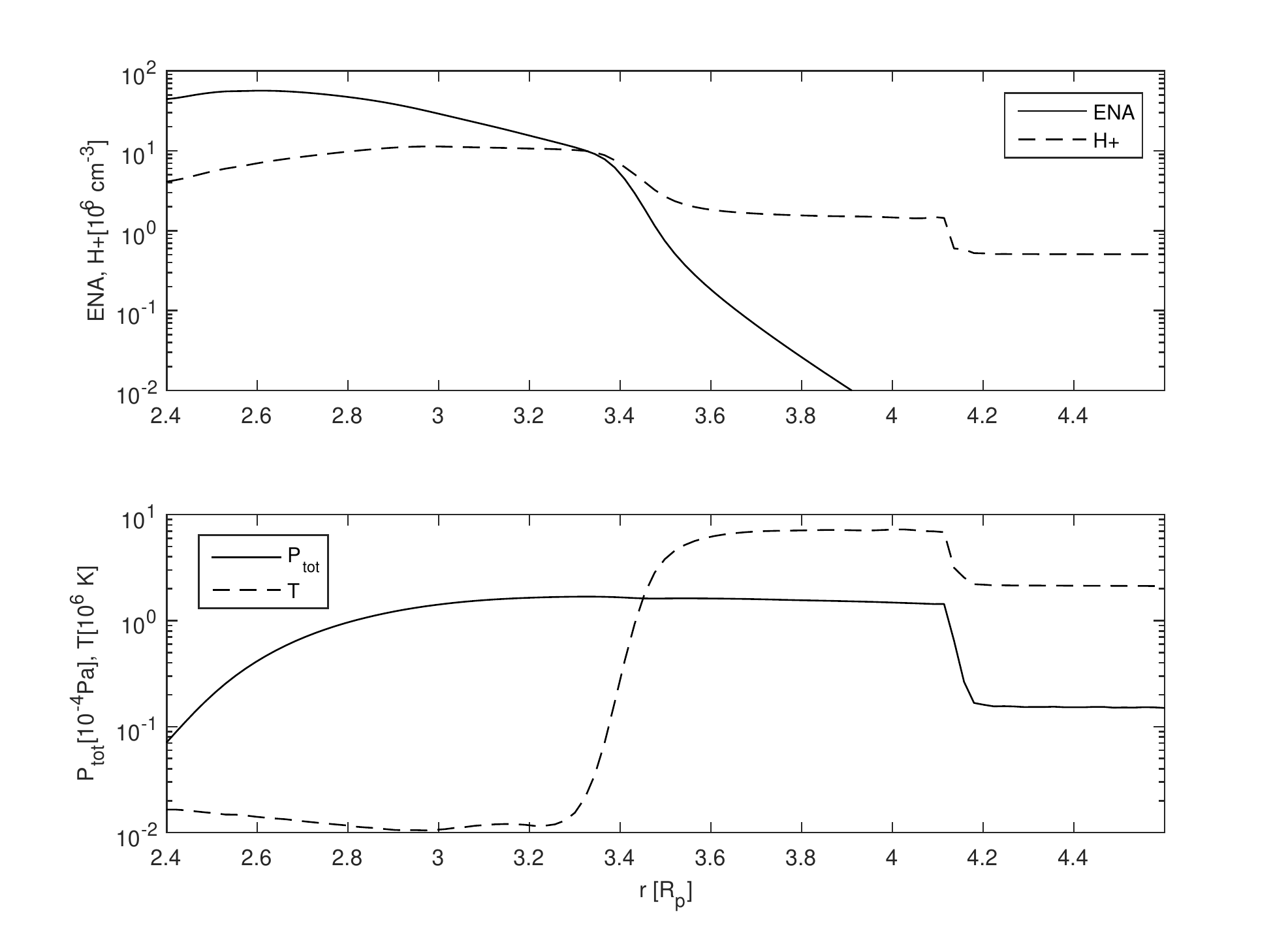}
\caption{Same as Fig.~\ref{fig:ena1}, but for the maximum stellar wind conditions.}
\label{fig:ena2}
\end{figure}

Figure~\ref{fig:ena1} shows the profiles of the ENA and ion densities along the $X$-axis between the magnetopause and the bow shock (top panel), as well as the profiles of the total pressure and temperature (bottom panel), both corresponding to the mean stellar wind. The total ion number density has a maximum of about $4\times10^4$\,cm$^{-3}$ due to charge exchange interaction between the stellar wind plasma and atmospheric neutral atoms, which are flowing through the magnetopause. In front of the magnetopause, the ion density and ENA density maxima are about $1.0\times10^7$\,cm$^{-3}$ and $2.3\times10^7$\,cm$^{-3}$, respectively. The ENAs form a layer around the magnetopause with a thickness of about $1.3R_\mathrm{p}$. Figure~\ref{fig:ena2} is similar to Fig.~\ref{fig:ena1}, but corresponds to the maximum stellar wind parameters. In this case, the ion density and ENA density maxima in front of the magnetopause are about $1.13\times10^7$\,cm$^{-3}$ and $5.6\times10^7$\,cm$^{-3}$, respectively. The ENA layer thickness around the magnetopause is about $1.2R_\mathrm{p}$.

\subsection{Impact of a coronal mass ejection}\label{sec:cme}
We also consider the case of a stellar CME to study the effect of a potential CME impact. Interaction with a CME could have provided the elevated plasma density levels necessary to produce the observed variability of the planetary Ly$\alpha$ transit signature. On the Sun, large flares are frequently accompanied by CMEs. Empirical models based on flare-CME relationships from the Sun predict high CME occurrence rates for active stars \citep[e.g.,][and references therein]{Odert17}, because of their high flare rates. The estimated X-ray flare energy of ${\sim}8\times10^{31}$\,erg (cf.,~Section~\ref{sec:flare}) corresponds to a flare that is (almost) always accompanied by a CME on the Sun \citep{Yashiro09}. If the flare was indeed accompanied by a CME, its estimated mass would be in the order of $10^{16}$\,g, and its velocity about 1300\,km\,s$^{-1}$, based on solar scalings \citep{Drake13, Odert17}. We applied an empirical CME prediction model to HD~189733, which calculates CME occurrence rates for Sun-like and cooler main-sequence stars based on their X-ray luminosities $L_\mathrm{X}$ \citep{Odert17}. Adopting $L_\mathrm{X}=1.67\times10^{28}$\,erg\,s$^{-1}$ (cf., Section~\ref{sec:params}), we obtain about $10{-}2000$ CMEs per day, depending on the power law index of the stellar flare energy distribution, $dN/dE{\propto}E^{-\alpha}$. We assumed $\alpha=1.5{-}2.5$, which corresponds to an observationally determined range typical for the Sun and other stars \citep{Guedel03}. The mass-loss rate from these CMEs are consistent with observations of stellar mass-loss rates \citep[e.g.,][and references therein]{Wood04}. However, recent modeling results suggest that CME rates may be lower than estimated from solar scalings due to the stronger magnetic fields on active stars \citep{Alvarado-Gomez18}, so the obtained numbers are possibly overestimates.

Knowledge of stellar CMEs is still sparse, and therefore the CME activity of HD~189733 is not constrained by observations either. One method to detect stellar CMEs is from transient blue-wing asymmetries or blue-shifted extra-emissions or absorptions (depending on viewing geometry) in Balmer lines, which probe the (partly) neutral prominence material embedded in the CME core \citep[e.g.,][and references therein]{Leitzinger14}. We analyzed archival optical spectroscopic observations of HD~189733 \citep[430 Stokes $I$ spectra from PolarBase;][]{Petit14} as described in \citet{Leitzinger20a}. The total on-source time of the spectra is about 122\,h, and the observations span the years 2006-2015. We find no signatures of CMEs in the Balmer lines. With this nondetection, we estimate ${<}0.6$ observable CMEs per day (95\% confidence). It is important to correct for the expected H$\alpha$ emission and geometrical constraints, since not all occurring CMEs are necessarily observable with a given method. We used a semiempirical model that takes into account the expected maximum possible CME core emission in the Balmer lines and geometrical constraints, considering the total on-source time and average signal-to-noise ratio of the spectra \citep{Odert20}. It predicts that $\lesssim$1\% of the intrinsically occurring CMEs would be observable in H$\alpha$ on HD~189733 for the given observational parameters. The estimated maximum observable CME rates are comparable to the upper limit from the analyzed observations, indicating that much longer observing times and/or higher quality of the spectra would be required to determine the CME rate of this star with the given method. We note that none of the analyzed H$\alpha$ observations were taken during the time of the studied flare and planetary transit in 2011. Due to the nondetection of CMEs on this star, we have to rely on reasonable extrapolations from the Sun to estimate plausible CME parameters for HD~189733.

We assumed a velocity of 1000\,km\,s$^{-1}$, which is commonly found for energetic CMEs already close to the Sun \citep{Yashiro04}, and which is also similar to the CME velocity derived above from the estimated flare energy. For the density, we used an enhancement factor of ten compared to the mean wind, which is also common for solar CMEs at separations of a few solar radii \citep{Schwenn06a}. We note that we cannot use solar CME density profiles as adopted, for instance, in \citet{Khodachenko07}, because the modeled stellar wind densities at HD~189733b's orbit are already higher than the solar CME densities at the same distance \citep{Khodachenko07}. Due to the lack of knowledge regarding stellar CME temperatures, we simply adopted the maximum wind value. For the magnetic field, we used 100~mG, similar to values of solar CMEs at around $10R_{\sun}$ from the Sun \citep{Patsourakos16}. The angle $\theta_B$ is assumed to be 90\degr, corresponding to the geometry of a centrally impacting ejected flux rope, as commonly observed in solar CMEs. With the assumed velocity, a CME would need about 1\,h to reach the planet's orbit. The typical duration of solar CMEs at a few solar radii is about 8\,h \citep{Lara04}. This means that if a CME occurred simultaneously with the flare, the planet may have still been exposed to a plasma environment dominated by the CME. The adopted CME parameters are summarized in Table~\ref{tab:sw}.

\begin{figure}
\includegraphics[width=\columnwidth]{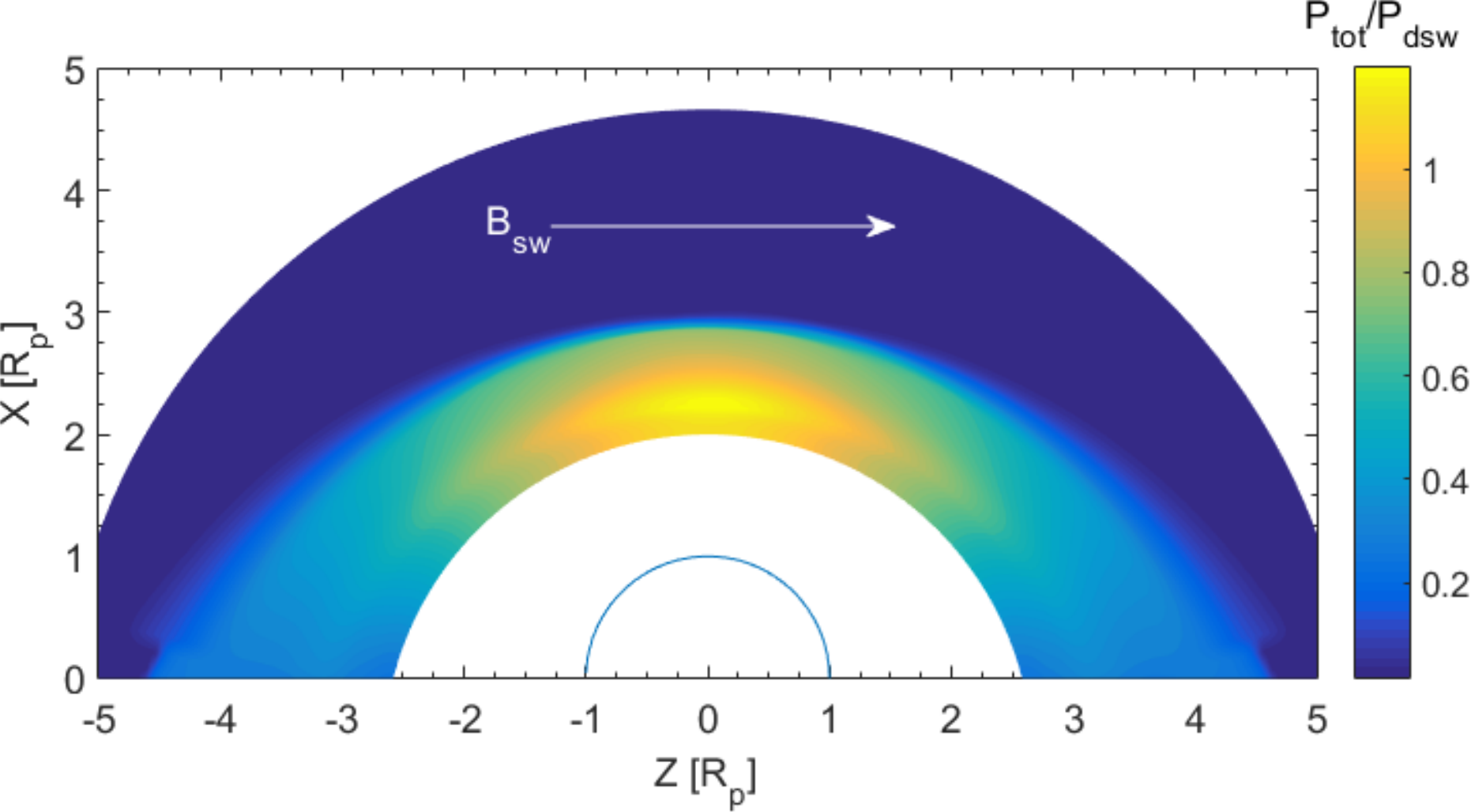}
\caption{Same as Fig.~\ref{fig:mhd1}, but for the CME conditions.}
\label{fig:mhd3}
\end{figure}

\begin{figure}
\includegraphics[width=\columnwidth]{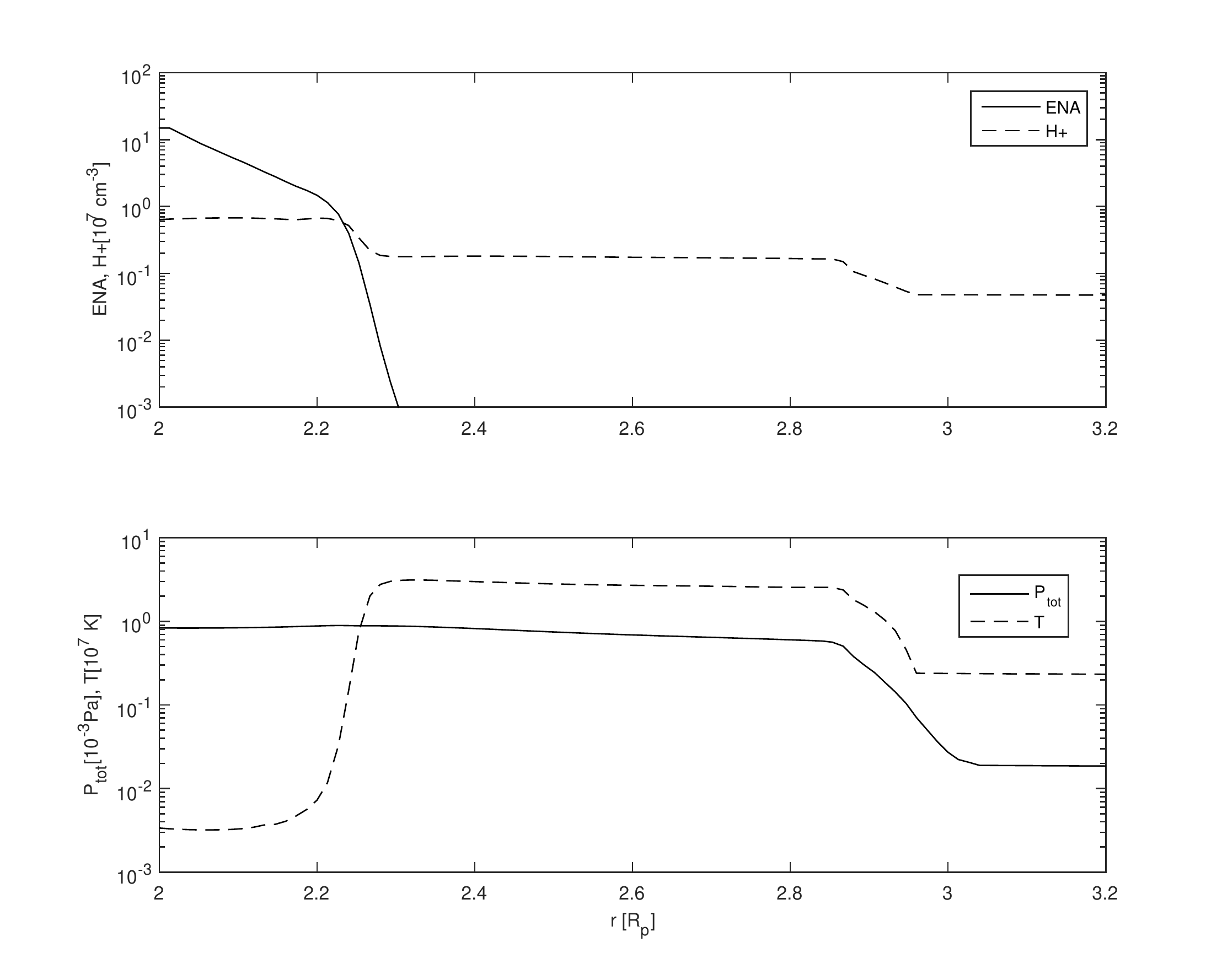}
\caption{Same as Fig.~\ref{fig:ena1}, but for the CME conditions.}
\label{fig:ena3}
\end{figure}

Figure~\ref{fig:mhd3} shows the spatial distribution of the total pressure obtained with the MHD flow model for the adopted CME parameters (Table~\ref{tab:sw}). In this case, the magnetopause is located closer to the planet compared to both wind cases because of the higher dynamic pressure, namely at a distance of ${\sim}2R_\mathrm{p}$. The corresponding atmospheric ion and neutral densities are about $1.8\times10^8$\,cm$^{-3}$ and $4.5\times10^7$\,cm$^{-3}$, respectively, and the temperature is $1.16\times10^4$\,K (cf. Fig.~\ref{fig:profiles}). The total pressure is again normalized to the CME dynamic pressure, $P_\mathrm{dsw} = 7.35\times10^{-3}$\,Pa. The ion density and ENA density maxima in front of the magnetopause (Fig.~\ref{fig:ena3}) are about $0.67\times10^7$\,cm$^{-3}$ and $1.5\times10^8$\,cm$^{-3}$, respectively. The ENA layer is much thinner in the CME case, with a thickness of about $0.25R_\mathrm{p}$.

Table~\ref{tab:summary} summarizes the results of the stellar wind interaction modeling. It compares the stand-off distances, the thickness of the ENA layers, and the maximum number densities of ENAs and ions in front of the magnetopause. Stronger winds or CMEs confine the planetary atmosphere to smaller regions because of the higher ram pressures. Higher stellar wind ram pressures also lead to more compressed, thinner ENA layers, but with higher maximum densities.

\begin{table}
\caption{Stand-off distances, ENA layer thickness, and maximum number densities of ENAs and ions in front of the magnetopause.}
\label{tab:summary}
\centering
\begin{tabular}{lllll}
\hline\hline
Wind & stand-off & ENA layer & ENA density & ion density \\
     & distance  & thickness & maximum     & maximum \\
     & $(R_\mathrm{p})$ & $(R_\mathrm{p})$ & (cm$^{-3}$) & (cm$^{-3}$) \\
\hline
mean & 2.6 & 1.3  & $2.3\times10^7$ & $1.0\times10^7$ \\
max  & 2.4 & 1.2  & $5.6\times10^7$ & $1.13\times10^7$ \\
CME  & 2   & 0.25 & $1.5\times10^8$ & $0.67\times10^7$ \\
\hline
\end{tabular}
\end{table}

\section{Modeling the Ly$\alpha$ absorption signature}\label{sec:abs}
To calculate the transmissivity in the Ly$\alpha$ line, we assumed that a hydrogen cloud surrounding the planet consists of two parts: a spherically symmetric lower atmosphere corresponding to the 1D atmospheric profile, and an upper exospheric part consisting of the ENAs calculated by the 3D MHD model. We considered one atmospheric profile (Fig.~\ref{fig:profiles}) and the three different stellar wind cases (mean, maximum, and CME; Table~\ref{tab:sw}). After a hydrogen cloud was simulated (Figs.~\ref{fig:mhd1}, \ref{fig:mhd2}, \ref{fig:mhd3}) and through knowledge of the positions and velocities of all hydrogen particles, we computed how these atoms attenuate the stellar Ly$\alpha$ radiation by using a post-processing software written in the Python programming language. To compute the transmissivity along the line of sight, we followed the approach of \citet{Semelin07}. The post-processing tool is described in detail in \citet{Kislyakova14b}. Here, we summarize its main features. 

Only neutral hydrogen atoms absorb in the Ly$\alpha$ line. One has to take into account spectral line broadening. Real spectral lines are subject to several broadening mechanisms: \textit{i)} natural broadening; \textit{ii)} collisional broadening; \textit{iii)} Doppler or thermal broadening. The ``natural line width'' is a result of quantum effects and arises due to the finite lifetime of an atom in a definite energy state. A photon emitted in a transition from this level to the ground state will have a range of possible frequencies $\Delta f{\sim}\Delta E/\hbar{\sim}1/\Delta t$, which can be approximated by a Lorentzian profile. Collisional broadening is caused by the collisions randomizing the phase of the emitted radiation. This effect can become very significant in a dense environment, yet above the exobase, it does not play a role and is important only in the lower parts of the atmosphere, therefore, we did not take it into account. 

The third type of broadening, which plays a significant role in the upper atmosphere of a hot exoplanet, is thermal broadening, which arises because the frequency of the absorption is shifted due to the Doppler effect. In the considered cases, an analytical solution for the absorption profile cannot be obtained, since it is not only thermal atoms that contribute to the broadening, but also ENAs. For this reason, we cannot use a Voigt profile (which can only be used for a pure Maxwellian distribution). We calculated the natural broadening for all atoms and bin it by velocity, which automatically gives us the Doppler broadening for a particular velocity distribution.

To compute the transmissivity along the line of sight, we followed the approach of \citet{Semelin07} and \citet{Kislyakova14b}. We used a Cartesian coordinate system with the $x$-axis pointing towards the star, the $y$-axis directed antiparallel to the planetary motion, and the $z$-axis completing the right-handed coordinate system. We calculated the relation between the observed intensity $I$ and the source intensity $I_0$ as a function of frequency $f$ of the stellar spectrum in the $yz$-plane by dividing the computational domain into a grid with $N_\mathrm{c}$ cells. For each cell in the grid along lines of sight in front of the star ($y^2 + z^2 < R_*^2$), the velocity spectrum of all hydrogen atoms in the column along the $x$-axis can be calculated. We accounted for the planetary inclination by shifting the cloud by $z = a\cos{i} = 3.67\times10^{10}$~cm relative to the center of the stellar disk at mid transit. Then, the transmissivity could be averaged over all columns in the $yz$-grid except for those particles that fell outside the projected limb of the star or inside the planetary disk.

We used the frequency-dependent cross-section, which depends on the normalized velocity spectrum, the Ly$\alpha$ resonance wavelength and the natural absorption cross-section in the rest frame of the scattered hydrogen atom \citep{Peebles93}. For lines of sight in front of the planet $(y - y_\mathrm{p} )^2 + (z - z_\mathrm{p} )^2 < R_\mathrm{p}^2$, where ($y_\mathrm{p}$,$z_\mathrm{p}$) is the planet center position, we set the transmissivity to zero.

\begin{figure}
\includegraphics[width=\columnwidth]{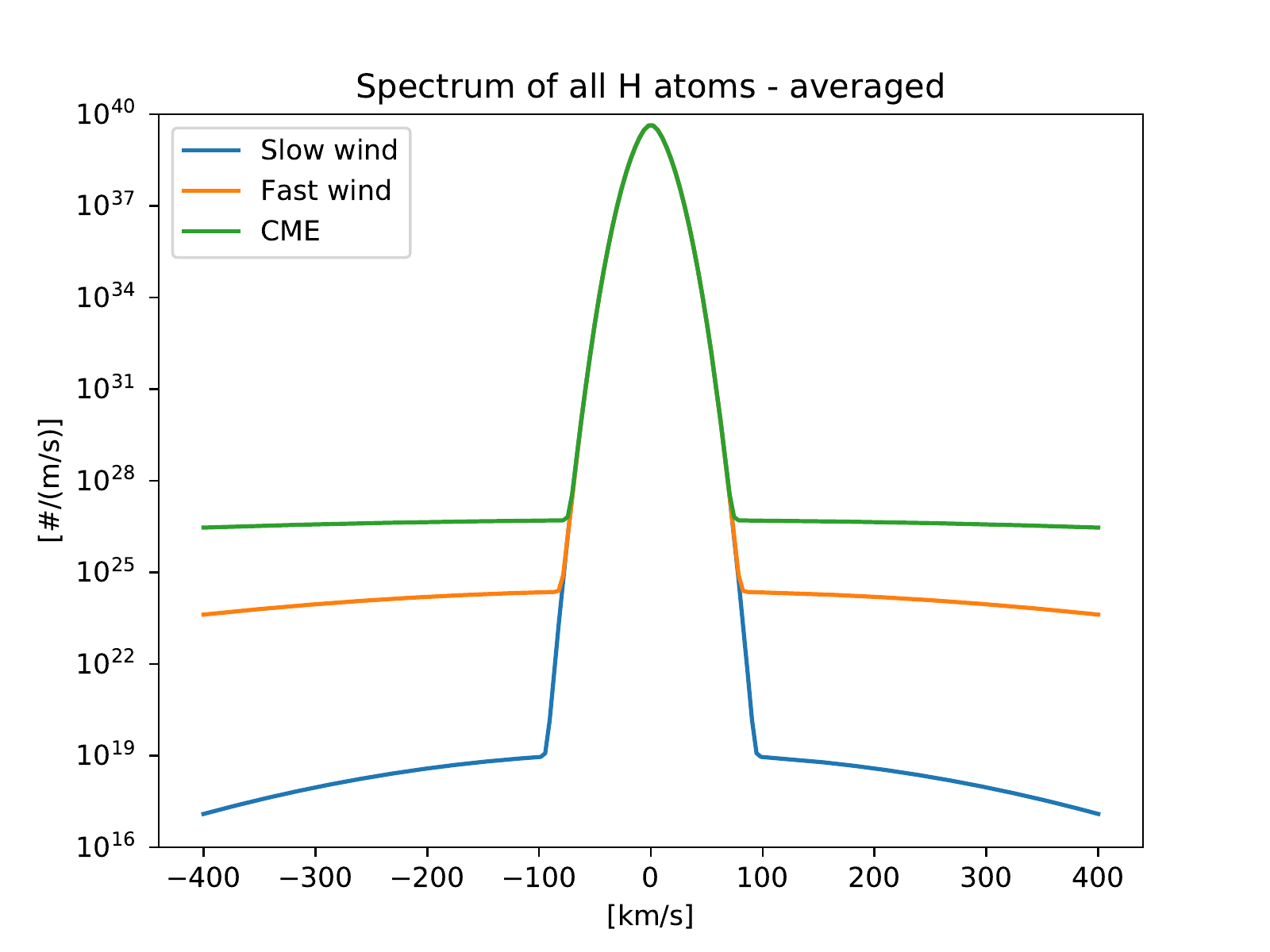}
\caption{Velocity spectra of neutral H atoms for the atmosphere of HD~189733b for mean (blue), maximum (orange), and CME conditions (green). The main peak represents the atmospheric neutrals, the low and wide wings are the ENAs with different bulk velocities and high temperatures.}
\label{f_avsp}
\end{figure}

\begin{figure}
\includegraphics[width=\columnwidth]{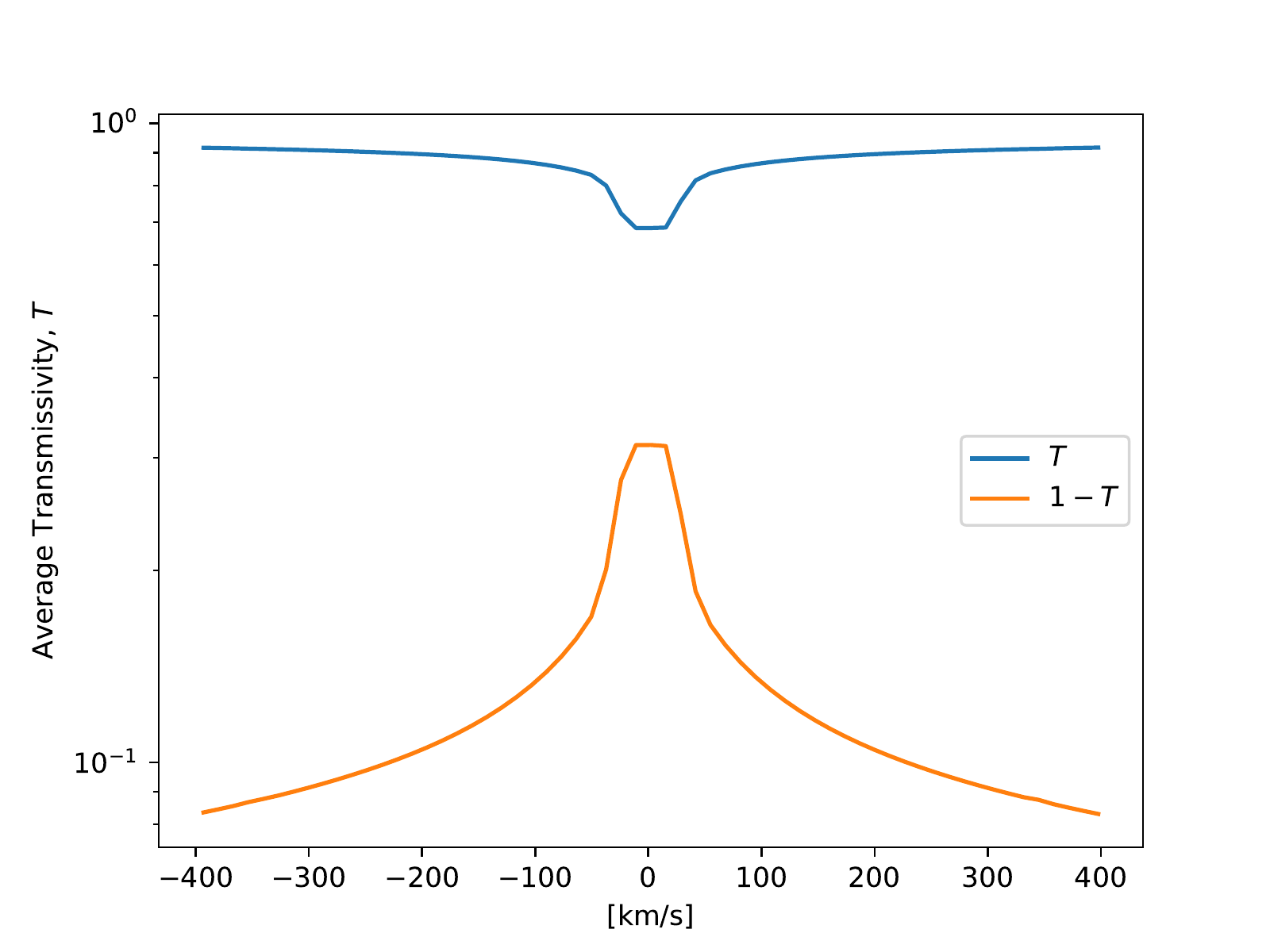}
\caption{Calculated transmissivity (blue) and corresponding absorption (orange) in the Ly$\alpha$ line as a function of velocity. The results for the mean wind parameters are shown, but the other two cases produce indistinguishable results due to the dominating contribution of atmospheric broadening.}
\label{f_transmissivity}
\end{figure}

\begin{figure}
\includegraphics[width=\columnwidth]{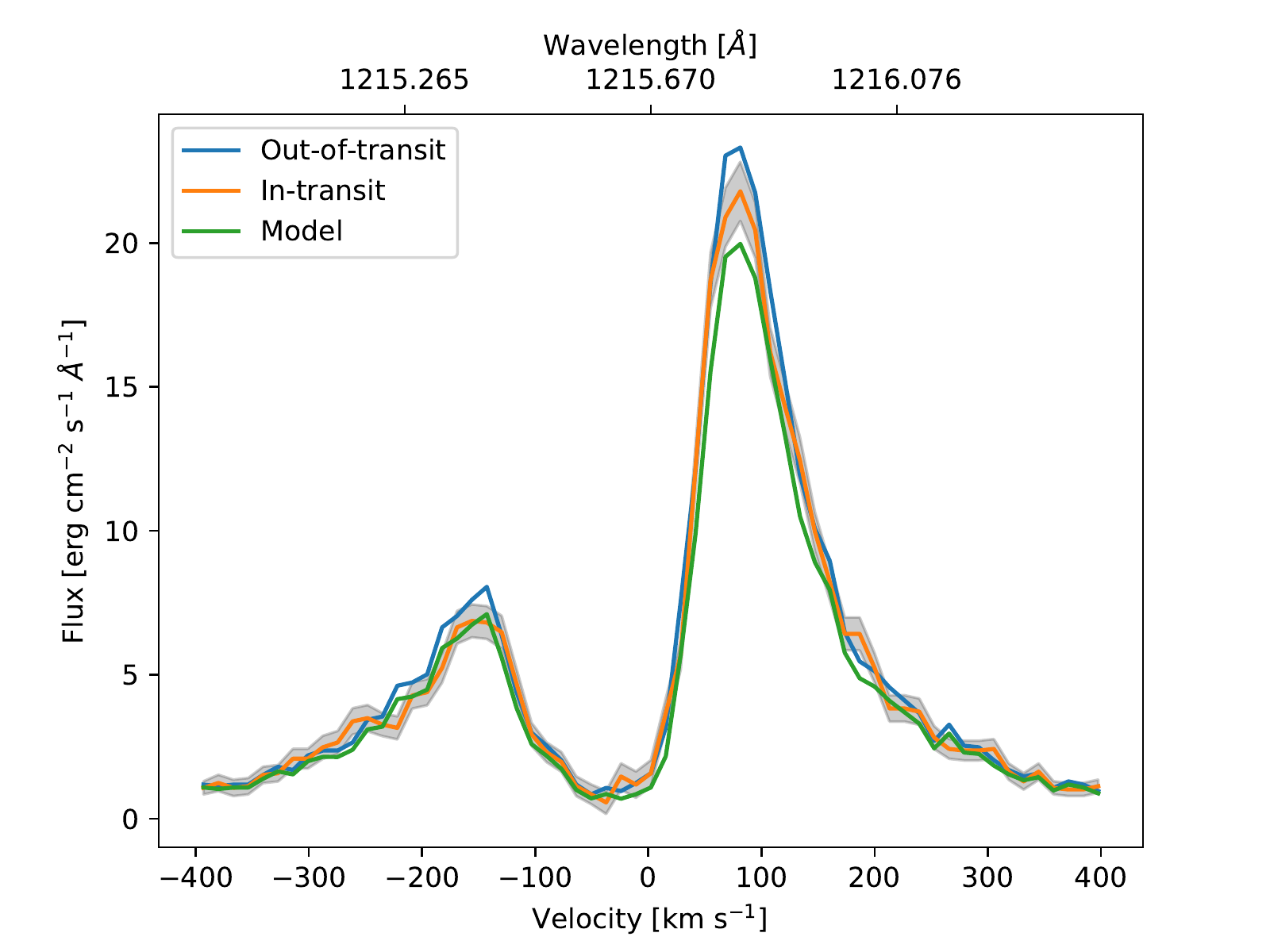}
\caption{Modeled absorption of HD~189733b compared to observations from 2011 \citep{LecavelierdesEtangs12}. The blue line shows the observed out-of-transit Ly$\alpha$ line profile of HD~189733. The green line shows the modeled in-transit absorption. The orange line shows the in-transit absorption observed in September 2011. The shaded region surrounding the orange line shows the observational errors. Our modeled spectrum reproduces the observed profile well (also in its red part) and is mostly within the error boundaries. Mean wind conditions were adopted, but the other cases produce identical results. The region between $-40$ and $+40$\,km\,s$^{-1}$ was affected by geocoronal emission in 2011 and should be ignored \citep{Bourrier13a}.} 
\label{f_absorption}
\end{figure}

To account for the contribution of the lower atmosphere, a Maxwellian velocity spectrum corresponding to a hydrogen gas with a specified column density and temperature is added to all pixels up to the outer extent of the atmosphere, according to the atmospheric profile calculated with the 1D hydrodynamic code. Similarly, to account for the ENAs, we added a Maxwellian spectrum corresponding to their temperatures with the central velocity located at their bulk velocity, which differs depending on their position relative to the magnetopause boundary. Due to the formation of a bow shock, ENAs are generated from decelerated stellar wind ions and are thus much slower than the stellar wind, but at the same time obtain a very high temperature, manifesting in a very broad spectrum spanning from very low up to very high velocities along the $x$-axis.

Figure~\ref{f_avsp} shows the velocity spectra of all neutral H atoms along the line of sight, including atmospheric particles (the main central peak in the plot) and the ENA population (seen as the wide ``wings'' on both sides of the main peak) for all three considered stellar wind cases. Positive velocities and negative velocities correspond to atoms flying toward and away from the star, respectively. The velocity spectrum of the atomic cloud can then be converted to frequencies via the relation $f = f_0 + v_x / \lambda_0$ with $f_0 = c / \lambda_0$, $\lambda_0 = 1215.65$\,\AA, and where $c$ is the speed of light. As one can see, ENAs form very wide wings around the central atmospheric peak. Different initial values for temperature, density, and velocity of the stellar wind influence the height and the flatness of the wings. However, one can see that all three cases produce very flat wings, which is due to a sharp temperature increase and wind deceleration near the planetary boundary.

Figure~\ref{f_transmissivity} shows the calculated average transmissivity and absorption for the mean wind spectrum shown in Fig.~\ref{f_avsp}. Despite differences in the velocity spectra in the ENA part, the transmissivity spectra are indistinguishable, so we do not show the other cases. This is due to the fact that the contribution from the ENAs produces very flat and low spectra, with the amount of particles in a given velocity bin not high enough to produce any significant contribution. Therefore, the absorption is mostly determined by the atmospheric broadening from the dense central peak, which is identical in all three cases, because we used the same atmospheric profile.

Finally, Fig.~\ref{f_absorption} compares the calculated absorption to the Ly$\alpha$ in-transit observations of HD~189733b from 2011. It was obtained by multiplying the out-of-transit spectrum (blue) by the transmissivity (Fig.~\ref{f_transmissivity}). This simple method may generate a bias if applied to the analyzed observations because of the combination of the broad line-spread function of STIS and the ISM absorption, yet this is limited to the region close to the line core where the ISM absorption is saturated, but should not affect the line wings where the planetary absorption is detected. In general, there is a good agreement between the calculated and observed spectra, although the absorption predicted by our model is slightly larger. Outside of the geocoronal emission region, the modeled blue-wing ($-400\ldots-40$\,km\,s$^{-1}$) absorption is 10.9\%, 1.7$\sigma$ larger than the observed value ($6.8\pm2.4$\%); for the red wing ($+40\ldots+400$\,km\,s$^{-1}$), we obtain, with 12.9\%, too much absorption by 5.5$\sigma$ (observed: $4.5\pm1.5$\%). This means that the model with our default parameters predicts neutral densities that are slightly too large. If we scaled the modeled absorption profile to match the blue-wing absorption, the red-wing absorption would be within 3$\sigma$ of the observations. The overabsorption could be due to limitations of our 1D hydrodynamic atmosphere model, as we used the atmospheric profiles along the star-planet line as the global input to the 3D MHD model. Recent 3D hydrodynamic models obtain more radially extended atmospheres along this direction compared to the terminator region \citep{Shaikhislamov18}.

As one could expect from Fig.~\ref{f_transmissivity}, we find insignificant differences of the absorption for the different wind scenarios due to the negligible contribution of ENAs in our model. This conclusion contradicts the one made by \citet{Kislyakova14b}, who were able to reproduce the Ly$\alpha$ observations of HD~209458b only assuming a specific wind configuration. This contradiction can be easily explained by the fact that the Direct Simulation Monte Carlo model by \citet{Kislyakova14b} did not account for the deceleration and temperature increase of the ENAs near the planetary obstacle. For this reason, their results only accounted for a Maxwellian spectrum according to the initial density, temperature, and velocity distribution. On the contrary, our results represent a better approximation of the wind properties near the planet, and show that different stellar wind conditions can produce similar Ly$\alpha$ signatures. Differences to previous studies are discussed in more detail in Section~\ref{sec:compare}.

One should keep in mind that we did not account for the compression and additional ionization of the planetary atmosphere by the stellar wind. Therefore, our results still represent an approximation, even though they present a significant improvement in comparison to earlier works by \citet{Holmstroem08} and \citet{Kislyakova14b}. However, these neglected processes could also be a reason for the slight overabsorption we obtained with our models.

\section{Discussion} \label{sec:disc}

\subsection{Comparison with other hydrodynamic models} \label{sec:otherhy}
As described above, our simulations yield a total mass-loss rate of $\dot{M}=2.5{-}5.4\times10^{10}$\,g\,s$^{-1}$, depending on the adopted XUV spectrum. Previous studies of this planet found $\dot{M}=4.8\times10^{10}{-}2\times10^{11}$\,g\,s$^{-1}$ for $F_\mathrm{XUV}=2\times10^4{-}10^5$\,erg\,cm$^{-2}$\,s$^{-1}$ \citep{Guo11}, $\dot{M} = 4.5\times10^{11}{-}9\times10^{11}$\,g\,s$^{-1}$ for $F_\mathrm{XUV}=24778$\,erg\,cm$^{-2}$\,s$^{-1}$ \citep[but different spectral energy distributions;][]{Guo16}, and $\dot{M} = 1.64\times10^{10}$\,g\,s$^{-1}$ for $F_\mathrm{XUV}=20893$\,erg\,cm$^{-2}$\,s$^{-1}$ \citep{Salz16a}\footnote{Their given value of $4.1\times10^{9}$\,g\,s$^{-1}$ corresponds to 1/4 of the isotropic mass-loss rate.}. Despite using similar XUV fluxes, the model of \citet{Salz16a} yielded a mass-loss rate lower by more than an order of magnitude compared to the other studies \citep{Guo11, Guo16}, which is likely because not all relevant radiative cooling processes were included in the latter models, leading to an overestimate of the escape rate. Our mass-loss rates are up to a factor of three higher than those of \citet{Salz16a}. Estimating the spectral shape with a ratio of fluxes like in \citet{Guo16} we find $\beta=F_{50-400\AA}/F_{50-900\AA}{\sim}F_{0-400\AA}/F_{0-912\AA}{\sim}0.8$ for both spectra, corresponding to a mass-loss rate of ${\sim}9\times10^{11}$\,g\,s$^{-1}$ in their model (see their Fig.~13), which is a factor of 15 higher than our results.

\begin{figure*}
$\begin{array}{cc}
\includegraphics[width=\columnwidth]{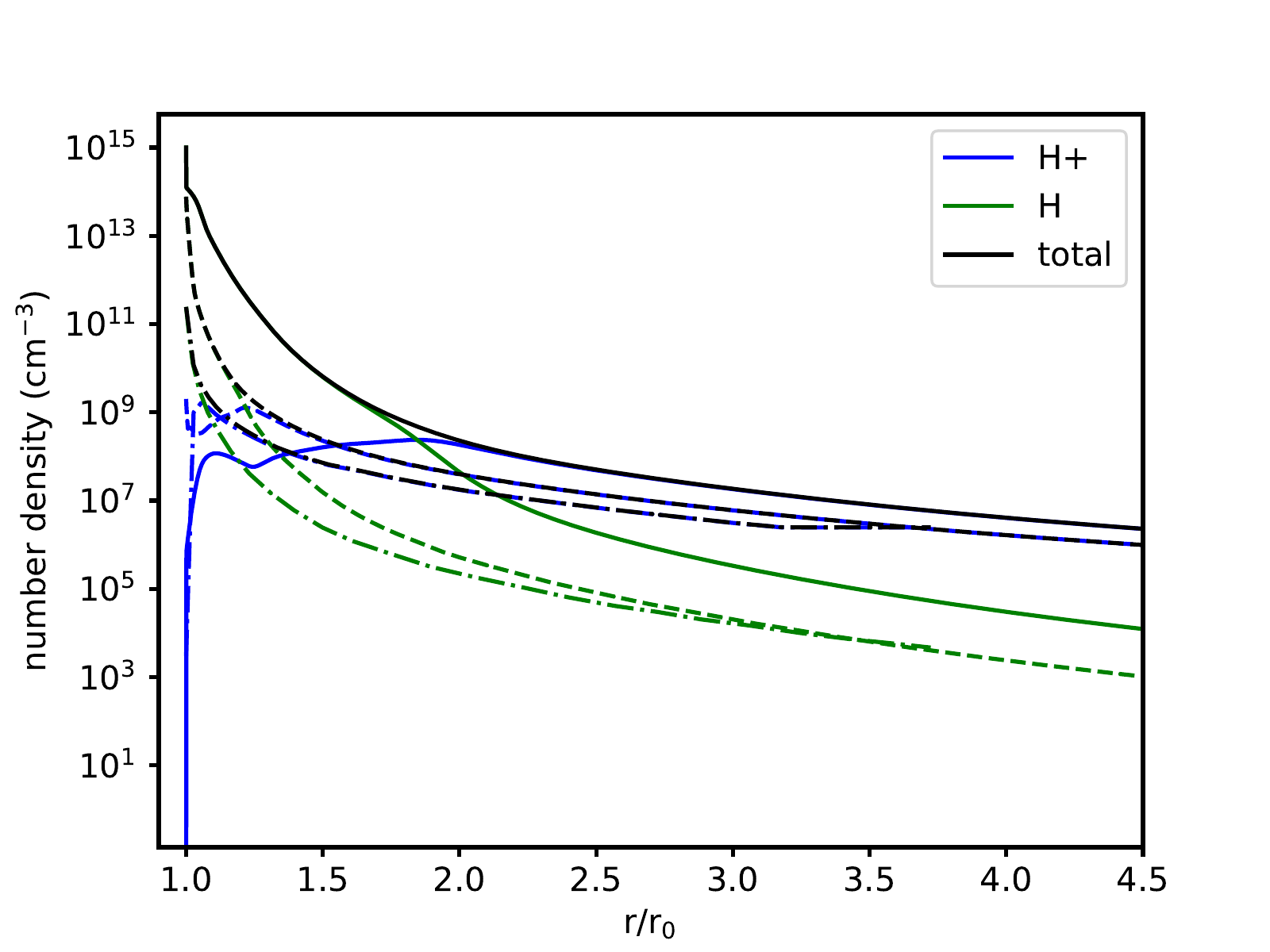} &
\includegraphics[width=\columnwidth]{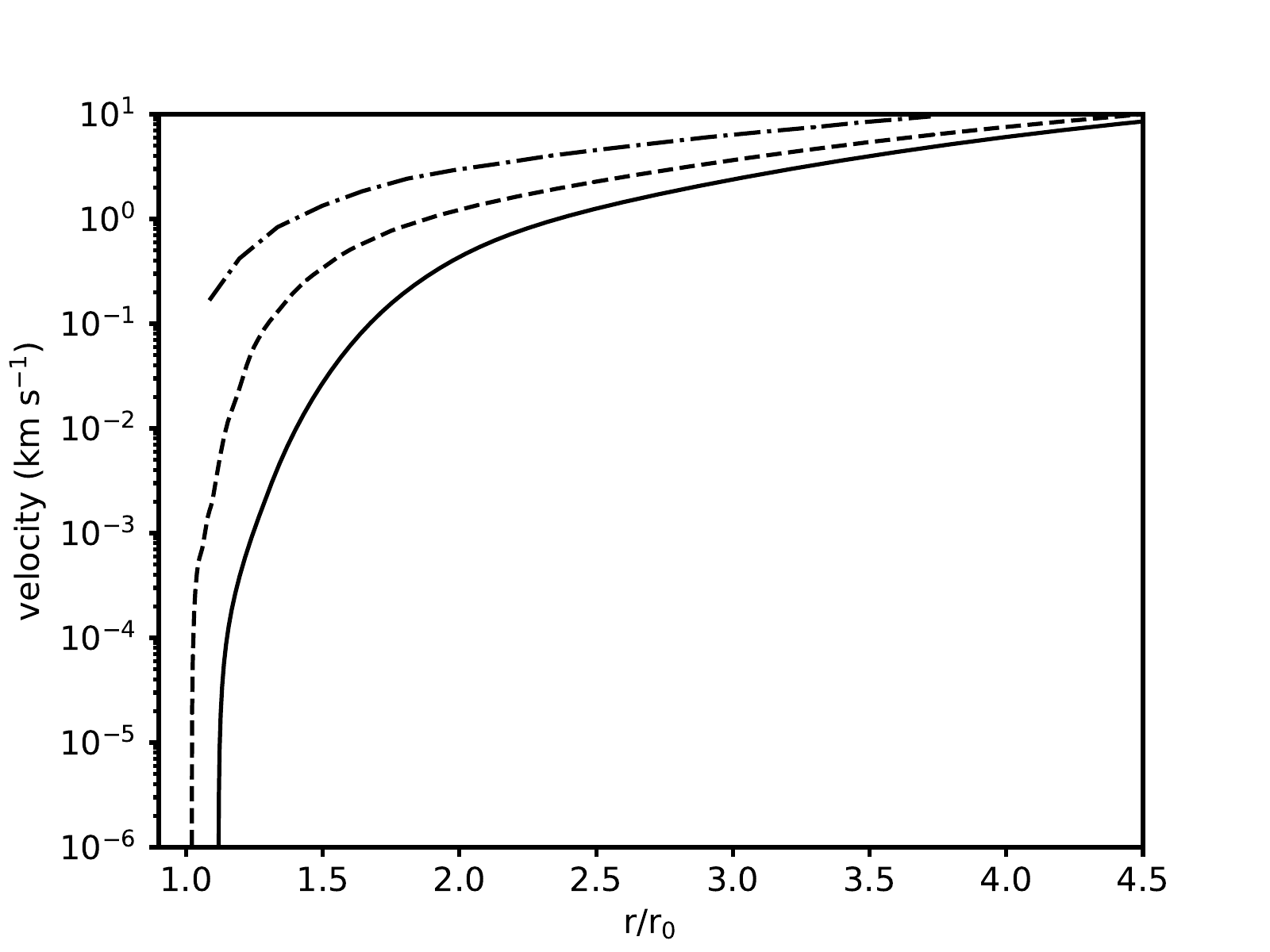} \\
\includegraphics[width=\columnwidth]{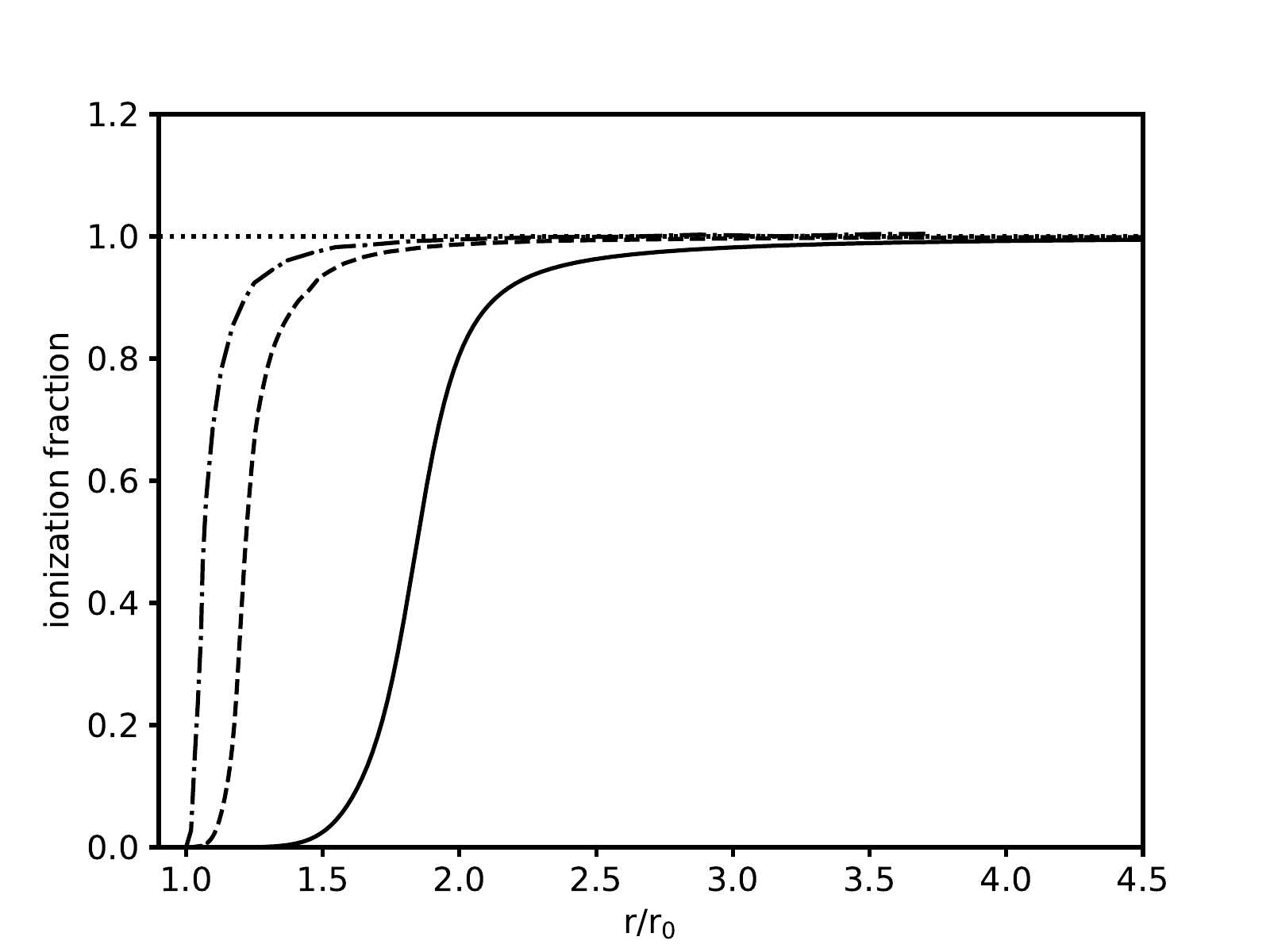} &
\includegraphics[width=\columnwidth]{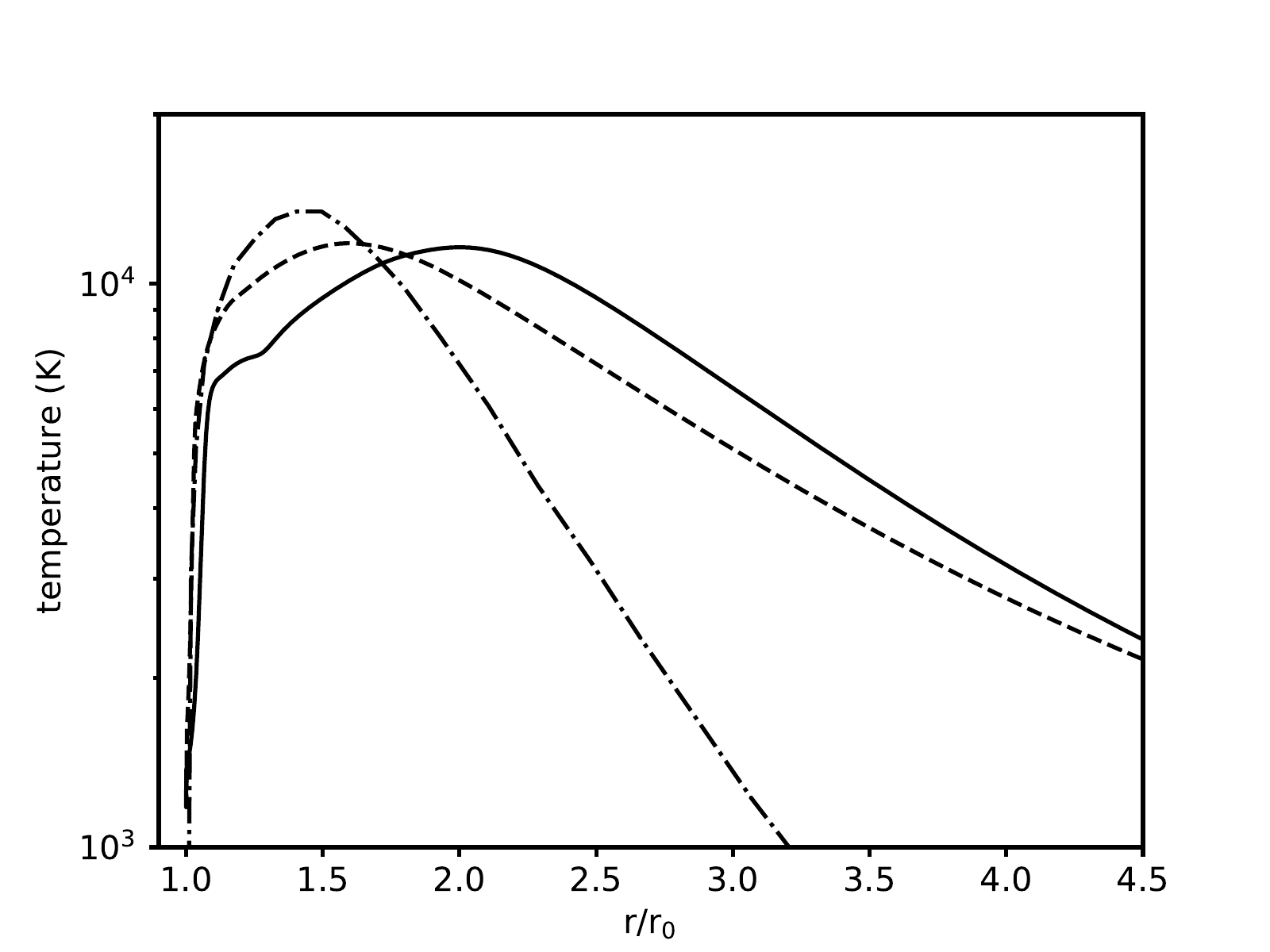}
\end{array}$
\caption{Atmospheric profiles for the \citetalias{Sanz-Forcada11} run (solid) compared with previous model results of \citep{Guo11} (dash-dotted) and \citep{Salz16a} (dashed).}
\label{fig:comp}
\end{figure*}

The atmospheric profiles are compared in Fig.~\ref{fig:comp}. For \citet{Guo11} we adopted their results for $F_\mathrm{XUV}=2\times10^4$\,erg\,cm$^{-2}$\,s$^{-1}$ (their Fig.~3). One can see that our model yields lower velocities, but higher densities throughout the computational domain. Moreover, ionization occurs at a larger height in our model compared to their results. This could be partly related to the much lower densities $n_0$ adopted in the other studies, as well as different XUV spectra. The temperature maximum is comparable to \citet{Salz16a}, but ours is located further out at about $2R_\mathrm{p}$ instead of $1.5R_\mathrm{p}$.

The main differences between the discussed models are: gray atmosphere approximation, no molecules, only H, only Ly$\alpha$ cooling in \citet{Guo11}; XUV spectra, molecules, H+He, but only H$_3^+$ cooling in \citet{Guo16}; XUV spectra, no molecules, H+He, all atomic radiative cooling processes in \citet{Salz16a}; XUV spectra, no molecules, only H, all atomic cooling processes in the present study. In addition, there are some differences in the adopted stellar and planetary parameters and lower boundary conditions, as well as usage of different solution methods of the hydrodynamic equations. Therefore, some differences in the mass-loss rates and atmospheric profiles are to be expected. We find the best qualitative agreement with \citet{Salz16} in that HD~189733b is an intermediate case between planets with high escape rates and stable planets in radiative equilibrium.

\subsection{Neglected processes}
There are several processes that are not considered in our present model that may have an impact on the planetary escape rates, and thus the modeled transit absorption. Firstly, usage of a 1D hydrodynamic code for the upper atmosphere limits the validity of the results to a region well within the Roche lobe. The average Roche lobe radius of HD~189733b is ${\sim}3.1R_\mathrm{p}$ \citep{Eggleton83}, smaller than the distance to the L1 point (cf., Section~\ref{sec:bc}). Close to and beyond the Roche lobe, the gas dynamics are dominated by 3D effects \citep[e.g.,][]{Bisikalo13a}. However, we find that even for mean stellar wind conditions, the pressure balance distance is at about $2.6\,R_\mathrm{p}$ (well within the Roche lobe), which justifies using the 1D results for the atmosphere up to this point. However, we note that the sonic point is reached close to the L1 point above $4\,R_\mathrm{p}$, which means that the dynamic pressure of the stellar wind at this point is larger than that of the planetary outflow. This means we could be overestimating the escape rates, because the stellar wind may confine the planetary mass-loss. However, it is possible that the confinement on the dayside is (partly) compensated by a stronger outflow on the nightside \citep{Murray-Clay09, Shaikhislamov16}. On the other hand, the absorbed XUV energy could also be radiated away by Ly$\alpha$ and free-free cooling due to the enhanced temperatures, which would lead to a reduction of the escape rates \citep{Salz16a}. Self-consistent modeling of planetary-stellar wind interaction would require a multidimensional multi-fluid model.

Due to the strong ionization of hot Jupiter atmospheres, an intrinsic planetary magnetic field may lead to a reduction of the escape rate \citep{Owen14, Trammell14, Khodachenko15a}. It is not known if HD~189733b possesses an intrinsic magnetic field strong enough to affect its mass loss. However, we do consider the generation of an induced magnetic field at the obstacle (magnetic barrier) in our MHD model \citep{Erkaev17}. An intrinsic planetary magnetic field may push the pressure balance distance between the planetary and stellar winds further out.

We include only hydrogen atoms and protons in our simulations. Neglecting H$_2$ and its related ions is justified for this planet, because their presence is confined to a very small region close to the optical radius \citep[see][and Appendix~\ref{sec:h2}]{Guo16}. This also means that cooling by H$_3^+$ is not important here due to its small number density. However, helium is likely to be present in the atmosphere with cosmic abundances. \citet{Salz16a} compared hydrodynamic simulations with and without including He and found mass-loss rates lower by a factor of two in the former case.

\subsection{Comparison with previous transit signature modeling}\label{sec:compare}
\citet{Bourrier13a} described the details of the 3D particle model, which they used to explain the Ly$\alpha$ observations of HD~189733b \citep{LecavelierdesEtangs12}. It depends on the planetary mass-loss rate of neutral H, ionizing flux (limiting the lifetime of H atoms), and stellar wind parameters at the planet's orbit ($V_\mathrm{sw}$, $N_\mathrm{sw}$, $T_\mathrm{sw}$). It includes radiation pressure, planetary and stellar gravities, charge exchange with stellar wind protons, and self-shielding of stellar photons and stellar wind protons by the H cloud. They found their best fits to the 2011 observations with $V_\mathrm{sw}=200$\,km\,s$^{-1}$, $T_\mathrm{sw}=3\times10^4$\,K, $N_\mathrm{sw}=10^3{-}3\times10^5$\,cm$^{-3}$, an ionizing flux ten times the solar one, and neutral H escape rates of $5\times10^8{-}1.5\times10^9$\,g\,s$^{-1}$. To reproduce the nondetection in 2010, for fixed wind parameters the H escape rates would need to be 5-20 times lower, whereas for a fixed ionizing flux and escape rate, the stellar wind proton densities would need to be about a factor of ten lower. This indicates that the 2011 observations can be explained with a higher H escape rate (but not much higher ionizing flux) or a denser wind compared to the conditions in 2010. \citet{LecavelierdesEtangs12} suggested that these discrepant observations could be due to a flare that occurred $\sim$8\,h prior to the transit in 2011.

The model of \citet{Bourrier13a} can explain the observed Ly$\alpha$ absorption in 2011 if the neutral H mass-loss rate at 2.95\,$R_\mathrm{p}$ (the lower boundary of their model) is in the order of $10^9$\,g\,s$^{-1}$. Our H mass-loss rate at this altitude is comparable to this requirement, but similar for average XUV and flaring cases (see Section~\ref{sec:flare}), even if ignoring the limited duration of the enhanced XUV flux during a flare. Hence, our neutral loss rates are consistent with their model for the 2011 observations, but they cannot explain the nondetection in 2010, because the XUV fluxes must have been very different. Either the fluxes were much lower, reducing the total mass-loss rate, or much higher, reducing the neutral H density due to enhanced ionization. This contradicts the necessity of similar ionizing fluxes at both epochs \citep{Bourrier13a}. Test simulations with significantly enhanced XUV fluxes show that even an increase by a factor of 1000 only decreases the neutral H mass-loss rate by a factor of three. For lower XUV fluxes, the total mass-loss rate decreases, but the neutral H mass-loss rate increases relative to our results (e.g., about a factor of two larger for a factor of five lower XUV). However, we stress that other hydrodynamic models of this planet find higher ionization (and therefore lower H densities) at 2.95\,$R_\mathrm{p}$, yielding much lower neutral loss rates \citep{Guo11, Salz16a, Chadney17}. We find that this is mainly related to the lower number densities assumed at the lower boundary in these studies. If we run our model with lower $n_0$ values, we also obtain neutral densities at 2.95\,$R_\mathrm{p}$ that are too low compared to those required in \citet{Bourrier13a}. This indicates that the neutral density at a given altitude does not just depend on the irradiating stellar XUV flux, but strongly depends on $n_0$, stronger than the total mass-loss rate. Higher values of $n_0$ shift the neutral-ion transition to larger heights. However, we find that lowering $n_0$ results in incomplete absorption of the XUV radiation in the computational domain, which makes the resulting mass-loss rates strongly dependent on the assumed $n_0$.

One may argue that if the true value of $n_0$ were indeed much lower than our assumptions, the absorption signal from the planetary atmosphere could be much weaker than what our results show and not compatible with the 2011 observations. However, it is unlikely that $n_0$ could be much smaller, because too much ionizing radiation would then reach layers close to the optical radius or even below, which is inconsistent with the observed value of the optical radius. The only possibility that the upper atmosphere could absorb most of the ionizing radiation above $R_\mathrm{p}$ despite a lower H density could be efficient absorption by non-hydrogen species (such as the absorption of X-rays by metals). Future studies should investigate whether this effect could be relevant for reasonable atmospheric metallicities.

The other possible explanation for the transit variability in the \citet{Bourrier13a} model is a higher stellar wind density in 2011 by a factor of about ten compared to 2010. Our results are not consistent with this picture, as we find a negligible influence of the stellar wind on the planetary absorption, even in the case of a putative CME that may have accompanied the flare. The absorption in our model is completely dominated by the contribution from the parts of the planetary atmosphere below the pressure balance distance (Section~\ref{sec:abs}). We also tested the effect of adopting stellar wind parameters more similar to the best fitting model of \citet{Bourrier13a} in our MHD model, namely $V_\mathrm{sw}=200$\,km\,s$^{-1}$, $T_\mathrm{sw}=3\times10^4$\,K, $N_\mathrm{sw}=10^6$\,cm$^{-3}$, and keeping the magnetic field parameters of our mean wind case. We find that the maximum ENA density is much smaller than in the cases from Table~\ref{tab:sw}, and that their bulk speed is low compared to the local thermal speed of the stellar wind. Therefore, such a wind does not produce a tail and a localized blue-shifted absorption signature in our model.

Previous modeling approaches employed Monte Carlo particle codes to model the generation and distribution of ENAs around exoplanets \citep{Holmstroem08, Bourrier13a, Kislyakova14b}. ENAs generated in such models by charge exchange of neutral planetary atoms with the stellar wind have a velocity distribution peaked at the stellar wind speed. After formation, ENAs are generally not coupled to the plasma flow due to lack of collisions; their motion is controlled by gravitational, centrifugal and Coriolis forces, as well as radiation pressure \citep[e.g.,][]{Kislyakova14b}. For HD~189733b, the situation is different. The high densities in the circumplanetary environment lead to an efficient coupling of ENAs and ions by collisions\footnote{Taking $n{\gtrsim}N_\mathrm{sw}$, $\sigma_\mathrm{col}{\sim}10^{-15}$\,cm$^{2}$ and $X{\sim}R_\mathrm{p}$ yields a Knudsen number $Kn{<}1$ in the modeling domain.}. This justifies modeling ENAs in the HD~189733 system with our MHD code. In our model, the ENAs are generated inside the bow shock. These ENAs thus have velocities much smaller than that of the stellar wind because they are formed from decelerated wind ions. This leads to ENA velocity spectra that peak at small velocities (Fig.~\ref{f_avsp}). Due to strong ionization by the intense XUV flux, the upper planetary atmosphere is almost completely ionized, so the generation of additional ENAs outside the bow shock (which would have velocities comparable to the stellar wind) can likely be neglected here due to the very small amount of available planetary neutrals.

As described in Section~\ref{sec:otherhy}, \citet{Guo16} used a hydrodynamic upper atmosphere model similar to ours in combination with different spectral energy distributions of the stellar XUV radiation. They are able to reproduce the 2011 observations with a spectral shape $\beta=0.76$, which is very similar to our adopted \citetalias{Sanz-Forcada11} spectrum ($\beta=0.82$) that also reproduces the transit absorption. The nondetection in 2010 would require a spectral shape of $\beta=0.38$ \citep{Guo16}. Our results are consistent with theirs in that the 2011 observations can be reproduced by solely considering the planetary atmosphere, ENAs being negligible. However, due to the difficulties in observing the EUV emission of stars, it is unknown if such strong temporal changes in spectral shape would be possible while leaving the total XUV emission rather similar.

\subsection{Effects of flares}
As described before, with our adopted mean XUV spectrum we can reproduce the transit observations from 2011. Thus, no transient stellar variability, like a flare, is needed to produce the planetary absorption. As discussed in Section~\ref{sec:flare}, the X-ray flare observed 8\,h before the planetary transit in 2011 was unlikely able to induce any measurable difference in planetary absorption. Stellar emissions higher by a factor of four, either in X-rays only or XUV, raise the total mass-loss rates by 60\% or 120\%, respectively. In the latter case, however, the neutral H mass-loss rate drops because of increased ionization. Using a realistic time evolution of a flare and not just a constant elevation of the irradiation would make the effects even smaller. Our findings are in agreement with \citet{Chadney17} who studied the effect of flares on the upper atmospheres of close-in exoplanets. They also found that a flare is likely not capable of producing the strong modulation of the transit absorption as seen in HD~189733b, but suggested that a stellar proton event could cause such differences.

\subsection{Stellar activity effects on the Ly$\alpha$ transit absorption}
Stellar activity can also affect the study of Ly$\alpha$ transit absorption in other ways than a modulation of the planet's atmospheric properties. To obtain the amount of absorption during the transit, the in-transit spectrum is usually compared to an out-of-transit spectrum. This can cause difficulties in the case of an active star for two main reasons. Firstly, the surface of a star may not be homogeneous due to the presence of active regions. Secondly, the star may be variable on time scales comparable to the planetary transit, so the stellar background emission during the transit could be different from the stellar spectrum recorded at some time before the transit.

The first effect was studied in \citet{Llama15}, who simulated hot Jupiter transits over solar X-ray, EUV, FUV, and optical images. Both occulted and unocculted active regions can modify the obtained absorption. It also depends if the active regions are brighter or darker than the inactive stellar disk in a given wavelength range. For bright active regions, unocculted ones lead to shallower transits, and occulted ones to deeper transits. If a star has an activity belt like the Sun, it depends on where a planet transits the disk if the observed absorption is deeper or shallower than if assuming a homogeneous disk. For the case of HD~189733b, \citet{Llama16} stated that other studies found no bumps in lightcurves, meaning no spot crossings, so the planet is unlikely to transit an active belt (in the case that the star actually has one). Since unocculted regions make a transit shallower, it is possible that the 2010 observations were affected by an increased number of unocculted active regions. However, we remind the reader that the planetary absorption in Ly$\alpha$ can only be studied in the line wings due to geocoronal emission and ISM absorption. The wings of the Ly$\alpha$ line show a much more homogeneous activity pattern than the core, since they are formed at lower temperatures deeper in the chromosphere \citep{Salz16a}.

For the second effect, \citet{Llama16} simulated transits in light curves of the disk-integrated solar Ly$\alpha$ line flux. Although the recovered planetary radii can be up to 50\% larger for solar-like variability, they found that the effect is unlikely to be responsible for the strong transit absorptions detected on stars like HD~189733b, even if the solar variability is scaled up in amplitude.

\section{Conclusions} \label{sec:conc}
We modeled the Ly$\alpha$ transit absorption of the hot Jupiter HD~189733b using a 1D hydrodynamic code for the upper atmosphere and a 3D MHD code for the planetary-stellar wind interaction and related production of ENAs. We find that the transit absorption observed in 2011 can be reproduced reasonably well with typical stellar XUV emissions ($F_\mathrm{XUV}\sim1.8\times10^4$\,erg\,cm$^{-2}$\,s$^{-1}$). We note that with our adopted default parameters, the model produces some extra-absorption across the Ly$\alpha$ line, which could be due to limitations of the 1D approach and/or the unaccounted compression of the atmosphere by the stellar wind. The influence of enhanced stellar irradiation during a flare similar to the one observed about 8\,h before the transit in 2011 is too small to significantly affect the neutral hydrogen outflow rates. Moreover, we find with our modeling approach that the absorption signature is dominated by the atmospheric neutral hydrogen and that the ENAs produced by charge-exchange with the stellar wind have a negligibly small effect, as they are produced inside the bow shock and have small velocities and high temperatures due to the stellar wind deceleration at the planetary obstacle. Moreover, we find no differences in planetary absorption for all tested stellar wind cases, including a CME event. This raises the question if the nondetection in 2010 was actually ``anomalous'', and not the absorption signature detected in 2011 as previously suggested. The variations between the two observations may be related to significant differences in stellar activity, by magnitude or spectral shape \citep{Guo16}, or its variability \citep{Llama16}. Observations of this system in Ly$\alpha$ at a further epoch, preferentially with simultaneous X-ray monitoring to constrain the stellar XUV emission, would help to clarify this issue.

\begin{acknowledgements}
We thank the referee for valuable comments and suggestions. We thank A. Vidotto for providing the modeled stellar wind parameters. This work was supported by the Austrian Science Fund (FWF) project P27256-N27 ``Characterizing Stellar and Exoplanetary Environments via Modeling of Lyman-$\alpha$ Transit Observations of Hot Jupiters''. The authors further acknowledge the support by the FWF NFN project S11601-N16 ``Pathways to Habitability: From Disks to Active Stars, Planets and Life'', and the related FWF NFN subproject S11607-N16 ``Particle/Radiative Interactions with Upper Atmospheres of Planetary Bodies Under Extreme Stellar Conditions'' (KK, HL, NVE), as well as FWF projects P30949-N36 (PO, ML) and I2939-N27. LF, DK and NVE acknowledge also the Austrian Forschungsf\"orderungsgesellschaft FFG project ``TAPAS4CHEOPS'' P853993. NVE and VAI acknowledge support by the Russian Science Foundation grant No 18-12-00080.
\end{acknowledgements}

%
\bibliographystyle{aa} 
\bibliography{hd189733b} 

\begin{thebibliography}{127}
\expandafter\ifx\csname natexlab\endcsname\relax\def\natexlab#1{#1}\fi

\bibitem[{{Adams}(2011)}]{Adams11}
{Adams}, F.~C. 2011, \apj, 730, 27

\bibitem[{{Alvarado-G{\'o}mez} {et~al.}(2018){Alvarado-G{\'o}mez}, {Drake},
  {Cohen}, {Moschou}, \& {Garraffo}}]{Alvarado-Gomez18}
{Alvarado-G{\'o}mez}, J.~D., {Drake}, J.~J., {Cohen}, O., {Moschou}, S.~P., \&
  {Garraffo}, C. 2018, \apj, 862, 93

\bibitem[{{Bakos} {et~al.}(2006){Bakos}, {P{\'a}l}, {Latham}, {Noyes}, \&
  {Stefanik}}]{Bakos06}
{Bakos}, G.~{\'A}., {P{\'a}l}, A., {Latham}, D.~W., {Noyes}, R.~W., \&
  {Stefanik}, R.~P. 2006, \apjl, 641, L57

\bibitem[{{Ballester} \& {Ben-Jaffel}(2015)}]{Ballester15}
{Ballester}, G.~E. \& {Ben-Jaffel}, L. 2015, \apj, 804, 116

\bibitem[{{Ballester} {et~al.}(2007){Ballester}, {Sing}, \&
  {Herbert}}]{Ballester07}
{Ballester}, G.~E., {Sing}, D.~K., \& {Herbert}, F. 2007, \nat, 445, 511

\bibitem[{{Barnes} {et~al.}(2016){Barnes}, {Haswell}, {Staab}, \&
  {Anglada-Escud{\'e}}}]{Barnes16a}
{Barnes}, J.~R., {Haswell}, C.~A., {Staab}, D., \& {Anglada-Escud{\'e}}, G.
  2016, \mnras, 462, 1012

\bibitem[{{Ben-Jaffel}(2007)}]{Ben-Jaffel07}
{Ben-Jaffel}, L. 2007, \apjl, 671, L61

\bibitem[{{Ben-Jaffel}(2008)}]{Ben-Jaffel08}
{Ben-Jaffel}, L. 2008, \apj, 688, 1352

\bibitem[{{Ben-Jaffel} \& {Ballester}(2013)}]{Ben-Jaffel13}
{Ben-Jaffel}, L. \& {Ballester}, G.~E. 2013, \aap, 553, A52

\bibitem[{{Ben-Jaffel} \& {Sona Hosseini}(2010)}]{Ben-Jaffel10}
{Ben-Jaffel}, L. \& {Sona Hosseini}, S. 2010, \apj, 709, 1284

\bibitem[{{Bisikalo} {et~al.}(2013){Bisikalo}, {Kaygorodov}, {Ionov},
  {Shematovich}, {Lammer}, \& {Fossati}}]{Bisikalo13a}
{Bisikalo}, D., {Kaygorodov}, P., {Ionov}, D., {et~al.} 2013, \apj, 764, 19

\bibitem[{{Bouchy} {et~al.}(2005){Bouchy}, {Udry}, {Mayor}, {Moutou}, {Pont},
  {Iribarne}, {da Silva}, {Ilovaisky}, {Queloz}, {Santos}, {S{\'e}gransan}, \&
  {Zucker}}]{Bouchy05}
{Bouchy}, F., {Udry}, S., {Mayor}, M., {et~al.} 2005, \aap, 444, L15

\bibitem[{{Bourrier} {et~al.}(2015){Bourrier}, {Ehrenreich}, \& {Lecavelier des
  Etangs}}]{Bourrier15}
{Bourrier}, V., {Ehrenreich}, D., \& {Lecavelier des Etangs}, A. 2015, \aap,
  582, A65

\bibitem[{{Bourrier} \& {Lecavelier des Etangs}(2013)}]{Bourrier13a}
{Bourrier}, V. \& {Lecavelier des Etangs}, A. 2013, \aap, 557, A124

\bibitem[{{Bourrier} {et~al.}(2013){Bourrier}, {Lecavelier des Etangs},
  {Dupuy}, {Ehrenreich}, {Vidal-Madjar}, {H{\'e}brard}, {Ballester},
  {D{\'e}sert}, {Ferlet}, {Sing}, \& {Wheatley}}]{Bourrier13}
{Bourrier}, V., {Lecavelier des Etangs}, A., {Dupuy}, H., {et~al.} 2013, \aap,
  551, A63

\bibitem[{{Bourrier} {et~al.}(2016){Bourrier}, {Lecavelier des Etangs},
  {Ehrenreich}, {Tanaka}, \& {Vidotto}}]{Bourrier16}
{Bourrier}, V., {Lecavelier des Etangs}, A., {Ehrenreich}, D., {Tanaka}, Y.~A.,
  \& {Vidotto}, A.~A. 2016, \aap, 591, A121

\bibitem[{{Cauley} {et~al.}(2017{\natexlab{a}}){Cauley}, {Redfield}, \&
  {Jensen}}]{Cauley17b}
{Cauley}, P.~W., {Redfield}, S., \& {Jensen}, A.~G. 2017{\natexlab{a}}, \aj,
  153, 217

\bibitem[{{Cauley} {et~al.}(2017{\natexlab{b}}){Cauley}, {Redfield}, \&
  {Jensen}}]{Cauley17a}
{Cauley}, P.~W., {Redfield}, S., \& {Jensen}, A.~G. 2017{\natexlab{b}}, \aj,
  153, 185

\bibitem[{{Cauley} {et~al.}(2016){Cauley}, {Redfield}, {Jensen}, \&
  {Barman}}]{Cauley16}
{Cauley}, P.~W., {Redfield}, S., {Jensen}, A.~G., \& {Barman}, T. 2016, \aj,
  152, 20

\bibitem[{{Cauley} {et~al.}(2015){Cauley}, {Redfield}, {Jensen}, {Barman},
  {Endl}, \& {Cochran}}]{Cauley15}
{Cauley}, P.~W., {Redfield}, S., {Jensen}, A.~G., {et~al.} 2015, \apj, 810, 13

\bibitem[{{Chadney} {et~al.}(2015){Chadney}, {Galand}, {Unruh}, {Koskinen}, \&
  {Sanz-Forcada}}]{Chadney15}
{Chadney}, J.~M., {Galand}, M., {Unruh}, Y.~C., {Koskinen}, T.~T., \&
  {Sanz-Forcada}, J. 2015, Icarus, 250, 357

\bibitem[{{Chadney} {et~al.}(2017){Chadney}, {Koskinen}, {Galand}, {Unruh}, \&
  {Sanz-Forcada}}]{Chadney17}
{Chadney}, J.~M., {Koskinen}, T.~T., {Galand}, M., {Unruh}, Y.~C., \&
  {Sanz-Forcada}, J. 2017, \aap, 608, A75

\bibitem[{{Christie} {et~al.}(2016){Christie}, {Arras}, \& {Li}}]{Christie16}
{Christie}, D., {Arras}, P., \& {Li}, Z.-Y. 2016, \apj, 820, 3

\bibitem[{{Cohen} {et~al.}(2011){Cohen}, {Kashyap}, {Drake}, {Sokolov},
  {Garraffo}, \& {Gombosi}}]{Cohen11}
{Cohen}, O., {Kashyap}, V.~L., {Drake}, J.~J., {et~al.} 2011, \apj, 733, 67

\bibitem[{{Davis}(1984)}]{Davis84}
{Davis}, S.~F. 1984, ICASE Report, 84-20, 1

\bibitem[{{de Kok} {et~al.}(2013){de Kok}, {Brogi}, {Snellen}, {Birkby},
  {Albrecht}, \& {de Mooij}}]{deKok13}
{de Kok}, R.~J., {Brogi}, M., {Snellen}, I.~A.~G., {et~al.} 2013, \aap, 554,
  A82

\bibitem[{Debrecht {et~al.}(2018)Debrecht, Carroll-Nellenback, Frank, Fossati,
  Blackman, \& Dobbs-Dixon}]{Debrecht18}
Debrecht, A., Carroll-Nellenback, J., Frank, A., {et~al.} 2018, \mnras, 478,
  2592

\bibitem[{{Drake} {et~al.}(2013){Drake}, {Cohen}, {Yashiro}, \&
  {Gopalswamy}}]{Drake13}
{Drake}, J.~J., {Cohen}, O., {Yashiro}, S., \& {Gopalswamy}, N. 2013, \apj,
  764, 170

\bibitem[{{Eggleton}(1983)}]{Eggleton83}
{Eggleton}, P.~P. 1983, \apj, 268, 368

\bibitem[{{Ehrenreich} {et~al.}(2012){Ehrenreich}, {Bourrier}, {Bonfils},
  {Lecavelier des Etangs}, {H{\'e}brard}, {Sing}, {Wheatley}, {Vidal-Madjar},
  {Delfosse}, {Udry}, {Forveille}, \& {Moutou}}]{Ehrenreich12}
{Ehrenreich}, D., {Bourrier}, V., {Bonfils}, X., {et~al.} 2012, \aap, 547, A18

\bibitem[{Ehrenreich {et~al.}(2015)Ehrenreich, Bourrier, Wheatley, des Etangs,
  H�brard, Udry, Bonfils, Delfosse, D�sert, Sing, \&
  Vidal-Madjar}]{Ehrenreich15}
Ehrenreich, D., Bourrier, V., Wheatley, P.~J., {et~al.} 2015, Nature, 522,, 459

\bibitem[{{Ehrenreich} {et~al.}(2008){Ehrenreich}, {Lecavelier Des Etangs},
  {H{\'e}brard}, {D{\'e}sert}, {Vidal-Madjar}, {McConnell}, {Parkinson},
  {Ballester}, \& {Ferlet}}]{Ehrenreich08}
{Ehrenreich}, D., {Lecavelier Des Etangs}, A., {H{\'e}brard}, G., {et~al.}
  2008, \aap, 483, 933

\bibitem[{{Ekenb{\"a}ck} {et~al.}(2010){Ekenb{\"a}ck}, {Holmstr{\"o}m}, {Wurz},
  {Grie{\ss}meier}, {Lammer}, {Selsis}, \& {Penz}}]{Ekenbaeck10}
{Ekenb{\"a}ck}, A., {Holmstr{\"o}m}, M., {Wurz}, P., {et~al.} 2010, \apj, 709,
  670

\bibitem[{{Erkaev} {et~al.}(2007){Erkaev}, {Kulikov}, {Lammer}, {Selsis},
  {Langmayr}, {Jaritz}, \& {Biernat}}]{Erkaev07}
{Erkaev}, N.~V., {Kulikov}, Y.~N., {Lammer}, H., {et~al.} 2007, \aap, 472, 329

\bibitem[{{Erkaev} {et~al.}(2016){Erkaev}, {Lammer}, {Odert}, {Kislyakova},
  {Johnstone}, {G{\"u}del}, \& {Khodachenko}}]{Erkaev16}
{Erkaev}, N.~V., {Lammer}, H., {Odert}, P., {et~al.} 2016, \mnras, 460, 1300

\bibitem[{{Erkaev} {et~al.}(2015){Erkaev}, {Lammer}, {Odert}, {Kulikov}, \&
  {Kislyakova}}]{Erkaev15}
{Erkaev}, N.~V., {Lammer}, H., {Odert}, P., {Kulikov}, Y.~N., \& {Kislyakova},
  K.~G. 2015, \mnras, 448, 1916

\bibitem[{{Erkaev} {et~al.}(2013){Erkaev}, {Lammer}, {Odert}, {Kulikov},
  {Kislyakova}, {Khodachenko}, {G{\"u}del}, {Hanslmeier}, \&
  {Biernat}}]{Erkaev13}
{Erkaev}, N.~V., {Lammer}, H., {Odert}, P., {et~al.} 2013, Astrobiology, 13,
  1011

\bibitem[{{Erkaev} {et~al.}(2017){Erkaev}, {Odert}, {Lammer}, {Kislyakova},
  {Fossati}, {Mezentsev}, {Johnstone}, {Kubyshkina}, {Shaikhislamov}, \&
  {Khodachenko}}]{Erkaev17}
{Erkaev}, N.~V., {Odert}, P., {Lammer}, H., {et~al.} 2017, \mnras, 470, 4330

\bibitem[{{Fares} {et~al.}(2010){Fares}, {Donati}, {Moutou}, {Jardine},
  {Grie{\ss}meier}, {Zarka}, {Shkolnik}, {Bohlender}, {Catala}, \& {Collier
  Cameron}}]{Fares10}
{Fares}, R., {Donati}, J.-F., {Moutou}, C., {et~al.} 2010, \mnras, 406, 409

\bibitem[{{Fossati} {et~al.}(2010){Fossati}, {Haswell}, {Froning}, {Hebb},
  {Holmes}, {Kolb}, {Helling}, {Carter}, {Wheatley}, {Cameron}, {Loeillet},
  {Pollacco}, {Street}, {Stempels}, {Simpson}, {Udry}, {Joshi}, {West},
  {Skillen}, \& {Wilson}}]{Fossati10}
{Fossati}, L., {Haswell}, C.~A., {Froning}, C.~S., {et~al.} 2010, \apjl, 714,
  L222

\bibitem[{{France} {et~al.}(2013){France}, {Froning}, {Linsky}, {Roberge},
  {Stocke}, {Tian}, {Bushinsky}, {D{\'e}sert}, {Mauas}, {Vieytes}, \&
  {Walkowicz}}]{France13}
{France}, K., {Froning}, C.~S., {Linsky}, J.~L., {et~al.} 2013, \apj, 763, 149

\bibitem[{{Gaia Collaboration} {et~al.}(2016){Gaia Collaboration}, {Brown},
  {Vallenari}, {Prusti}, {de Bruijne}, {Mignard}, {Drimmel}, {Babusiaux},
  {Bailer-Jones}, {Bastian}, {Biermann}, {Evans}, {Eyer}, {Jansen}, {Jordi},
  {Katz}, {Klioner}, {Lammers}, {Lindegren}, {Luri}, {O'Mullane}, {Panem},
  {Pourbaix}, {Randich}, {Sartoretti}, {Siddiqui}, {Soubiran}, {Valette}, {van
  Leeuwen}, {Walton}, {Aerts}, {Arenou}, {Cropper}, {H{\o}g}, {Lattanzi},
  {Grebel}, {Holland}, {Huc}, {Passot}, {Perryman}, {Bramante}, {Cacciari},
  {Casta{\~n}eda}, {Chaoul}, {Cheek}, {De Angeli}, {Fabricius}, {Guerra},
  {Hern{\'a}ndez}, {Jean-Antoine-Piccolo}, {Masana}, {Messineo}, {Mowlavi},
  {Nienartowicz}, {Ord{\'o}{\~n}ez-Blanco}, {Panuzzo}, {Portell}, {Richards},
  {Riello}, {Seabroke}, {Tanga}, {Th{\'e}venin}, {Torra}, {Els},
  {Gracia-Abril}, {Comoretto}, {Garcia-Reinaldos}, {Lock}, {Mercier},
  {Altmann}, {Andrae}, {Astraatmadja}, {Bellas-Velidis}, {Benson}, {Berthier},
  {Blomme}, {Busso}, {Carry}, {Cellino}, {Clementini}, {Cowell}, {Creevey},
  {Cuypers}, {Davidson}, {De Ridder}, {de Torres}, {Delchambre}, {Dell'Oro},
  {Ducourant}, {Fr{\'e}mat}, {Garc{\'{\i}}a-Torres}, {Gosset}, {Halbwachs},
  {Hambly}, {Harrison}, {Hauser}, {Hestroffer}, {Hodgkin}, {Huckle}, {Hutton},
  {Jasniewicz}, {Jordan}, {Kontizas}, {Korn}, {Lanzafame}, {Manteiga},
  {Moitinho}, {Muinonen}, {Osinde}, {Pancino}, {Pauwels}, {Petit},
  {Recio-Blanco}, {Robin}, {Sarro}, {Siopis}, {Smith}, {Smith}, {Sozzetti},
  {Thuillot}, {van Reeven}, {Viala}, {Abbas}, {Abreu Aramburu}, {Accart},
  {Aguado}, {Allan}, {Allasia}, {Altavilla}, {{\'A}lvarez}, {Alves},
  {Anderson}, {Andrei}, {Anglada Varela}, {Antiche}, {Antoja}, {Ant{\'o}n},
  {Arcay}, {Bach}, {Baker}, {Balaguer-N{\'u}{\~n}ez}, {Barache}, {Barata},
  {Barbier}, {Barblan}, {Barrado y Navascu{\'e}s}, {Barros}, {Barstow},
  {Becciani}, {Bellazzini}, {Bello Garc{\'{\i}}a}, {Belokurov}, {Bendjoya},
  {Berihuete}, {Bianchi}, {Bienaym{\'e}}, {Billebaud}, {Blagorodnova},
  {Blanco-Cuaresma}, {Boch}, {Bombrun}, {Borrachero}, {Bouquillon}, {Bourda},
  {Bouy}, {Bragaglia}, {Breddels}, {Brouillet}, {Br{\"u}semeister},
  {Bucciarelli}, {Burgess}, {Burgon}, {Burlacu}, {Busonero}, {Buzzi}, {Caffau},
  {Cambras}, {Campbell}, {Cancelliere}, {Cantat-Gaudin}, {Carlucci},
  {Carrasco}, {Castellani}, {Charlot}, {Charnas}, {Chiavassa}, {Clotet},
  {Cocozza}, {Collins}, {Costigan}, {Crifo}, {Cross}, {Crosta}, {Crowley},
  {Dafonte}, {Damerdji}, {Dapergolas}, {David}, {David}, {De Cat}, {de Felice},
  {de Laverny}, {De Luise}, {De March}, {de Martino}, {de Souza}, {Debosscher},
  {del Pozo}, {Delbo}, {Delgado}, {Delgado}, {Di Matteo}, {Diakite},
  {Distefano}, {Dolding}, {Dos Anjos}, {Drazinos}, {Duran}, {Dzigan},
  {Edvardsson}, {Enke}, {Evans}, {Eynard Bontemps}, {Fabre}, {Fabrizio},
  {Faigler}, {Falc{\~a}o}, {Farr{\`a}s Casas}, {Federici}, {Fedorets},
  {Fern{\'a}ndez-Hern{\'a}ndez}, {Fernique}, {Fienga}, {Figueras}, {Filippi},
  {Findeisen}, {Fonti}, {Fouesneau}, {Fraile}, {Fraser}, {Fuchs}, {Gai},
  {Galleti}, {Galluccio}, {Garabato}, {Garc{\'{\i}}a-Sedano}, {Garofalo},
  {Garralda}, {Gavras}, {Gerssen}, {Geyer}, {Gilmore}, {Girona}, {Giuffrida},
  {Gomes}, {Gonz{\'a}lez-Marcos}, {Gonz{\'a}lez-N{\'u}{\~n}ez},
  {Gonz{\'a}lez-Vidal}, {Granvik}, {Guerrier}, {Guillout}, {Guiraud},
  {G{\'u}rpide}, {Guti{\'e}rrez-S{\'a}nchez}, {Guy}, {Haigron},
  {Hatzidimitriou}, {Haywood}, {Heiter}, {Helmi}, {Hobbs}, {Hofmann}, {Holl},
  {Holland}, {Hunt}, {Hypki}, {Icardi}, {Irwin}, {Jevardat de Fombelle},
  {Jofr{\'e}}, {Jonker}, {Jorissen}, {Julbe}, {Karampelas}, {Kochoska},
  {Kohley}, {Kolenberg}, {Kontizas}, {Koposov}, {Kordopatis}, {Koubsky},
  {Krone-Martins}, {Kudryashova}, {Kull}, {Bachchan}, {Lacoste-Seris}, {Lanza},
  {Lavigne}, {Le Poncin-Lafitte}, {Lebreton}, {Lebzelter}, {Leccia}, {Leclerc},
  {Lecoeur-Taibi}, {Lemaitre}, {Lenhardt}, {Leroux}, {Liao}, {Licata},
  {Lindstr{\o}m}, {Lister}, {Livanou}, {Lobel}, {L{\"o}ffler}, {L{\'o}pez},
  {Lorenz}, {MacDonald}, {Magalh{\~a}es Fernandes}, {Managau}, {Mann},
  {Mantelet}, {Marchal}, {Marchant}, {Marconi}, {Marinoni}, {Marrese},
  {Marschalk{\'o}}, {Marshall}, {Mart{\'{\i}}n-Fleitas}, {Martino}, {Mary},
  {Matijevi{\v c}}, {Mazeh}, {McMillan}, {Messina}, {Michalik}, {Millar},
  {Miranda}, {Molina}, {Molinaro}, {Molinaro}, {Moln{\'a}r}, {Moniez},
  {Montegriffo}, {Mor}, {Mora}, {Morbidelli}, {Morel}, {Morgenthaler},
  {Morris}, {Mulone}, {Muraveva}, {Musella}, {Narbonne}, {Nelemans},
  {Nicastro}, {Noval}, {Ord{\'e}novic}, {Ordieres-Mer{\'e}}, {Osborne},
  {Pagani}, {Pagano}, {Pailler}, {Palacin}, {Palaversa}, {Parsons}, {Pecoraro},
  {Pedrosa}, {Pentik{\"a}inen}, {Pichon}, {Piersimoni}, {Pineau}, {Plachy},
  {Plum}, {Poujoulet}, {Pr{\v s}a}, {Pulone}, {Ragaini}, {Rago}, {Rambaux},
  {Ramos-Lerate}, {Ranalli}, {Rauw}, {Read}, {Regibo}, {Reyl{\'e}}, {Ribeiro},
  {Rimoldini}, {Ripepi}, {Riva}, {Rixon}, {Roelens}, {Romero-G{\'o}mez},
  {Rowell}, {Royer}, {Ruiz-Dern}, {Sadowski}, {Sagrist{\`a} Sell{\'e}s},
  {Sahlmann}, {Salgado}, {Salguero}, {Sarasso}, {Savietto}, {Schultheis},
  {Sciacca}, {Segol}, {Segovia}, {Segransan}, {Shih}, {Smareglia}, {Smart},
  {Solano}, {Solitro}, {Sordo}, {Soria Nieto}, {Souchay}, {Spagna}, {Spoto},
  {Stampa}, {Steele}, {Steidelm{\"u}ller}, {Stephenson}, {Stoev}, {Suess},
  {S{\"u}veges}, {Surdej}, {Szabados}, {Szegedi-Elek}, {Tapiador}, {Taris},
  {Tauran}, {Taylor}, {Teixeira}, {Terrett}, {Tingley}, {Trager}, {Turon},
  {Ulla}, {Utrilla}, {Valentini}, {van Elteren}, {Van Hemelryck}, {van
  Leeuwen}, {Varadi}, {Vecchiato}, {Veljanoski}, {Via}, {Vicente}, {Vogt},
  {Voss}, {Votruba}, {Voutsinas}, {Walmsley}, {Weiler}, {Weingrill}, {Wevers},
  {Wyrzykowski}, {Yoldas}, {{\v Z}erjal}, {Zucker}, {Zurbach}, {Zwitter},
  {Alecu}, {Allen}, {Allende Prieto}, {Amorim}, {Anglada-Escud{\'e}},
  {Arsenijevic}, {Azaz}, {Balm}, {Beck}, {Bernstein}, {Bigot}, {Bijaoui},
  {Blasco}, {Bonfigli}, {Bono}, {Boudreault}, {Bressan}, {Brown}, {Brunet},
  {Bunclark}, {Buonanno}, {Butkevich}, {Carret}, {Carrion}, {Chemin},
  {Ch{\'e}reau}, {Corcione}, {Darmigny}, {de Boer}, {de Teodoro}, {de Zeeuw},
  {Delle Luche}, {Domingues}, {Dubath}, {Fodor}, {Fr{\'e}zouls}, {Fries},
  {Fustes}, {Fyfe}, {Gallardo}, {Gallegos}, {Gardiol}, {Gebran}, {Gomboc},
  {G{\'o}mez}, {Grux}, {Gueguen}, {Heyrovsky}, {Hoar}, {Iannicola}, {Isasi
  Parache}, {Janotto}, {Joliet}, {Jonckheere}, {Keil}, {Kim}, {Klagyivik},
  {Klar}, {Knude}, {Kochukhov}, {Kolka}, {Kos}, {Kutka}, {Lainey}, {LeBouquin},
  {Liu}, {Loreggia}, {Makarov}, {Marseille}, {Martayan}, {Martinez-Rubi},
  {Massart}, {Meynadier}, {Mignot}, {Munari}, {Nguyen}, {Nordlander}, {Ocvirk},
  {O'Flaherty}, {Olias Sanz}, {Ortiz}, {Osorio}, {Oszkiewicz}, {Ouzounis},
  {Palmer}, {Park}, {Pasquato}, {Peltzer}, {Peralta}, {P{\'e}turaud},
  {Pieniluoma}, {Pigozzi}, {Poels}, {Prat}, {Prod'homme}, {Raison}, {Rebordao},
  {Risquez}, {Rocca-Volmerange}, {Rosen}, {Ruiz-Fuertes}, {Russo}, {Sembay},
  {Serraller Vizcaino}, {Short}, {Siebert}, {Silva}, {Sinachopoulos}, {Slezak},
  {Soffel}, {Sosnowska}, {Strai{\v z}ys}, {ter Linden}, {Terrell}, {Theil},
  {Tiede}, {Troisi}, {Tsalmantza}, {Tur}, {Vaccari}, {Vachier}, {Valles}, {Van
  Hamme}, {Veltz}, {Virtanen}, {Wallut}, {Wichmann}, {Wilkinson}, {Ziaeepour},
  \& {Zschocke}}]{GaiaCollaboration16a}
{Gaia Collaboration}, {Brown}, A.~G.~A., {Vallenari}, A., {et~al.} 2016, \aap,
  595, A2

\bibitem[{{Garc{\'{\i}}a Mu{\~n}oz}(2007)}]{GarciaMunoz07}
{Garc{\'{\i}}a Mu{\~n}oz}, A. 2007, \planss, 55, 1426

\bibitem[{{Glaister}(1991{\natexlab{a}})}]{Glaister91a}
{Glaister}, P. 1991{\natexlab{a}}, Computers Math. Applic., 21, 39

\bibitem[{{Glaister}(1991{\natexlab{b}})}]{Glaister91}
{Glaister}, P. 1991{\natexlab{b}}, Computers Math. Applic., 22, 45

\bibitem[{{Glover} \& {Jappsen}(2007)}]{Glover07}
{Glover}, S.~C.~O. \& {Jappsen}, A.-K. 2007, \apj, 666, 1

\bibitem[{{Gray} {et~al.}(2003){Gray}, {Corbally}, {Garrison}, {McFadden}, \&
  {Robinson}}]{Gray03}
{Gray}, R.~O., {Corbally}, C.~J., {Garrison}, R.~F., {McFadden}, M.~T., \&
  {Robinson}, P.~E. 2003, \aj, 126, 2048

\bibitem[{{G{\"u}del} {et~al.}(2003){G{\"u}del}, {Audard}, {Kashyap}, {Drake},
  \& {Guinan}}]{Guedel03}
{G{\"u}del}, M., {Audard}, M., {Kashyap}, V.~L., {Drake}, J.~J., \& {Guinan},
  E.~F. 2003, \apj, 582, 423

\bibitem[{{Guo}(2011)}]{Guo11}
{Guo}, J.~H. 2011, \apj, 733, 98

\bibitem[{{Guo}(2013)}]{Guo13}
{Guo}, J.~H. 2013, \apj, 766, 102

\bibitem[{{Guo} \& {Ben-Jaffel}(2016)}]{Guo16}
{Guo}, J.~H. \& {Ben-Jaffel}, L. 2016, \apj, 818, 107

\bibitem[{{Haswell} {et~al.}(2012){Haswell}, {Fossati}, {Ayres}, {France},
  {Froning}, {Holmes}, {Kolb}, {Busuttil}, {Street}, {Hebb}, {Collier Cameron},
  {Enoch}, {Burwitz}, {Rodriguez}, {West}, {Pollacco}, {Wheatley}, \&
  {Carter}}]{Haswell12}
{Haswell}, C.~A., {Fossati}, L., {Ayres}, T., {et~al.} 2012, \apj, 760, 79

\bibitem[{{Henry} \& {Winn}(2008)}]{Henry08}
{Henry}, G.~W. \& {Winn}, J.~N. 2008, \aj, 135, 68

\bibitem[{{Holmstr{\"o}m} {et~al.}(2008){Holmstr{\"o}m}, {Ekenb{\"a}ck},
  {Selsis}, {Penz}, {Lammer}, \& {Wurz}}]{Holmstroem08}
{Holmstr{\"o}m}, M., {Ekenb{\"a}ck}, A., {Selsis}, F., {et~al.} 2008, \nat,
  451, 970

\bibitem[{{Huebner} \& {Mukherjee}(2015)}]{Huebner15}
{Huebner}, W.~F. \& {Mukherjee}, J. 2015, \planss, 106, 11

\bibitem[{{H{\"u}nsch} {et~al.}(1999){H{\"u}nsch}, {Schmitt}, {Sterzik}, \&
  {Voges}}]{Huensch99}
{H{\"u}nsch}, M., {Schmitt}, J.~H.~M.~M., {Sterzik}, M.~F., \& {Voges}, W.
  1999, \aaps, 135, 319

\bibitem[{{Ionov} \& {Shematovich}(2015)}]{Ionov15}
{Ionov}, D. \& {Shematovich}, V. 2015, Solar System Research, 49, 339

\bibitem[{{Jensen} {et~al.}(2012){Jensen}, {Redfield}, {Endl}, {Cochran},
  {Koesterke}, \& {Barman}}]{Jensen12}
{Jensen}, A.~G., {Redfield}, S., {Endl}, M., {et~al.} 2012, \apj, 751, 86

\bibitem[{{K{\"a}ppeli} \& {Mishra}(2016)}]{Kaeppeli16}
{K{\"a}ppeli}, R. \& {Mishra}, S. 2016, \aap, 587, A94

\bibitem[{{Khodachenko} {et~al.}(2007){Khodachenko}, {Lammer}, {Lichtenegger},
  {Langmayr}, {Erkaev}, {Grie{\ss}meier}, {Leitner}, {Penz}, {Biernat},
  {Motschmann}, \& {Rucker}}]{Khodachenko07}
{Khodachenko}, M.~L., {Lammer}, H., {Lichtenegger}, H.~I.~M., {et~al.} 2007,
  \planss, 55, 631

\bibitem[{{Khodachenko} {et~al.}(2017){Khodachenko}, {Shaikhislamov}, {Lammer},
  {Kislyakova}, {Fossati}, {Johnstone}, {Arkhypov}, {Berezutsky},
  {Miroshnichenko}, \& {Posukh}}]{Khodachenko17}
{Khodachenko}, M.~L., {Shaikhislamov}, I.~F., {Lammer}, H., {et~al.} 2017,
  \apj, 847, 126

\bibitem[{{Khodachenko} {et~al.}(2015){Khodachenko}, {Shaikhislamov}, {Lammer},
  \& {Prokopov}}]{Khodachenko15a}
{Khodachenko}, M.~L., {Shaikhislamov}, I.~F., {Lammer}, H., \& {Prokopov},
  P.~A. 2015, \apj, 813, 50

\bibitem[{{Kislyakova} {et~al.}(2014){Kislyakova}, {Holmstr{\"o}m}, {Lammer},
  {Odert}, \& {Khodachenko}}]{Kislyakova14b}
{Kislyakova}, K.~G., {Holmstr{\"o}m}, M., {Lammer}, H., {Odert}, P., \&
  {Khodachenko}, M.~L. 2014, Science, 346, 981

\bibitem[{Kohl {et~al.}(2018)Kohl, Salz, Czesla, \& Schmitt}]{Kohl18}
Kohl, S., Salz, M., Czesla, S., \& Schmitt, J. H. M.~M. 2018, \aap, 619, A96

\bibitem[{{Koskinen} {et~al.}(2013){Koskinen}, {Harris}, {Yelle}, \&
  {Lavvas}}]{Koskinen13}
{Koskinen}, T.~T., {Harris}, M.~J., {Yelle}, R.~V., \& {Lavvas}, P. 2013,
  \icarus, 226, 1678

\bibitem[{{Kulow} {et~al.}(2014){Kulow}, {France}, {Linsky}, \&
  {Loyd}}]{Kulow14}
{Kulow}, J.~R., {France}, K., {Linsky}, J., \& {Loyd}, R.~O.~P. 2014, \apj,
  786, 132

\bibitem[{{Lara} {et~al.}(2004){Lara}, {Gonzalez-Esparza}, \&
  {Gopalswamy}}]{Lara04}
{Lara}, A., {Gonzalez-Esparza}, J.~A., \& {Gopalswamy}, N. 2004, Geofisica
  Internacional, 43, 75

\bibitem[{{Lavie} {et~al.}(2017){Lavie}, {Ehrenreich}, {Bourrier}, {Lecavelier
  des Etangs}, {Vidal-Madjar}, {Delfosse}, {Gracia Berna}, {Heng}, {Thomas},
  {Udry}, \& {Wheatley}}]{Lavie17}
{Lavie}, B., {Ehrenreich}, D., {Bourrier}, V., {et~al.} 2017, \aap, 605, L7

\bibitem[{{Lecavelier des Etangs} {et~al.}(2012){Lecavelier des Etangs},
  {Bourrier}, {Wheatley}, {Dupuy}, {Ehrenreich}, {Vidal-Madjar}, {H{\'e}brard},
  {Ballester}, {D{\'e}sert}, {Ferlet}, \& {Sing}}]{LecavelierdesEtangs12}
{Lecavelier des Etangs}, A., {Bourrier}, V., {Wheatley}, P.~J., {et~al.} 2012,
  \aap, 543, L4

\bibitem[{{Lecavelier des Etangs} {et~al.}(2010){Lecavelier des Etangs},
  {Ehrenreich}, {Vidal-Madjar}, {Ballester}, {D{\'e}sert}, {Ferlet},
  {H{\'e}brard}, {Sing}, {Tchakoumegni}, \& {Udry}}]{LecavelierdesEtangs10}
{Lecavelier des Etangs}, A., {Ehrenreich}, D., {Vidal-Madjar}, A., {et~al.}
  2010, \aap, 514, A72

\bibitem[{{Leitzinger} {et~al.}(2014){Leitzinger}, {Odert}, {Greimel},
  {Korhonen}, {Guenther}, {Hanslmeier}, {Lammer}, \&
  {Khodachenko}}]{Leitzinger14}
{Leitzinger}, M., {Odert}, P., {Greimel}, R., {et~al.} 2014, \mnras, 443, 898

\bibitem[{{Leitzinger} {et~al.}(2020){Leitzinger}, {Odert}, {Greimel}, {Vida},
  {Kriskovics}, {Guenther}, {Korhonen}, {Koller}, {Hanslmeier},
  {K{\H{o}}v{\'a}ri}, \& {Lammer}}]{Leitzinger20a}
{Leitzinger}, M., {Odert}, P., {Greimel}, R., {et~al.} 2020, \mnras, 493, 4570

\bibitem[{Lindsay \& Stebbings(2005)}]{Lindsay05}
Lindsay, B.~G. \& Stebbings, R.~F. 2005, Journal of Geophysical Research (Space
  Physics), 110, A12213

\bibitem[{{Linsky} {et~al.}(2014){Linsky}, {Fontenla}, \& {France}}]{Linsky14}
{Linsky}, J.~L., {Fontenla}, J., \& {France}, K. 2014, \apj, 780, 61

\bibitem[{{Linsky} {et~al.}(2010){Linsky}, {Yang}, {France}, {Froning},
  {Green}, {Stocke}, \& {Osterman}}]{Linsky10}
{Linsky}, J.~L., {Yang}, H., {France}, K., {et~al.} 2010, \apj, 717, 1291

\bibitem[{{Llama} \& {Shkolnik}(2015)}]{Llama15}
{Llama}, J. \& {Shkolnik}, E.~L. 2015, \apj, 802, 41

\bibitem[{{Llama} \& {Shkolnik}(2016)}]{Llama16}
{Llama}, J. \& {Shkolnik}, E.~L. 2016, \apj, 817, 81

\bibitem[{{Llama} {et~al.}(2013){Llama}, {Vidotto}, {Jardine}, {Wood}, {Fares},
  \& {Gombosi}}]{Llama13}
{Llama}, J., {Vidotto}, A.~A., {Jardine}, M., {et~al.} 2013, \mnras, 436, 2179

\bibitem[{{MacCormack}(1969)}]{MacCormack69}
{MacCormack}, R.~W. 1969, AIAA, Paper 69-354,

\bibitem[{{MacCormack}(1971)}]{MacCormack71}
{MacCormack}, R.~W. 1971, in Lecture Notes in Physics, Berlin Springer Verlag,
  Vol.~8, Numerical Methods in Fluid DynamicsNumerical Methods in Fluid
  Dynamics, ed. M.~{Holt}, 151--163

\bibitem[{{Marin} \& {Grosso}(2017)}]{Marin17}
{Marin}, F. \& {Grosso}, N. 2017, \apj, 835, 283

\bibitem[{{Matsakos} {et~al.}(2015){Matsakos}, {Uribe}, \&
  {K{\"o}nigl}}]{Matsakos15}
{Matsakos}, T., {Uribe}, A., \& {K{\"o}nigl}, A. 2015, \aap, 578, A6

\bibitem[{{Menager} {et~al.}(2013){Menager}, {Barth{\'e}lemy}, {Koskinen},
  {Lilensten}, {Ehrenreich}, \& {Parkinson}}]{Menager13}
{Menager}, H., {Barth{\'e}lemy}, M., {Koskinen}, T., {et~al.} 2013, \icarus,
  226, 1709

\bibitem[{{Murray-Clay} {et~al.}(2009){Murray-Clay}, {Chiang}, \&
  {Murray}}]{Murray-Clay09}
{Murray-Clay}, R.~A., {Chiang}, E.~I., \& {Murray}, N. 2009, \apj, 693, 23

\bibitem[{{Odert} {et~al.}(2020){Odert}, {Leitzinger}, {Guenther}, \&
  {Heinzel}}]{Odert20}
{Odert}, P., {Leitzinger}, M., {Guenther}, E.~W., \& {Heinzel}, P. 2020,
  \mnras, 494, 3766

\bibitem[{{Odert} {et~al.}(2017){Odert}, {Leitzinger}, {Hanslmeier}, \&
  {Lammer}}]{Odert17}
{Odert}, P., {Leitzinger}, M., {Hanslmeier}, A., \& {Lammer}, H. 2017, \mnras,
  472, 876

\bibitem[{{Owen} \& {Adams}(2014)}]{Owen14}
{Owen}, J.~E. \& {Adams}, F.~C. 2014, \mnras, 444, 3761

\bibitem[{{Patsourakos} \& {Georgoulis}(2016)}]{Patsourakos16}
{Patsourakos}, S. \& {Georgoulis}, M.~K. 2016, \aap, 595, A121

\bibitem[{Peebles(1993)}]{Peebles93}
Peebles, P. J.~E. 1993, Principles of Physical Cosmology (Princeton University
  Press)

\bibitem[{{Penz} {et~al.}(2008){Penz}, {Erkaev}, {Kulikov}, {Langmayr},
  {Lammer}, {Micela}, {Cecchi-Pestellini}, {Biernat}, {Selsis}, {Barge},
  {Deleuil}, \& {L{\'e}ger}}]{Penz08b}
{Penz}, T., {Erkaev}, N.~V., {Kulikov}, Y.~N., {et~al.} 2008, \planss, 56, 1260

\bibitem[{{Petit} {et~al.}(2014){Petit}, {Louge}, {Th{\'e}ado}, {Paletou},
  {Manset}, {Morin}, {Marsden}, \& {Jeffers}}]{Petit14}
{Petit}, P., {Louge}, T., {Th{\'e}ado}, S., {et~al.} 2014, \pasp, 126, 469

\bibitem[{{Pillitteri} {et~al.}(2011){Pillitteri}, {G{\"u}nther}, {Wolk},
  {Kashyap}, \& {Cohen}}]{Pillitteri11}
{Pillitteri}, I., {G{\"u}nther}, H.~M., {Wolk}, S.~J., {Kashyap}, V.~L., \&
  {Cohen}, O. 2011, \apjl, 741, L18

\bibitem[{{Pillitteri} {et~al.}(2015){Pillitteri}, {Maggio}, {Micela},
  {Sciortino}, {Wolk}, \& {Matsakos}}]{Pillitteri15}
{Pillitteri}, I., {Maggio}, A., {Micela}, G., {et~al.} 2015, \apj, 805, 52

\bibitem[{{Pillitteri} {et~al.}(2010){Pillitteri}, {Wolk}, {Cohen}, {Kashyap},
  {Knutson}, {Lisse}, \& {Henry}}]{Pillitteri10}
{Pillitteri}, I., {Wolk}, S.~J., {Cohen}, O., {et~al.} 2010, \apj, 722, 1216

\bibitem[{{Pillitteri} {et~al.}(2014){Pillitteri}, {Wolk}, {Lopez-Santiago},
  {G{\"u}nther}, {Sciortino}, {Cohen}, {Kashyap}, \& {Drake}}]{Pillitteri14}
{Pillitteri}, I., {Wolk}, S.~J., {Lopez-Santiago}, J., {et~al.} 2014, \apj,
  785, 145

\bibitem[{{Poppenhaeger} {et~al.}(2013){Poppenhaeger}, {Schmitt}, \&
  {Wolk}}]{Poppenhaeger13}
{Poppenhaeger}, K., {Schmitt}, J.~H.~M.~M., \& {Wolk}, S.~J. 2013, \apj, 773,
  62

\bibitem[{{Poppenhaeger} \& {Wolk}(2014)}]{Poppenhaeger14}
{Poppenhaeger}, K. \& {Wolk}, S.~J. 2014, \aap, 565, L1

\bibitem[{Route(2019)}]{Route19}
Route, M. 2019, \apj, 872, 79

\bibitem[{{Salz} {et~al.}(2016{\natexlab{a}}){Salz}, {Czesla}, {Schneider}, \&
  {Schmitt}}]{Salz16a}
{Salz}, M., {Czesla}, S., {Schneider}, P.~C., \& {Schmitt}, J.~H.~M.~M.
  2016{\natexlab{a}}, \aap, 586, A75

\bibitem[{{Salz} {et~al.}(2016{\natexlab{b}}){Salz}, {Schneider}, {Czesla}, \&
  {Schmitt}}]{Salz16}
{Salz}, M., {Schneider}, P.~C., {Czesla}, S., \& {Schmitt}, J.~H.~M.~M.
  2016{\natexlab{b}}, \aap, 585, L2

\bibitem[{{Sanz-Forcada} {et~al.}(2010){Sanz-Forcada},
  {Garc{\'{\i}}a-{\'A}lvarez}, {Velasco}, {Solano}, {Ribas}, {Micela}, \&
  {Pollock}}]{Sanz-Forcada10a}
{Sanz-Forcada}, J., {Garc{\'{\i}}a-{\'A}lvarez}, D., {Velasco}, A., {et~al.}
  2010, in IAU Symposium, Vol. 264, IAU Symposium, ed. .~J.-P.~R.
  A.~G.~Kosovichev, A.~H.~Andrei, 478--483

\bibitem[{{Sanz-Forcada} {et~al.}(2011){Sanz-Forcada}, {Micela}, {Ribas},
  {Pollock}, {Eiroa}, {Velasco}, {Solano}, \&
  {Garc{\'{\i}}a-{\'A}lvarez}}]{Sanz-Forcada11}
{Sanz-Forcada}, J., {Micela}, G., {Ribas}, I., {et~al.} 2011, \aap, 532, A6

\bibitem[{{Schwartz} {et~al.}(2017){Schwartz}, {Kashner}, {Jovmir}, \&
  {Cowan}}]{Schwartz17}
{Schwartz}, J.~C., {Kashner}, Z., {Jovmir}, D., \& {Cowan}, N.~B. 2017, \apj,
  850, 154

\bibitem[{{Schwenn} {et~al.}(2006){Schwenn}, {Raymond}, {Alexander},
  {Ciaravella}, {Gopalswamy}, {Howard}, {Hudson}, {Kaufmann}, {Klassen},
  {Maia}, {Munoz-Martinez}, {Pick}, {Reiner}, {Srivastava}, {Tripathi},
  {Vourlidas}, {Wang}, \& {Zhang}}]{Schwenn06a}
{Schwenn}, R., {Raymond}, J.~C., {Alexander}, D., {et~al.} 2006, \ssr, 123, 127

\bibitem[{{Sekiya} {et~al.}(1980){Sekiya}, {Nakazawa}, \&
  {Hayashi}}]{Sekiya80a}
{Sekiya}, M., {Nakazawa}, K., \& {Hayashi}, C. 1980, Prog. Theor. Phys., 64,
  1968

\bibitem[{Semelin {et~al.}(2007)Semelin, Combes, \& Baek}]{Semelin07}
Semelin, B., Combes, F., \& Baek, S. 2007, \aap, 474, 365

\bibitem[{Shaikhislamov {et~al.}(2018)Shaikhislamov, Khodachenko, Lammer,
  Berezutsky, Miroshnichenko, \& Rumenskikh}]{Shaikhislamov18}
Shaikhislamov, I.~F., Khodachenko, M.~L., Lammer, H., {et~al.} 2018, \mnras,
  481, 5315

\bibitem[{{Shaikhislamov} {et~al.}(2016){Shaikhislamov}, {Khodachenko},
  {Lammer}, {Kislyakova}, {Fossati}, {Johnstone}, {Prokopov}, {Berezutsky},
  {Zakharov}, \& {Posukh}}]{Shaikhislamov16}
{Shaikhislamov}, I.~F., {Khodachenko}, M.~L., {Lammer}, H., {et~al.} 2016,
  \apj, 832, 173

\bibitem[{{Shaikhislamov} {et~al.}(2014){Shaikhislamov}, {Khodachenko},
  {Sasunov}, {Lammer}, {Kislyakova}, \& {Erkaev}}]{Shaikhislamov14}
{Shaikhislamov}, I.~F., {Khodachenko}, M.~L., {Sasunov}, Y.~L., {et~al.} 2014,
  \apj, 795, 132

\bibitem[{{Shematovich} {et~al.}(2014){Shematovich}, {Ionov}, \&
  {Lammer}}]{Shematovich14}
{Shematovich}, V.~I., {Ionov}, D.~E., \& {Lammer}, H. 2014, \aap, 571, A94

\bibitem[{{Stassun} {et~al.}(2017){Stassun}, {Collins}, \& {Gaudi}}]{Stassun17}
{Stassun}, K.~G., {Collins}, K.~A., \& {Gaudi}, B.~S. 2017, \aj, 153, 136

\bibitem[{{Stone} \& {Proga}(2009)}]{Stone09}
{Stone}, J.~M. \& {Proga}, D. 2009, \apj, 694, 205

\bibitem[{{Tian} {et~al.}(2005){Tian}, {Toon}, {Pavlov}, \& {De
  Sterck}}]{Tian05}
{Tian}, F., {Toon}, O.~B., {Pavlov}, A.~A., \& {De Sterck}, H. 2005, \apj, 621,
  1049

\bibitem[{{Tilley} {et~al.}(2016){Tilley}, {Harnett}, \& {Winglee}}]{Tilley16}
{Tilley}, M.~A., {Harnett}, E.~M., \& {Winglee}, R.~M. 2016, \apj, 827, 77

\bibitem[{{Trammell} {et~al.}(2011){Trammell}, {Arras}, \& {Li}}]{Trammell11}
{Trammell}, G.~B., {Arras}, P., \& {Li}, Z.-Y. 2011, \apj, 728, 152

\bibitem[{{Trammell} {et~al.}(2014){Trammell}, {Li}, \& {Arras}}]{Trammell14}
{Trammell}, G.~B., {Li}, Z.-Y., \& {Arras}, P. 2014, \apj, 788, 161

\bibitem[{{Tremblin} \& {Chiang}(2013)}]{Tremblin13}
{Tremblin}, P. \& {Chiang}, E. 2013, \mnras, 428, 2565

\bibitem[{{Verner} {et~al.}(1996){Verner}, {Ferland}, {Korista}, \&
  {Yakovlev}}]{Verner96a}
{Verner}, D.~A., {Ferland}, G.~J., {Korista}, K.~T., \& {Yakovlev}, D.~G. 1996,
  \apj, 465, 487

\bibitem[{{Vidal-Madjar} {et~al.}(2004){Vidal-Madjar}, {D{\'e}sert},
  {Lecavelier des Etangs}, {H{\'e}brard}, {Ballester}, {Ehrenreich}, {Ferlet},
  {McConnell}, {Mayor}, \& {Parkinson}}]{Vidal-Madjar04}
{Vidal-Madjar}, A., {D{\'e}sert}, J., {Lecavelier des Etangs}, A., {et~al.}
  2004, \apjl, 604, L69

\bibitem[{{Vidal-Madjar} {et~al.}(2013){Vidal-Madjar}, {Huitson}, {Bourrier},
  {D{\'e}sert}, {Ballester}, {Lecavelier des Etangs}, {Sing}, {Ehrenreich},
  {Ferlet}, {H{\'e}brard}, \& {McConnell}}]{Vidal-Madjar13}
{Vidal-Madjar}, A., {Huitson}, C.~M., {Bourrier}, V., {et~al.} 2013, \aap, 560,
  A54

\bibitem[{{Vidal-Madjar} {et~al.}(2008){Vidal-Madjar}, {Lecavelier des Etangs},
  {D{\'e}sert}, {Ballester}, {Ferlet}, {H{\'e}brard}, \&
  {Mayor}}]{Vidal-Madjar08}
{Vidal-Madjar}, A., {Lecavelier des Etangs}, A., {D{\'e}sert}, J., {et~al.}
  2008, \apjl, 676, L57

\bibitem[{{Vidal-Madjar} {et~al.}(2003){Vidal-Madjar}, {Lecavelier des Etangs},
  {D{\'e}sert}, {Ballester}, {Ferlet}, {H{\'e}brard}, \&
  {Mayor}}]{Vidal-Madjar03}
{Vidal-Madjar}, A., {Lecavelier des Etangs}, A., {D{\'e}sert}, J.-M., {et~al.}
  2003, \nat, 422, 143

\bibitem[{{Waite} {et~al.}(1983){Waite}, {Cravens}, {Kozyra}, {Nagy}, {Atreya},
  \& {Chen}}]{Waite83}
{Waite}, J.~H., {Cravens}, T.~E., {Kozyra}, J., {et~al.} 1983, \jgr, 88, 6143

\bibitem[{{Wood}(2004)}]{Wood04}
{Wood}, B.~E. 2004, Living Rev. Sol. Phys., 1, 2

\bibitem[{{Yashiro} \& {Gopalswamy}(2009)}]{Yashiro09}
{Yashiro}, S. \& {Gopalswamy}, N. 2009, in IAU Symposium, Vol. 257, Universal
  Heliophysical Processes, ed. N.~{Gopalswamy} \& D.~F. {Webb}, 233--243

\bibitem[{{Yashiro} {et~al.}(2004){Yashiro}, {Gopalswamy}, {Michalek},
  {St.~Cyr}, {Plunkett}, {Rich}, \& {Howard}}]{Yashiro04}
{Yashiro}, S., {Gopalswamy}, N., {Michalek}, G., {et~al.} 2004, \jgr, 109,
  A07105

\bibitem[{{Yelle}(2004)}]{Yelle04}
{Yelle}, R.~V. 2004, Icarus, 170, 167

\end{thebibliography}
%


\begin{appendix} 
\section{MacCormack TVD scheme} \label{sec:tvd}
The hydrodynamic model, as described in \citet{Erkaev16}, solves the set of 1D hydrodynamic equations using the MacCormack scheme \citep{MacCormack69}, which is second-order accurate in space and time. This scheme is well suited for supersonic outflows, but may experience instabilities and oscillations if the outflows are largely subsonic, as in the case of the high-gravity planet HD~189733b. Several studies developed corrections to the original MacCormack scheme to give it total variation diminishing (TVD) properties \citep[e.g.,][]{Davis84}. This suppresses spurious numerical oscillations, which may occur at steep gradients and destabilize the code. Specifically, we adopted the method of \citet{Glaister91} designed for nonuniform grids \citep{Glaister91a}.

The system of hydrodynamic equations (Eqs.~\ref{eq:hd1}-\ref{eq:hd3}) was first normalized as described in \citet{Erkaev16} and then written in vector form
\begin{equation}
\frac{\partial U}{\partial t} + \frac{\partial F}{\partial r} = S,
\end{equation}
where $U$ is the vector of the variables, $F$ the fluxes, and $S$ the source terms. Here,
\begin{equation}\label{eq:vec}
U=\begin{pmatrix}\rho r^2\\\rho u r^2\\E_\mathrm{th} r^2\end{pmatrix}, F=\begin{pmatrix}\rho u r^2\\\left(\rho u^2+p \right)r^2\\E_\mathrm{th} u r^2\end{pmatrix}, S=\begin{pmatrix}0\\g\rho r^2 + 2pr\\Q_\mathrm{net}r^2 - p\frac{\partial ur^2}{\partial r}\end{pmatrix}.
\end{equation}
The MacCormack method consists of a predictor step:
\begin{equation}\label{eq:mc1}
U_i^{n+1/2} = U_i - \Delta t \frac{F_{i+1}^n-F_i^n}{r_{i+1}-r_i} + \Delta t S_i^n
,\end{equation}
and a corrector step:
\begin{equation}\label{eq:mc2}
U_i^{n+1} = \frac{1}{2}\left(U_i^n+U_i^{n+1/2}\right) - \frac{\Delta t}{2}\frac{F_i^{n+1/2}-F_{i-1}^{n+1/2}}{r_i-r_{i-1}} + \frac{\Delta t}{2}S_i^{n+1/2},
\end{equation}
where the indices $i$ and $n$ refer to spatial and temporal steps, respectively. The superscript $n+1/2$ refers to quantities evaluated after the predictor step. The time step was updated according to the Courant-Friedrichs-Lewy condition $\Delta t = C_\mathrm{CFL}\Delta x_\mathrm{min}/(|u|+c)_\mathrm{max}$, where $\Delta x_\mathrm{min}$ is the smallest grid size, $c$ is the adiabatic sound speed, $(|u|+c)_\mathrm{max}$ is the maximum wave speed on the grid, and $C_\mathrm{CFL}\le1$ is the Courant-Friedrichs-Lewy number, which we took to be 0.8.

Here, we newly implemented the improved scheme of \citet{MacCormack71} to solve the momentum equation. This is related to the destabilizing effect of cases where the velocities $u_i{<}0$ and $u_{i+1}{>}0$. Since the momentum flux $\rho u^2$ loses the information on the sign of $u$, \citet{MacCormack71} suggested, in such cases, replacing the momentum flux differences $(\rho u^2)_{i+1}-(\rho u^2)_{i}$ with $[(\rho u)_{i+1}-(\rho u)_{i}] (u_{i+1}+u_i)/2$ in Eq.~\ref{eq:mc1}, and with correspondingly adjusted indices in Eq.~\ref{eq:mc2}. This preserves the sign of the flux and removes instabilities.

The TVD corrections from \citet{Glaister91} are implemented as follows. After every corrector step, TVD correction terms are added to the solution,
\begin{equation}
\begin{aligned}
U_{i,\mathrm{TVD}}^{n+1} = U_i^{n+1} 
&+ \left[\hat{K}_{i+1/2}^+ + \hat{K}_{i+1/2}^-\right] \frac{\Delta U_{i+1/2}^n}{\Delta r_i} \\
&- \left[\hat{K}_{i-1/2}^+ + \hat{K}_{i-1/2}^-\right] \frac{\Delta U_{i-1/2}^n}{\Delta r_i},
\end{aligned}
\end{equation}
where $\Delta U_{i+1/2}^n = U_{i+1}^n-U_i^n$, $\Delta U_{i-1/2}^n = U_{i}^n-U_{i-1}^n$, and $\Delta r_i = (r_{i+1}-r_{i-1})/2$. The coefficients $\hat{K}$ are determined by
\begin{equation}
\hat{K}_{i\pm1/2}^{\pm} = \frac{1}{2} |\lambda_{i\pm1/2}|_\mathrm{max}\Delta t\left(1-|\lambda_{i\pm1/2}|_\mathrm{max}\frac{\Delta t}{\Delta r_{i\pm1/2}}\right) \left(1-\psi(M_{i\pm1/2}^{\pm})\right),
\end{equation}
where $|\lambda_{i\pm1/2}|_\mathrm{max} = |u_{i\pm1/2}|+c_{i\pm1/2}$ is the maximum local wave speed and $\psi(M)=\max(0,\min(1,M))$ is the minmod limiter. The coefficients $M^\pm$ are calculated via
\begin{equation}
M_{i\pm1/2}^{\pm} = \frac{\left(y_{i\pm1/2-s}^{\pm} - |\lambda_{i\pm1/2-s}|_\mathrm{max}\frac{\Delta t}{\Delta r_{i\pm1/2-s}}\right)}{\left(1- |\lambda_{i\pm1/2}|_\mathrm{max}\frac{\Delta t}{\Delta r_{i\pm1/2}}\right)} \frac{\left(\Delta U_{i\pm1/2-s}^n \cdot \Delta U_{i\pm1/2}^n\right)}{\left(\Delta U_{i\pm1/2}^n \cdot \Delta U_{i\pm1/2}^n\right)},
\end{equation}
where $s$ denotes the sign function of the superscript\footnote{Meaning $s=1$ for superscript ``$+$'' and $s=-1$ for superscript ``$-$''.} and the expression $(*\cdot*)$ refers to the inner product of the difference vectors $\Delta U$. The parameter
\begin{equation}
y_{i\pm1/2}^{\pm} = \frac{\Delta r_{i\pm1/2+s}}{\Delta r_{i\pm1/2}}
\end{equation}
is the ratio of successive grid spacings.

\subsection{Modification of the momentum equation}
Due to the highly subsonic nature of the outflow for planets with high gravity, the solution of the velocity profile is affected by unphysical behavior (e.g., negative values, $u$ increasing towards the planet) close to the lower boundary of the computational domain. The situation improves if the code is run for very long times (because of the small time steps), but this is inconvenient. The problem is (partly) due to the solution method. We tried to solve the momentum equation in a nonconservative form, treating both the pressure and gravity terms as source terms, meaning the vector components (Eq.~\ref{eq:vec}) are modified to $F_2=\rho u^2r^2$ and $S_2=-\rho r^2\partial \Phi/\partial r-r^2\partial p/\partial r$. In the limit of $u\rightarrow0$, hydrostatic equilibrium should be reached, and the two terms in $S_2$ should cancel each other out. Schemes that maintain hydrostatic equilibrium in this limit are called well-balanced schemes. We modified the discretization of the source terms according to the suggestions of \citet{Kaeppeli16}. Firstly, the derivatives of $\Phi$ and $p$ in $S_2$ are discretized the same way as the respective advection terms in Eqs.~\ref{eq:mc1} and \ref{eq:mc2}; secondly, the prefactors of these derivatives ($\rho r^2$, $r^2$) are calculated as averages of their values at the grid points involved in the derivatives. For Eq.~\ref{eq:mc1}, this yields
\begin{equation}\label{eq:wb}
-\frac{(\rho r^2)_{i+1}+(\rho r^2)_{i}}{2} \frac{\Phi_{i+1}-\Phi_{i}}{r_{i+1}-r_{i}} - \frac{(r^2)_{i+1}+(r^2)_{i}}{2} \frac{p_{i+1}-p_{i}}{r_{i+1}-r_{i}},
\end{equation}
and similar for Eq.~\ref{eq:mc2} with correspondingly adjusted indices. This procedure improved the resulting velocity profiles. We checked that this implementation did not affect the results by comparing runs for a lower gravity planet (where this modification is not necessary) using both the conservative and the modified formulation of the momentum equation.

\subsection{Mass and energy conservation}\label{sec:cons}
\begin{figure*}
\includegraphics[width=\columnwidth]{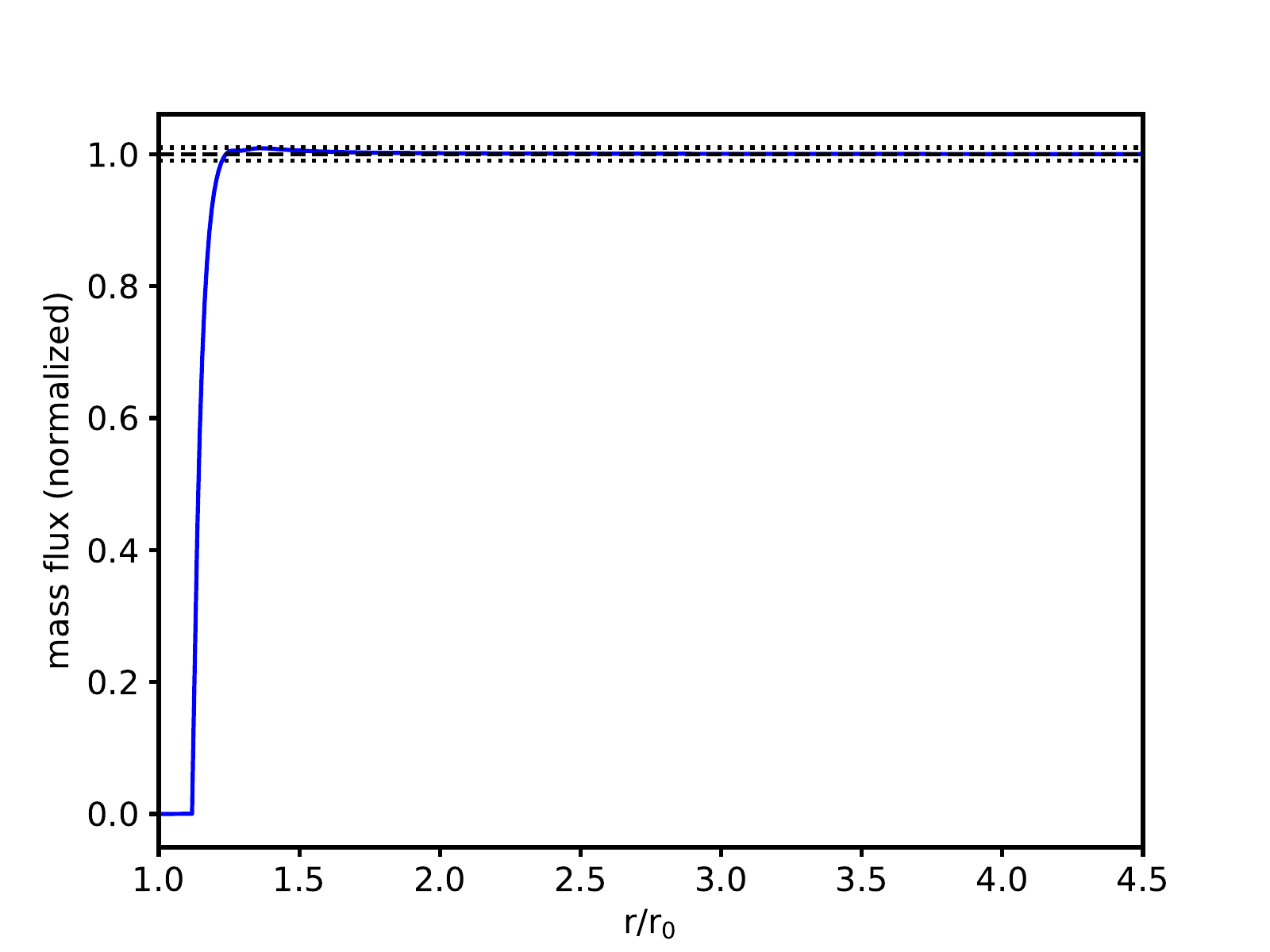}
\includegraphics[width=\columnwidth]{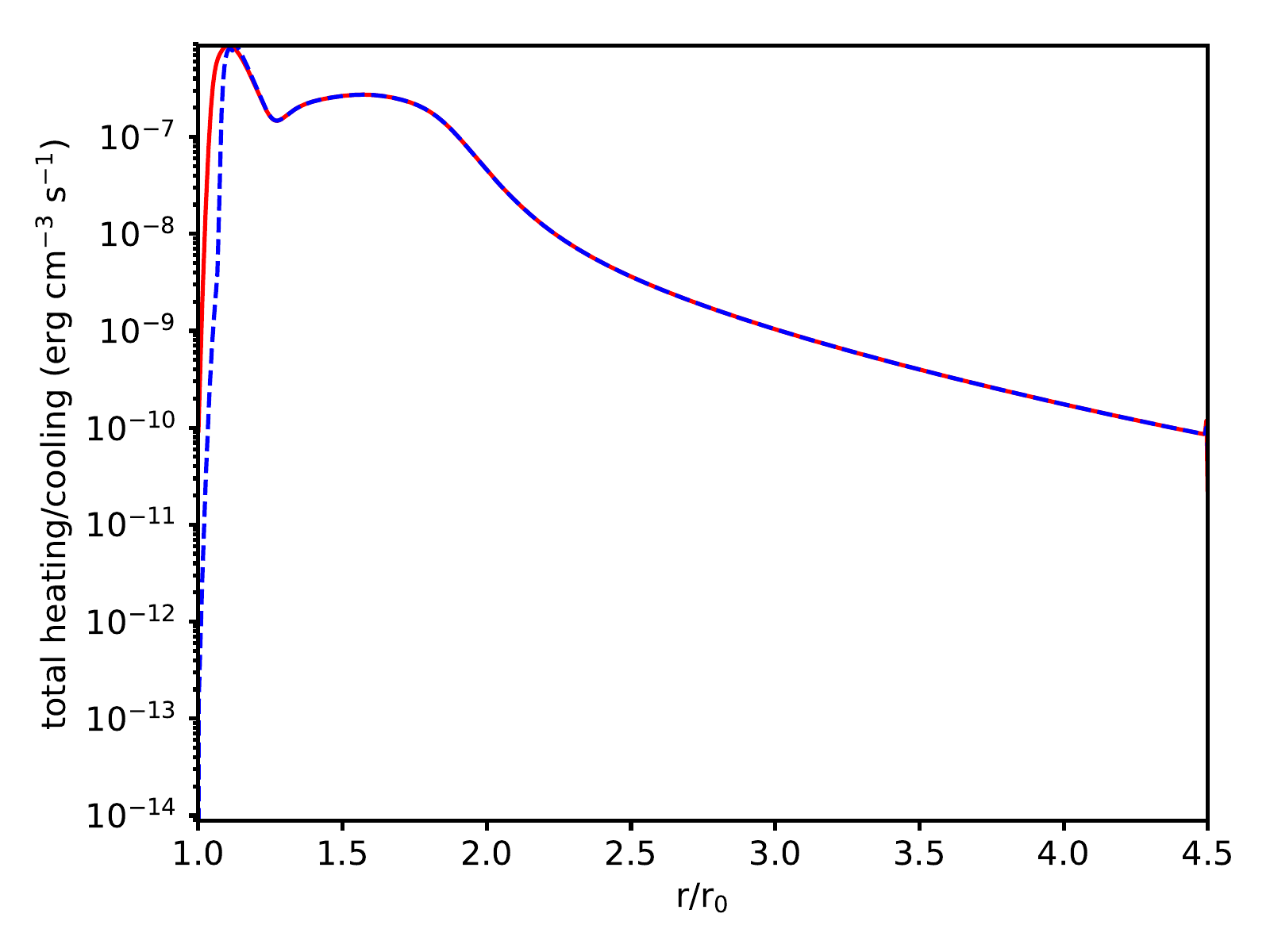}
\caption{Demonstration of the conservation properties our code in a converged solution (\citetalias{Sanz-Forcada11} run). Left: normalized mass flux (blue), the dashed and dotted black lines indicate unity and $\pm$1\%, respectively. Right: total heating (red) and cooling rates (dashed blue).}
\label{fig:cons}
\end{figure*}

Figure~\ref{fig:cons} shows the mass flux $\rho u r^2$ normalized to its value at the upper boundary as a function of $r$, as well as the total heating and cooling rates for the \citetalias{Sanz-Forcada11} run. One can see that both mass and energy are well conserved in our model. We adopted the convergence criterion that the momentum flux should be conserved to 1\% throughout the computational domain. This is only violated close to the lower boundary where the highly subsonic velocities introduce numerical difficulties for this planet. Also, the total heating and cooling rates agree well, indicating good energy conservation properties of the code, except close to the planet where the numerical errors lead to slightly lower adiabatic cooling than expected. 

\section{Comparison of 1D and 2D heating}\label{sec:1d2d}
\begin{figure}
\includegraphics[width=\columnwidth]{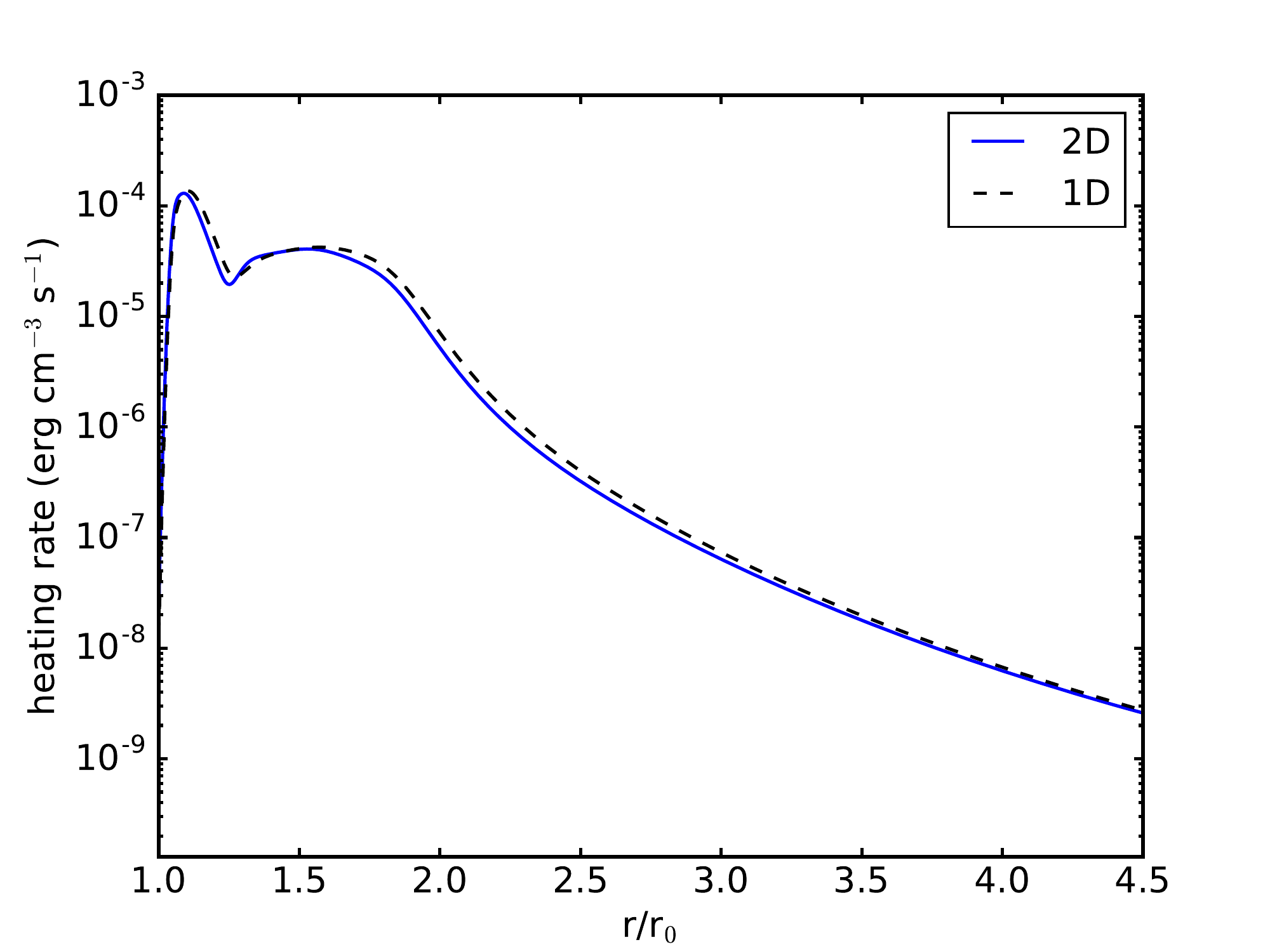}
\caption{Comparison of 2D and 1D XUV volume heating rates.}
\label{fig:1d2d}
\end{figure}

Figure~\ref{fig:1d2d} shows a comparison between Eq.~\ref{eq:qxuv} and the 2D heating function \citep[e.g.,][]{Erkaev16}. For the latter, it was assumed that the density profile along the star-planet line is valid for all other directions. This is approximately valid if the Roche lobe is far from the planet, and the planet is not tidally locked so that the stellar XUV flux is distributed efficiently over the whole atmosphere. However, for a hot Jupiter, these assumptions are likely not valid, and a 3D model would be needed. Moreover, since we introduced a wavelength-dependent heating function, which raises the computational demand, we used a more simple 1D heating function. Adopting the neutral hydrogen density profile from the \citetalias{Sanz-Forcada11} run, we calculated the 2D volume heating rate a posteriori and compared it with Eq.~\ref{eq:qxuv}. One can see that for the adopted planetary and stellar parameters, $\alpha=4$ represents the best match and was consequently used in all runs.

\section{Molecular hydrogen}\label{sec:h2}
\begin{figure}
\includegraphics[width=\columnwidth]{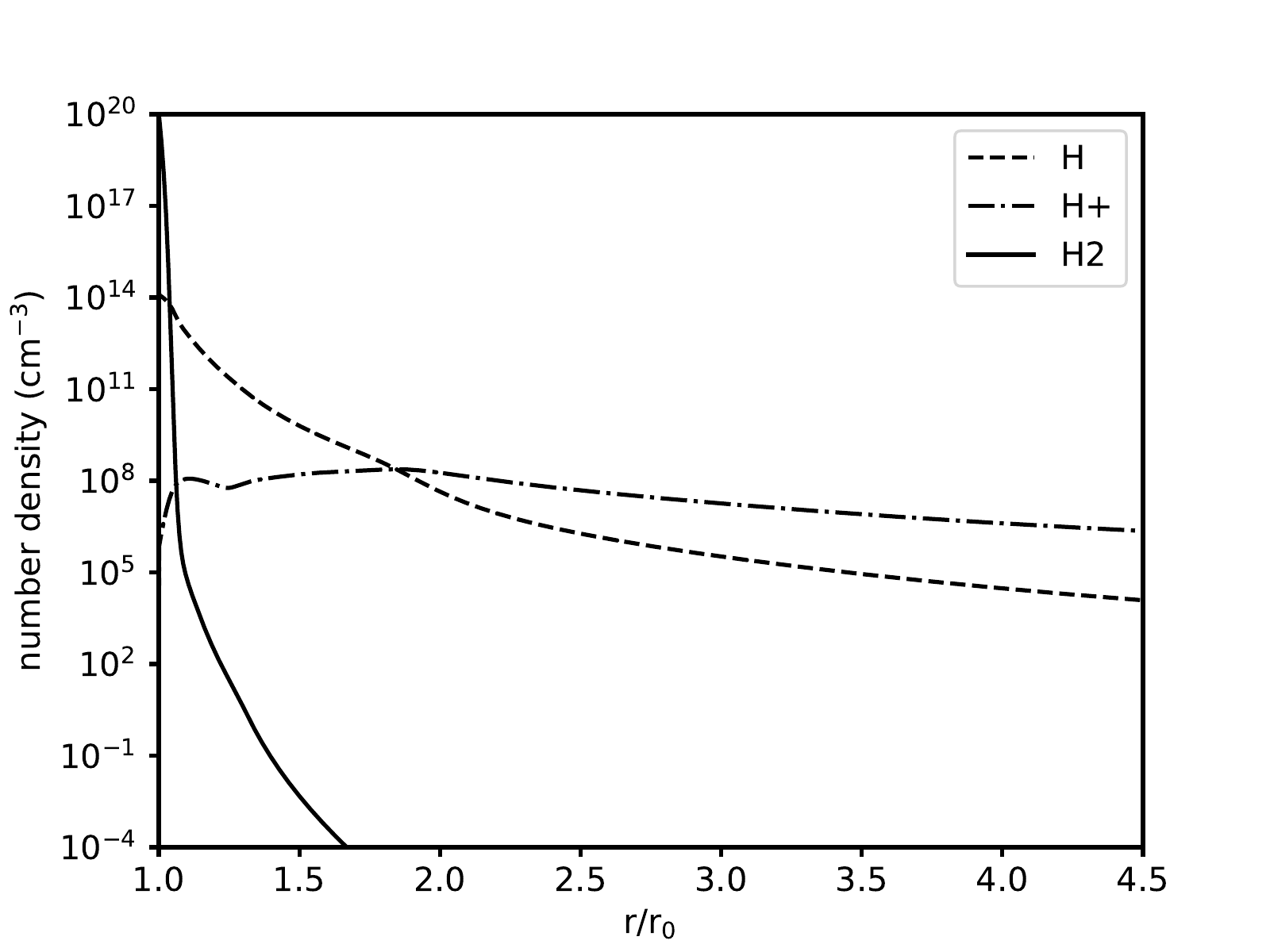}
\caption{Number densities of H$_2$, H, and H$^+$, where the latter two are taken from the \citetalias{Sanz-Forcada11} run and the H$_2$ density is calculated with Eq.~\ref{eq:h2}.}
\label{fig:h2}
\end{figure}

Here, we test if neglecting molecular hydrogen could have an impact on the results. We calculated the equilibrium abundance of H$_2$, which can be justified considering that it will only be abundant close to $R_\mathrm{p}$ where velocities are very small. The H$_2$ density can be written as
\begin{equation}\label{eq:h2}
n_\mathrm{H_2} = \frac{\gamma_\mathrm{H} n n_\mathrm{H}^2}{\nu_\mathrm{H_2} + \nu_\mathrm{diss}n},
\end{equation}
where $\gamma_\mathrm{H}=8\times10^{-33}(300/T)^{0.6}$ is the rate of the reaction producing H$_2$ (H+H$\rightarrow$H$_2$) and $\nu_\mathrm{diss}=1.5\times10^{-9}\exp(-49000/T)$ the thermal dissociation rate \citep{Yelle04, Erkaev16}. The photoionization rate $\nu_\mathrm{H_2}$ is calculated after Eq.~\ref{eq:phion}, but replacing $\sigma_\mathrm{ion}$ with that for H$_2$ \citep{Huebner15}. The resulting number density profile of H$_2$, along with the densities of H atoms and protons for the \citetalias{Sanz-Forcada11} run are shown in Fig.~\ref{fig:h2}. One can see that H$_2$ would only be the dominating species at distances ${\lesssim}1.04R_\mathrm{p}$, very close to the optical radius. The main part of the upper atmosphere is composed of H and H$^+$. This is consistent with \citet{Guo16}, who found the same H$_2$/H transition radius with their hydrodynamic model (which does include hydrogen chemistry) and that the planetary wind is dominated by H and H$^+$. Therefore, we do not expect that inclusion of H$_2$ would noticeably change our results.

\section{Stellar wind parameters}
\begin{figure*}
$\begin{array}{cc}
\includegraphics[width=\columnwidth]{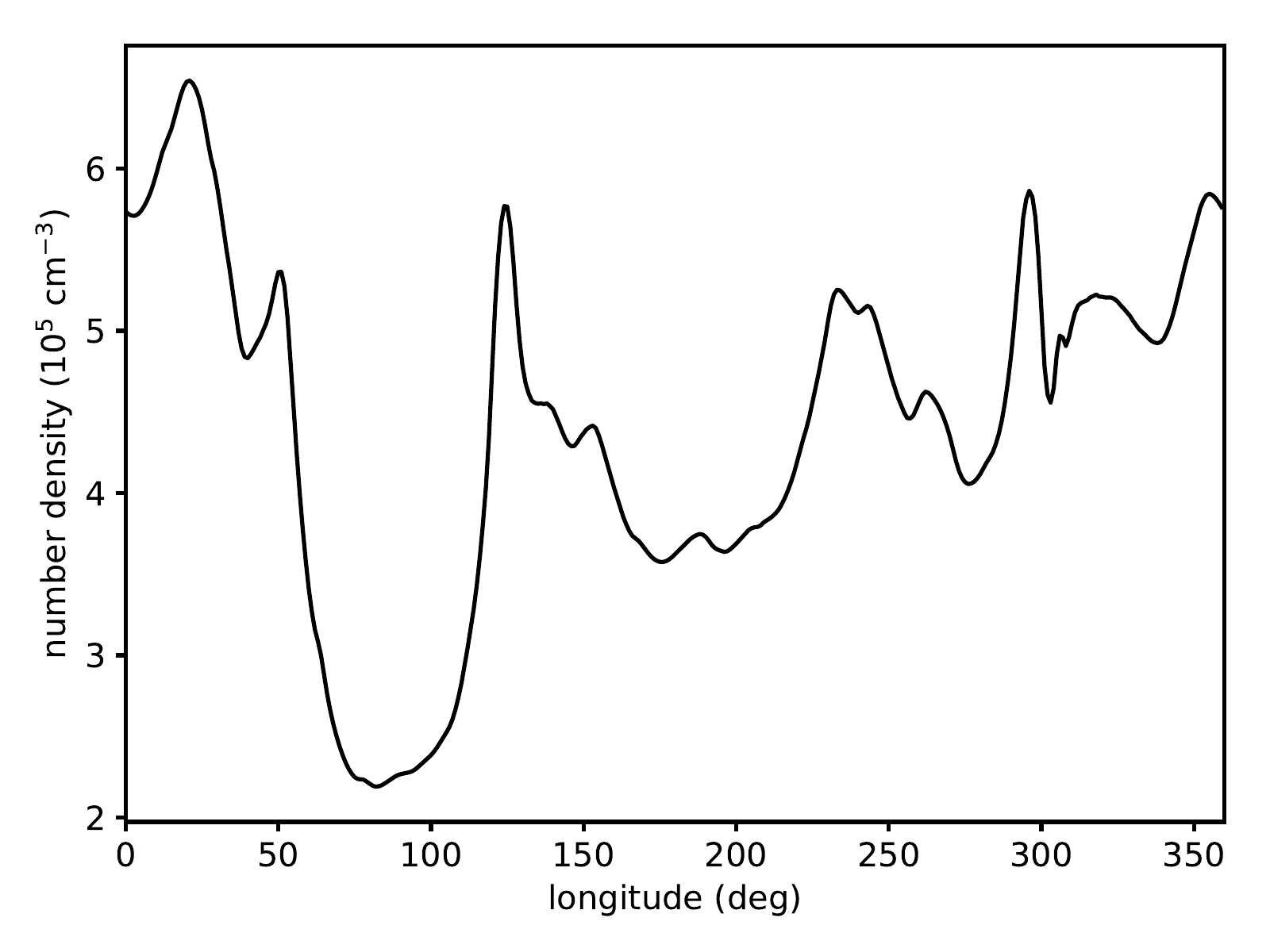} &
\includegraphics[width=\columnwidth]{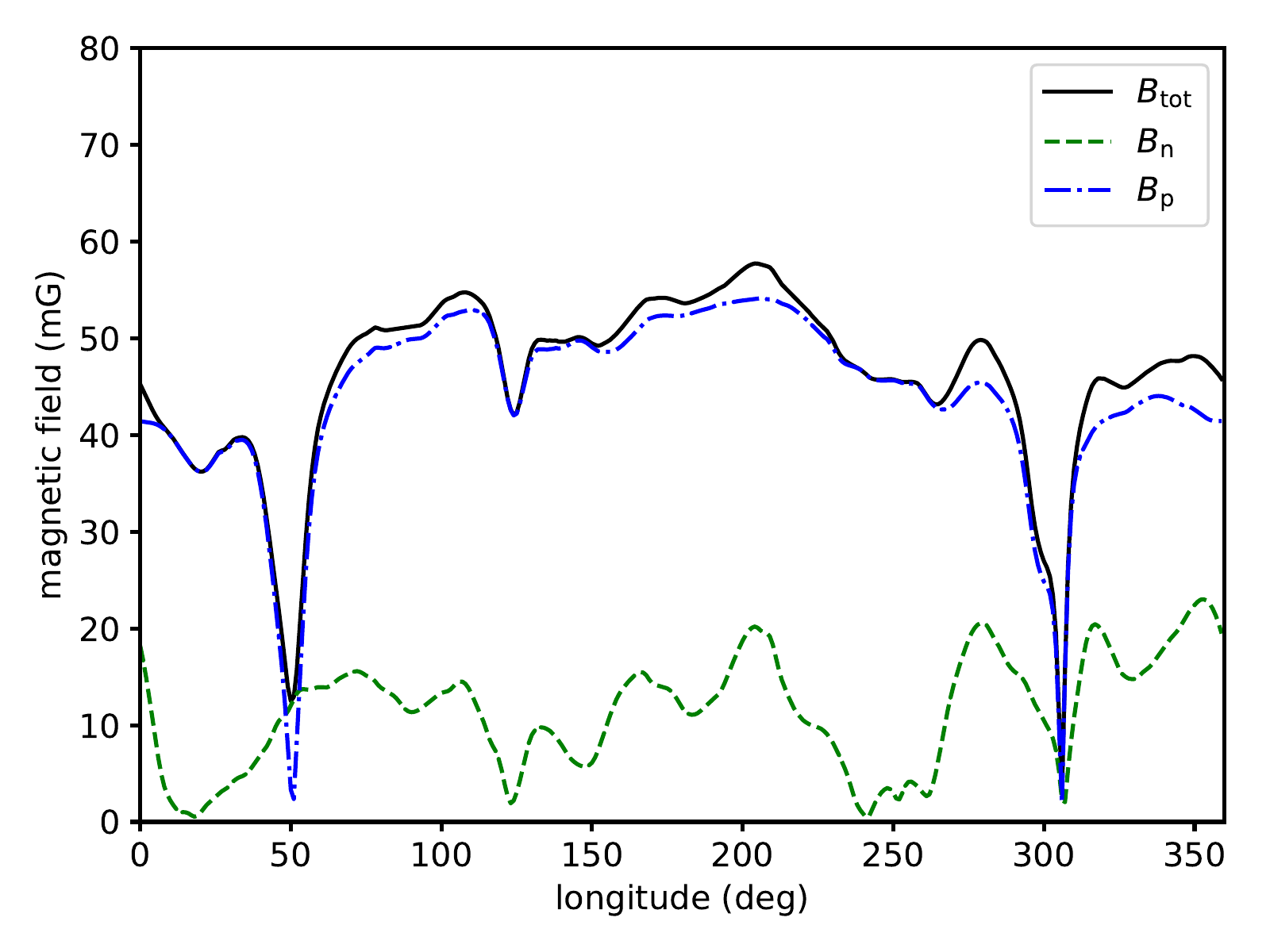} \\
\includegraphics[width=\columnwidth]{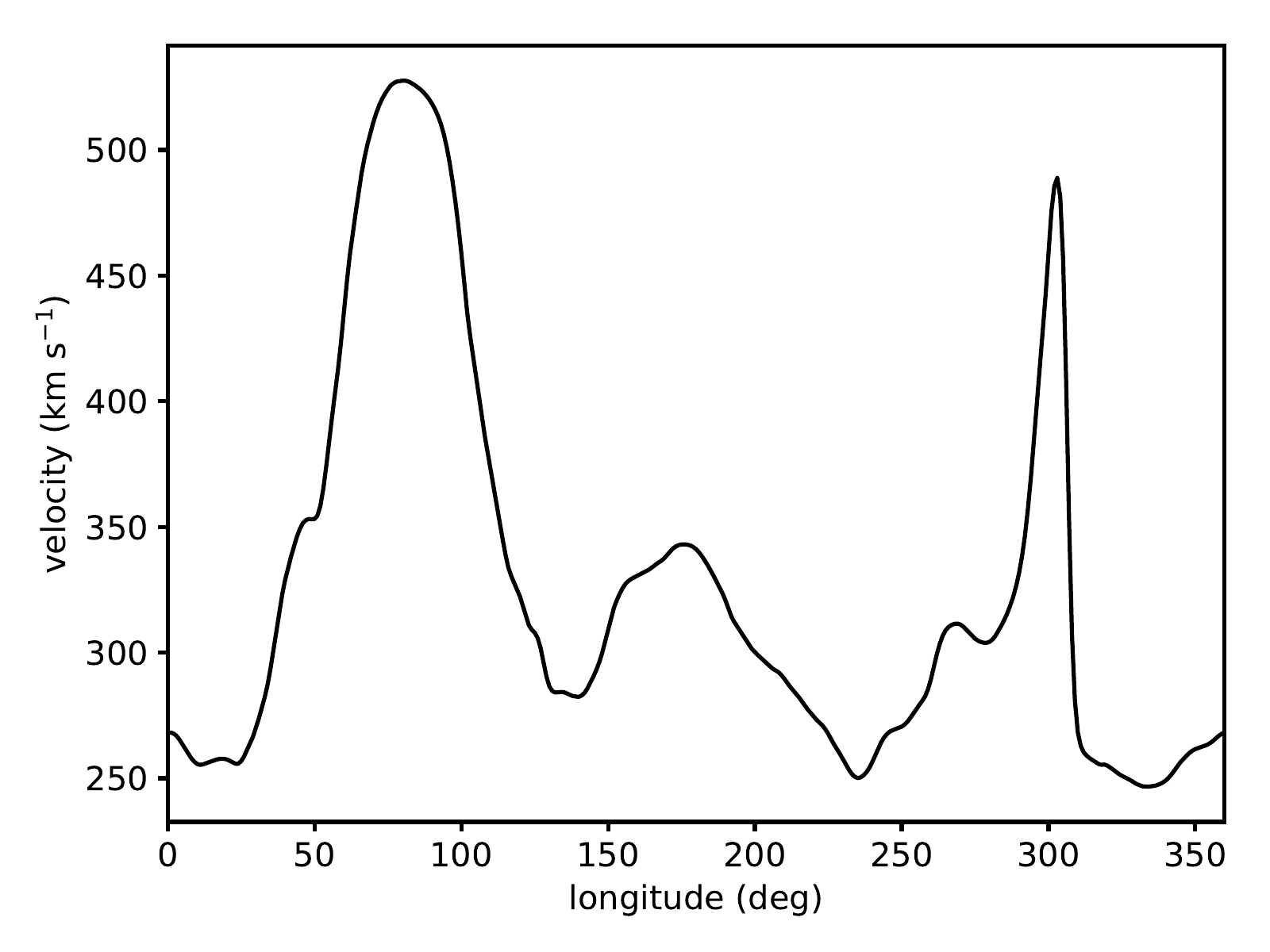} &
\includegraphics[width=\columnwidth]{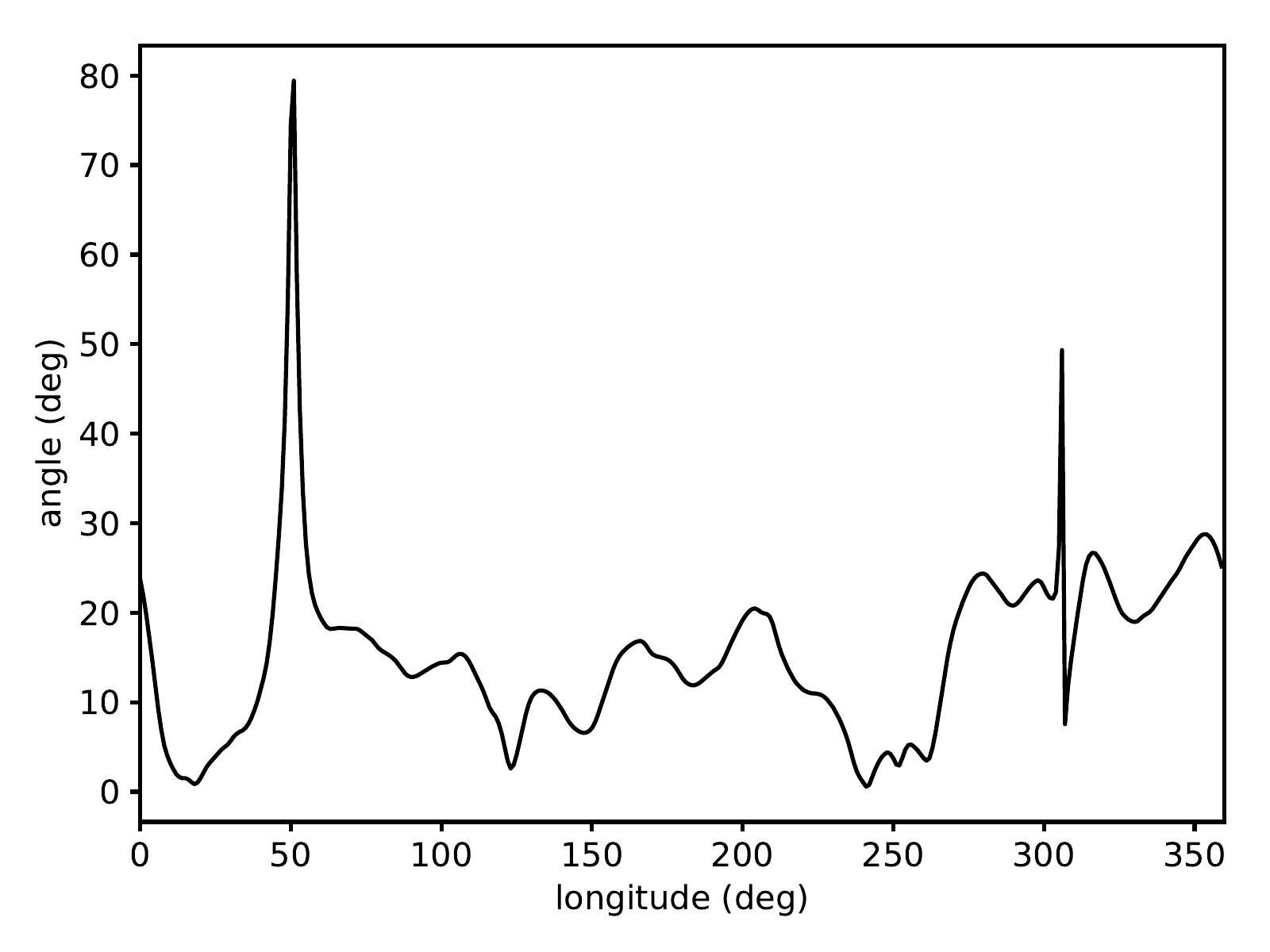} \\
\includegraphics[width=\columnwidth]{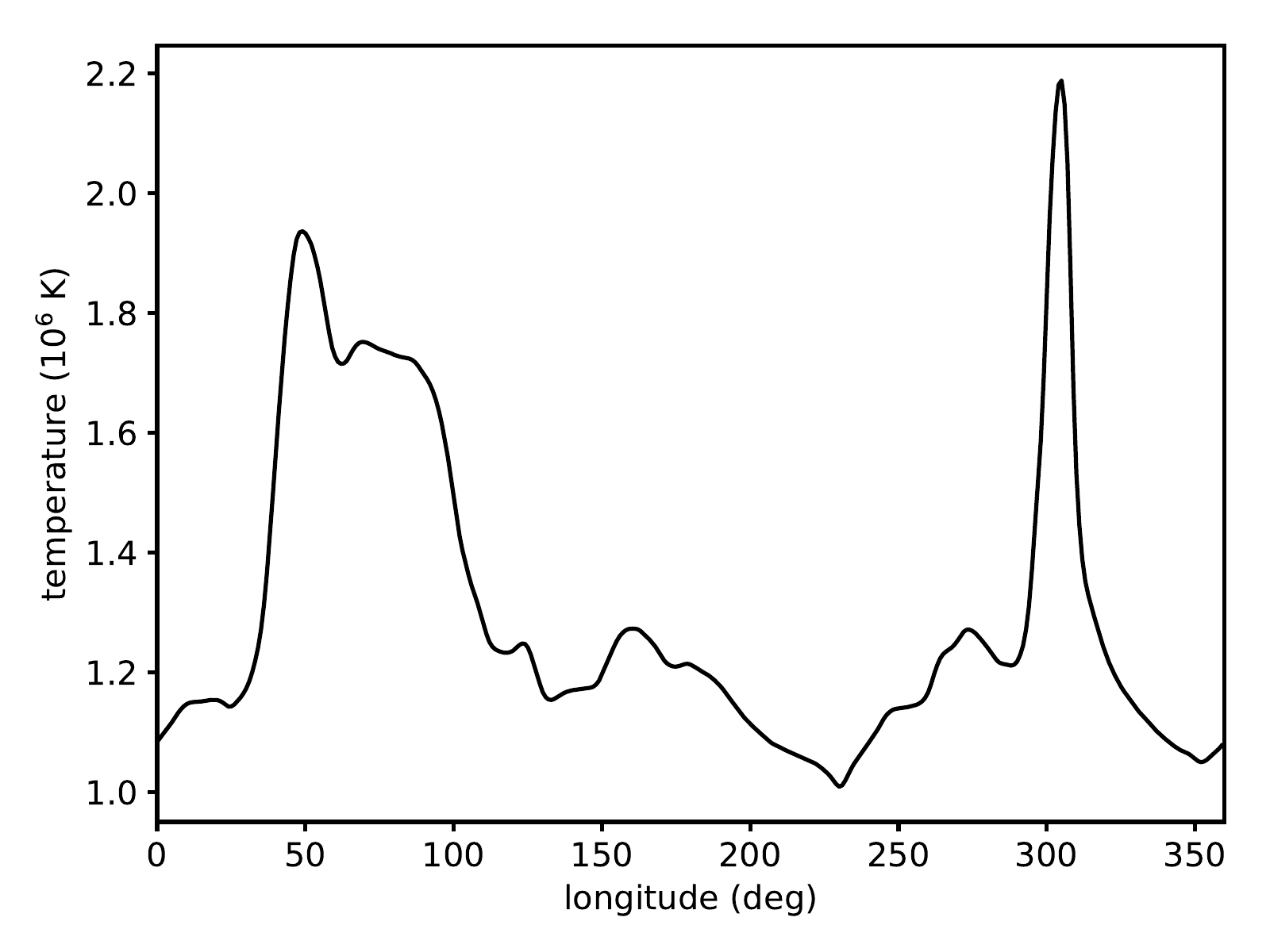} &
\includegraphics[width=\columnwidth]{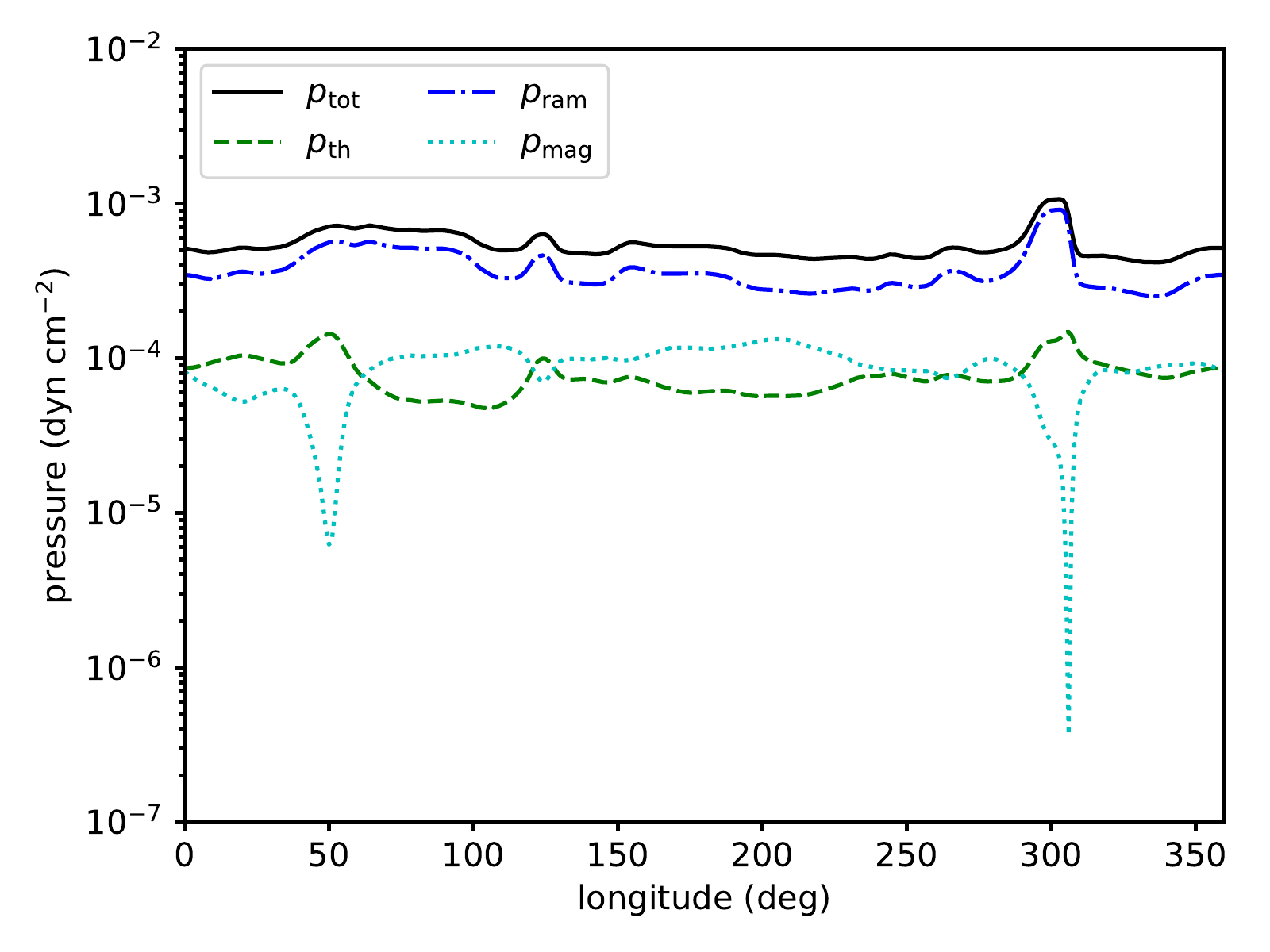}
\end{array}$
\caption{Variation of the stellar wind parameters along the orbit of HD~189733b. Shown are the number density, velocity, temperature, strength of the total magnetic field and its components (parallel and normal to $V_\mathrm{sw}$), the angle between the magnetic field and the star-planet line, and the total pressure with its components (thermal, magnetic, and ram pressures).}
\label{fig:sw}
\end{figure*}

In Fig.~\ref{fig:sw}, we show the stellar wind parameters along the orbit based on the measured magnetic field map \citep{Fares10} and a 3D wind model \citep{Llama13}.

\end{appendix}
\end{document}